\documentclass[aps,prb,reprint]{revtex4-1}
\usepackage{amsfonts,amssymb,amsmath,bm,graphicx,color}
\usepackage[colorlinks=true,citecolor=blue,linkcolor=blue,urlcolor=blue]{hyperref}
\allowdisplaybreaks[4]
\newcommand\cc{{\rm c.c.}}
\newcommand\hb[1]{\hat {\bm #1}}
\DeclareMathOperator\diag{diag}
\DeclareMathOperator\polyln{Li}
\DeclareMathOperator\tr{tr}
\begin{document}
\title{Anomalous Thermal Hall Effect in a Disordered Weyl Ferromagnet}
\author{Atsuo Shitade}
\affiliation{RIKEN Center for Emergent Matter Science, 2-1 Hirosawa, Wako, Saitama 351-0198, Japan}
\date{\today}
\begin{abstract}
  We investigate the electric and thermal transport properties in a disordered Weyl ferromagnet on an equal footing by using the Keldysh formalism in curved spacetime.
  In particular, we calculate the anomalous thermal Hall conductivity, which consists of the Kubo formula and the heat magnetization, without relying on the Wiedemann-Franz law.
  We take nonmagnetic impurities into account within the self-consistent $T$-matrix approximation
  and reproduce the Wiedemann-Franz law for the extrinsic Fermi-surface and intrinsic Fermi-sea terms, respectively.
  This is the first step towards a unified theory of the anomalous Hall effect at finite temperature, where we should take into account both disorder and interactions.
\end{abstract}
\maketitle
\section{Introduction} \label{sec:intro}
The thermal Hall effect (THE) is a heat analog of the Hall effect,
namely, the heat current flows perpendicular to a temperature gradient when the time-reversal symmetry is broken.
Recently, the THE has been experimentally studied in ferromagnetic metals,~\cite{PhysRevLett.100.016601,PhysRevB.81.054414,PhysRevB.79.100404}
ferromagnetic insulators,~\cite{Onose16072010,PhysRevB.85.134411} and frustrated magnets in a magnetic field.~\cite{Hirschberger03042015,PhysRevLett.115.106603,Watanabe02082016}
In ferromagnetic metals,~\cite{PhysRevLett.100.016601,PhysRevB.81.054414,PhysRevB.79.100404}
the THE is utilized in order to investigate effects of inelastic scattering on the anomalous Hall effect (AHE).
According to the Wiedemann-Franz law,
the Lorenz ratio $L^{ij} \equiv \kappa^{ij}/T \sigma^{ij}$ coincides with the universal Lorenz number $L_0 \equiv \pi^2 k_{\rm B}^2/3 e^2$ at $T = 0$,
but decreases as temperature increases and inelastic scattering dominates elastic scattering.
Here $\sigma^{ij}$ and $\kappa^{ij}$ are the electric and thermal (Hall) conductivities, respectively, and $T$ is temperature.
Therefore the Lorenz ratio is a quantitative indicator of inelastic scattering.
In ferromagnetic insulators~\cite{Onose16072010,PhysRevB.85.134411} and frustrated magnets,~\cite{Hirschberger03042015,PhysRevLett.115.106603,Watanabe02082016}
the THE is a powerful probe of chargeless spin excitations.
Further experimental progress in the THE is expected in the future.

Theoretically, it is difficult to calculate the thermal Hall conductivity owing to the fundamental reasons below.
First, the Kubo formula is justified by introducing Luttinger's gravitational potential since a temperature gradient is not mechanical but statistical force.~\cite{PhysRev.135.A1505}
Second, the heat magnetization correction to the Kubo formula is necessary because the heat current as well as the density matrix is perturbed
by a gravitational potential.~\cite{0022-3719-10-12-021,PhysRevB.55.2344,PhysRevLett.106.197202,PhysRevB.84.184406,PhysRevLett.107.236601,Shitade01122014}
Once the Berry-phase formula has been established,~\cite{PhysRevLett.106.197202,PhysRevB.84.184406,PhysRevLett.107.236601,Shitade01122014}
we can easily calculate the thermal Hall conductivity in the clean noninteracting limit.
However, as mentioned above, it is highly desired to calculate the thermal Hall conductivity in disordered or interacting systems.
The formula derived in the celebrated paper by Smr\v{c}ka and St\v{r}eda,
who first pointed out the necessity of the magnetization correction, can be applied to generic disordered systems in principle,~\cite{0022-3719-10-12-021} but not easily in practice.
Practically, we take into account disorder or interactions by a perturbation theory, e.g., Feynman diagrams.
In this context, the ordinary Hall effect in disordered metals was studied with Feynman diagrams.~\cite{Fukuyama01091969,Fukuyama01121969}
Recently, we found a gravitational vector potential by gauging the time translation symmetry and derived the Keldysh formalism in curved spacetime.~\cite{Shitade01122014}
With this formalism, we can calculate the Kubo formula and the heat magnetization in disordered or interacting systems within a perturbation theory.

In this paper, we first apply the Keldysh formalism in curved spacetime to a three-dimensional disordered Weyl ferromagnet, which is known to exhibit the AHE.
The main purpose is to show the Wiedemann-Franz law by a practical perturbation theory with respect to the impurity strength.
We take nonmagnetic impurities into account within the self-consistent $T$-matrix approximation and calculate $\sigma^{ij}$ and $T \kappa^{ij}$ on an equal footing.
We begin with brief reviews of gauging the time translation symmetry and the Keldysh formalism in curved spacetime.
The details of calculation are given in Appendices.
This paper serves as the first step towards a unified theory of the AHE at finite temperature, in which we should take into account both disorder and interactions.

Before going to the main part, let us review some theoretical aspects of the AHE.~\cite{RevModPhys.82.1539}
The AHE occurs in ferromagnets with spin-orbit coupling.
In the first microscopic theory by Karplus and Luttinger,~\cite{PhysRev.95.1154} the AHE was attributed to the anomalous velocity, which is now understood as the Berry curvature.~\cite{RevModPhys.82.1959}
Since this mechanism works in the absence of disorder and depends only on the band structure, leading to $\sigma^{xy} \propto (\sigma^{xx})^0$, it is called the intrinsic mechanism.
Soon later, it was revealed that the skew scattering, i.e., the antisymmetric part of the transition probability due to disorder, also gives rise to the AHE.~\cite{Smit1955877,Smit195839}
An important consequence of the skew-scattering mechanism is $\sigma^{xy} \propto \sigma^{xx}$.
Another mechanism due to disorder was proposed by Berger.~\cite{PhysRevB.2.4559,PhysRevB.5.1862}
This side-jump mechanism results in $\sigma^{xy} \propto (\sigma^{xx})^0$, which is the same as the intrinsic mechanism.
The most systematic way to deal with all these mechanisms on an equal footing is based on the Keldysh formalism.~\cite{PTP.116.61,PTP.117.415}
One important advantage of this formalism is that we can go beyond a finite-order perturbation theory with respect to the impurity strength with preserving the gauge invariance.
By using this formalism, some authors investigated a two-dimensional disordered Rashba ferromagnet within the self-consistent $T$-matrix approximation
and found the crossover between the superclean and moderately dirty regimes,
where the skew-scattering and intrinsic contributions are dominant, respectively.~\cite{PhysRevLett.97.126602,PhysRevB.77.165103,PhysRevB.79.195129}
We also note that the nontrivial scaling relation $\sigma^{xy} \propto (\sigma^{xx})^{1.6}$ was found in the dirty regime.~\cite{PhysRevLett.97.126602,PhysRevB.77.165103,PhysRevB.79.195129}

Let us also review three-dimensional Weyl semimetals~\cite{Hosur2013857,0953-8984-27-11-113201} which we investigate in this paper.
Weyl semimetals have pairs of Weyl nodes at which two bands touch each other.
Typically, Weyl semimetals are realized from Dirac semimetals by breaking the time-reversal~\cite{PhysRevB.83.205101,PhysRevLett.107.127205} or inversion symmetry.~\cite{1367-2630-9-9-356,PhysRevB.85.035103}
In particular, Weyl semimetals without the time-reversal symmetry exhibit the AHE.~\cite{PhysRevLett.107.127205,PhysRevB.86.115133,PhysRevB.88.245107,PhysRevLett.113.187202}
When the Fermi level is sufficiently close to Weyl nodes, the AHE is purely intrinsic,~\cite{PhysRevLett.113.187202}
which results from chiral anomaly.~\cite{PhysRevB.86.115133,PhysRevB.88.245107}
On the other hand, when the Fermi level is away from Weyl nodes, the extrinsic mechanisms due to disorder can be relevant.
In this paper, we mean such doped Weyl semimetals without the time-reversal symmetry by Weyl ferromagnets
because we are interested in both the extrinsic and intrinsic mechanisms of the AHE and THE.
Although the THE in Weyl ferromagnets was studied with the semiclassical Boltzmann theory,~\cite{PhysRevB.90.165115,PhysRevB.93.035116} we go beyond the relaxation time approximation used in the literature.

Hereafter we assign the Latin ($a, b, \dots = {\hat 0}, {\hat 1}, \dots, {\hat d}$) and Greek ($\mu, \nu, \dots = 0, 1, \dots, d$) alphabets
to locally flat and global coordinates, respectively.
We follow the Einstein convention, which implies summation over the spacetime dimension $D = d + 1$ when an index appears twice in a single term.
The Minkowski metric is taken as $\eta_{ab} = \diag (-1, +1, \dots, +1)$.
The Planck constant and the charge are denoted by $\hbar$ and $q = -e$, respectively, and the speed of light and the Boltzmann constant are put to $c = k_{\rm B} = 1$.
The upper or lower signs in equations correspond to boson or fermion.

\section{Gauging the Time Translation Symmetry} \label{sec:gauge}
First, we review the idea of gravitational vector potential by gauging the time translation symmetry.~\cite{Shitade01122014}
It is well known that electromagnetic scalar and vector potentials are introduced by gauging the U($1$) symmetry.
By gauging the time translation symmetry as well as the U($1$) symmetry, we can calculate the electric and thermal (Hall) conductivities on an equal footing.
Furthermore, the heat magnetization is defined by the magnetic component of the field strength induced from a gravitational vector potential.

According to Noether's first theorem, a global continuous symmetry is associated with the conservation law of a current.
In the presence of the global time translation symmetry, namely,
when the matter action is invariant under global time translation $x^{\prime 0} = x^0 + \epsilon^0$, $x^{\prime i} = x^i$, and $\psi^{\prime}(x^{\prime}) = \psi(x)$,
the conserved energy current density $T^{\mu}_{\phantom{\mu} 0} = (H, J_{\rm E}^i)$ is constructed as
\begin{equation}
  T^{\mu}_{\phantom{\mu} 0}
  = \frac{\partial L}{\partial (\partial_{\mu} \psi)} \partial_0 \psi + \partial_0 \psi^{\dag} \frac{\partial L}{\partial (\partial_{\mu} \psi^{\dag})} - \delta^{\mu}_{\phantom{\mu} 0} L, \label{eq:en1}
\end{equation}
with $L[\psi^{(\dag)}(x), \partial_{\mu} \psi^{(\dag)}(x)]$ being the matter Lagrangian density in flat spacetime.

We impose the local time translation symmetry by following the gauge principle.
Indeed, the local symmetries of time translation, space translation, and Lorentz transformations are imposed by the general covariance principle.
Under local time translation $x^{\prime 0} = x^0 + \epsilon^0(x)$, $x^{\prime i} = x^i$, and $\psi^{\prime}(x^{\prime}) = \psi(x)$, the matter action is no longer invariant
because $\partial_{\mu} \psi$ transforms as $\partial_{\mu}^{\prime} \psi^{\prime}(x^{\prime}) = (\partial x^{\nu}/\partial x^{\prime \mu}) \partial_{\nu} \psi(x)$.
To preserve the local time translation symmetry of the total action,
we introduce a gauge field called vielbein $e^{\hat 0}_{\phantom{\hat 0} \mu}$,
which transforms as $e^{\prime {\hat 0}}_{\phantom{\prime {\hat 0}} \mu}(x^{\prime}) = (\partial x^{\nu}/\partial x^{\prime \mu}) e^{\hat 0}_{\phantom{\hat 0} \nu}(x)$,
by replacing $\partial_a$ with the covariant derivative $D_a \equiv e_a^{\phantom{a} \mu} (\partial_{\mu} - i q A_{\mu}/\hbar)$.
Here $e_a^{\phantom{a} \mu}$ is the inverse of $e^a_{\phantom{a} \mu} = (e^{\hat 0}_{\mu}, e^{\hat k}_{\phantom{\hat k} \mu} = \delta^{\hat k}_{\phantom{\hat k} \mu})$.
We also introduce a U($1$) gauge field $A_{\mu}$ by gauging the U($1$) symmetry.
According to Noether's second theorem, the energy current density is also expressed by
\begin{equation}
  T^{\mu}_{\phantom{\mu} {\hat 0}}
  = -\frac{\partial L}{\partial e^{\hat 0}_{\phantom{\hat 0} \mu}}. \label{eq:en2}
\end{equation}
Note that $L[\psi^{(\dag)}(x), \partial_{\mu} \psi^{(\dag)}(x), e^{\hat 0}_{\phantom{\hat 0} \mu}(x)]$ is the matter Lagrangian density in curved spacetime
and contains the determinant of vielbein $e \equiv \det e^a_{\phantom{a} \mu}$,
which comes from the covariant volume element $d^D x e$.
From Eq.~\eqref{eq:en2}, $e^{\hat 0}_{\phantom{\hat 0} 0} = 1 + \phi_{\rm g}$ is coupled to the Hamiltonian density
and is identified as Luttinger's gravitational potential $\phi_{\rm g}$ which describes nonuniform temperature.~\cite{PhysRev.135.A1505}
$e^{\hat 0}_{\phantom{\hat 0} i}$ is coupled to the energy current density and is called gravitational vector potential.~\cite{Shitade01122014}

Vielbein induces a field strength called torsion.
In analogy to electromagnetic fields induced from a U($1$) gauge field, we define torsional electromagnetic fields by
\begin{subequations} \begin{align}
  T^{\hat 0}_{\phantom{\hat 0} j0}
  = & \partial_j e^{\hat 0}_{\phantom{\hat 0} 0} - \partial_0 e^{\hat 0}_{\phantom{\hat 0} j}, \label{eq:torsion1a} \\
  T^{\hat 0}_{\phantom{\hat 0} ij}
  = & \partial_i e^{\hat 0}_{\phantom{\hat 0} j} - \partial_j e^{\hat 0}_{\phantom{\hat 0} i}. \label{eq:torsion1b}
\end{align} \label{eq:torsion1}\end{subequations}
A torsional electric field describes a temperature gradient.
In fact, if we choose a particular gauge of $\partial_0 e^{\hat 0}_{\phantom{\hat 0} j} = 0$,
Eq.~\eqref{eq:torsion1a} is reduced to $T^{\hat 0}_{\phantom{\hat 0} j0} = \partial_j e^{\hat 0}_{\phantom{\hat 0} 0} = \partial_j \phi_{\rm g}$.
On the other hand, a torsional magnetic field is conjugate to the heat magnetization as a magnetic field is conjugate to the orbital magnetization.
Thus, the Kubo formula for the thermal (Hall) conductivity and the heat magnetization are calculated by~\cite{Shitade01122014}
\begin{subequations} \begin{align}
  T {\tilde \kappa}^{{\hat \imath} {\hat \jmath}}
  \equiv & \frac{\partial J_{\rm Q}^{\hat \imath}}{\partial (-T^{\hat 0}_{\phantom{\hat 0} {\hat \jmath} {\hat 0}})}, \label{eq:def1a} \\
  \beta M_{{\rm Q} {\hat k}}
  \equiv & -\frac{1}{2} \epsilon_{{\hat \imath} {\hat \jmath} {\hat k}}
  \frac{\partial \Omega}{\partial (-\beta^{-1} T^{\hat 0}_{\phantom{\hat 0} {\hat \imath} {\hat \jmath}})}, \label{eq:def1b}
\end{align} \label{eq:def}\end{subequations}
in which $\Omega \equiv E - T S - \mu N$ is the grand potential.
Note that Eq.~\eqref{eq:def} is defined by $T^{\hat 0}_{\phantom{\hat 0} ab} \equiv e_a^{\phantom{a} \mu} e_b^{\phantom{b} \nu} T^{\hat 0}_{\phantom{\hat 0} \mu \nu}$
but not by $T^{\hat 0}_{\phantom{\hat 0} \mu \nu}$,
because $T^{\hat 0}_{\phantom{\hat 0} ab}$ is invariant under local time translation $x^{\prime 0} = x^0 + \epsilon^0(x)$ and $x^{\prime i} = x^i$.
This is obvious since $e_a^{\phantom{a} \mu}$ and $T^{\hat 0}_{\phantom{\hat 0} \mu \nu}$ transform as
$e_{\phantom{\prime} a}^{\prime \phantom{a} \mu}(x^{\prime}) = (\partial x^{\prime \mu}/\partial x^{\nu}) e_a^{\phantom{a} \nu}(x)$
and $T^{\prime {\hat 0}}_{\phantom{\prime {\hat 0}} \mu \nu}(x^{\prime})
= (\partial x^{\lambda}/\partial x^{\prime \mu}) (\partial x^{\sigma}/\partial x^{\prime \nu}) T^{\hat 0}_{\phantom{\hat 0} \lambda \sigma}(x)$, respectively.
Nonetheless, we explicitly show the invariance of $T^{\hat 0}_{\phantom{\hat 0} ab}$ in Appendix~\ref{app:inv}.
Therefore, Eq.~\eqref{eq:def1a} is gauge-invariant extension of Luttinger's idea.~\cite{PhysRev.135.A1505}
Equation~\eqref{eq:def1b} is in analogy with the thermodynamic definition of the orbital magnetization.~\cite{PhysRevLett.99.197202,PhysRevB.84.205137,PhysRevB.86.214415,PhysRevB.90.125132}
Regarding the factor $\beta$, we follow Ref.~\onlinecite{PhysRevLett.107.236601} and below find that it is consistent with the proper thermal Hall conductivity.
We also note that the spatial component of vielbein $e^{\hat k}_{\phantom{\hat k} \mu}$ is introduced by gauging the space translation symmetry,
and the current responses to torsion $T^{\hat k}_{\phantom{\hat k} \mu \nu}$ were discussed
in Refs.~\onlinecite{PhysRevLett.107.075502,Hidaka01012013,PhysRevD.88.025040,PhysRevD.90.105004,PhysRevB.90.134510,PhysRevLett.116.166601}.

\section{Keldysh Formalism in Curved Spacetime} \label{sec:keldysh}
Next, we review the Keldysh formalism in curved spacetime,~\cite{Shitade01122014} which is natural extension of that in electromagnetic fields.~\cite{PTP.116.61,PTP.117.415}
The Keldysh formalism is a powerful quantum-mechanical tool to treat nonlinear and nonequilibrium phenomena.~\cite{9780521874991,9780521760829}
In the Wigner representation in terms of the center-of-mass coordinate $X$ and the relative canonical momentum $p$, convolution in the Dyson equation is represented by the Moyal product.
In the presence of electromagnetic fields and torsion, the Keldysh Green's function depends on $X$ through the mechanical momentum $\pi_a(X, p) \equiv e_a^{\phantom{a} \mu}(X) [p_{\mu} - q A_{\mu}(X)]$.
By changing the variables $(X, p) \to (X, \pi)$, the Moyal product is replaced by the star product which contains electromagnetic fields $F_{ab}$~\cite{PTP.116.61,PTP.117.415}
and torsion $T^{\hat 0}_{\phantom{\hat 0} ab}$.~\cite{Shitade01122014}
Thus we can develop a systematic perturbation theory with respect to electromagnetic fields and torsion.

We start from the Keldysh formalism in flat spacetime.~\cite{9780521874991,9780521760829}
The Keldysh Green's function ${\hat G}$ is defined on the closed time path and satisfies the Dyson equation.
In terms of real time, it is expressed in the matrix form by
\begin{equation}
  {\hat G}
  = \begin{bmatrix}
        G^{\rm R} & 2 G^< \\
        0 & G^{\rm A}
      \end{bmatrix}, \label{eq:green1}
\end{equation}
in which $G^{\rm R}$, $G^{\rm A}$, and $G^<$ are the retarded, advanced, and lesser Green's functions,
\begin{subequations} \begin{align}
  G^{\rm R}(x_1, x_2)
  \equiv & -i/\hbar \theta(t_1 - t_2) \left\langle [\psi(x_1), \psi^{\dag}(x_2)]_{\mp} \right\rangle, \label{eq:green2a} \\
  G^{\rm A}(x_1, x_2)
  \equiv & +i/\hbar \theta(t_2 - t_1) \left\langle [\psi(x_1), \psi^{\dag}(x_2)]_{\mp} \right\rangle, \label{eq:green2b} \\
  G^<(x_1, x_2)
  \equiv & \mp i/\hbar \left\langle \psi^{\dag}(x_2) \psi(x_1) \right\rangle, \label{eq:green2c}
\end{align} \label{eq:green2}\end{subequations}
respectively.
The Dyson equation is expressed by
\begin{equation}
  {\cal L}(x_1) {\hat G}(x_1, x_2) - {\hat \Sigma} \ast {\hat G}(x_1, x_2)
  = \delta(x_1 - x_2), \label{eq:dyson1}
\end{equation}
where ${\cal L}$ and ${\hat \Sigma}$ are the Lagrangian density and the self-energy, respectively, and $\ast$ indicates convolution.

In the presence of vielbein, the volume element $d^D x$ is replaced by the covariant volume element $d^D x e$.
Correspondingly, the Dyson equation is modified as~\cite{9780521278584,9780521877879}
\begin{equation}
  {\cal L}(x_1) {\hat G}(x_1, x_2) - {\hat \Sigma} \ast^e {\hat G}(x_1, x_2)
  = \delta^e(x_1, x_2), \label{eq:dyson2}
\end{equation}
in which the modified convolution and $\delta$ function are given by
\begin{subequations} \begin{align}
  {\hat A} \ast^e {\hat B}(x_1, x_2)
  \equiv & \int d^D x_3 e(x_3) {\hat A}(x_1, x_3) {\hat B}(x_3, x_2), \label{eq:det1a} \\
  \delta^e(x_1, x_2)
  \equiv & e^{-1/2}(x_1) \delta(x_1 - x_2) e^{-1/2}(x_2), \label{eq:det1b}
\end{align} \label{eq:det1}\end{subequations}
respectively.
Although Eq.~\eqref{eq:dyson2} looks complicated,
by introducing a weight-one tensor density ${\hb A}(x_1, x_2) \equiv e^{1/2}(x_1) {\hat A}(x_1, x_2) e^{1/2}(x_2)$ from a tensor ${\hat A}(x_1, x_2)$,
it is symbolically reduced to the Dyson equation in flat spacetime as
\begin{equation}
  {\cal L}(x_1) {\hb G}(x_1, x_2) - {\hb \Sigma} \ast {\hb G}(x_1, x_2)
  = \delta(x_1 - x_2). \label{eq:dyson3}
\end{equation}
Do not confuse a tensor density with a vector in three dimensions, which is indicated by an arrow below.

In order to deal with convolution, we move to the Wigner representation~\cite{9780521874991,9780521760829}
\begin{equation}
  {\hat A}(X, p)
  \equiv \int d^D x e^{-i p_a x^a/\hbar} {\hat A}(X + x/2, X - x/2), \label{eq:wigner}
\end{equation}
in terms of the center-of-mass coordinate $X = (x_1 + x_2)/2$ and the relative canonical momentum $p$ conjugate to $x = x_1 - x_2$.
In this representation, convolution is represented
by the noncommutative Moyal product as ${\hat A} \ast {\hat B}(X, p) = {\hat A}(X, p) e^{i \hbar {\cal F}_0/2} {\hat B}(X, p)$.
${\cal F}_0 = \partial_{X^a} \otimes \partial_{p_a} - \partial_{p_a} \otimes \partial_{X^a}$ is the Poisson bracket in flat spacetime known in analytical mechanics.
Note that the partial derivative on the left (right) side of $\otimes$ acts on the left (right) only.
See Appendix~\ref{app:moyal} for derivation.
As a result, the Dyson equation Eq.~\eqref{eq:dyson3} is simply represented by
\begin{equation}
  ({\cal L} - {\hb \Sigma}) \ast {\hb G}(X, p)
  = 1. \label{eq:dyson4}
\end{equation}

Now we take into account electromagnetic fields~\cite{PTP.116.61,PTP.117.415} and torsion.~\cite{Shitade01122014}
These gauge fields are introduced through the covariant derivative $D_a = e_a^{\phantom{a} \mu} (\partial_{\mu} - i q A_{\mu}/\hbar)$,
which corresponds to the so-called Peierls substitution $\pi_a(X, p) \equiv e_a^{\phantom{a} \mu}(X) [p_{\mu} - q A_{\mu}(X)]$ in the Wigner representation.
Thus the Dyson equation in curved spacetime is given by
\begin{equation}
  ({\cal L} - {\hb \Sigma}) \ast {\hb G}(X, \pi(X, p))
  = 1. \label{eq:dyson5}
\end{equation}
However, since $\pi(X, p)$ is a complicated function of $X$, it is better to change variables $(X, p) \to (X, \pi)$,
which leads to replacing the Moyal product $\ast = e^{i \hbar {\cal F}_0/2}$ by the star product $\star \simeq 1 + i \hbar {\cal F}/2$.
By using the chain rule of partial derivatives, the Poisson bracket in curved spacetime is given by
\begin{align}
  {\cal F}
  = & e_a^{\phantom{a} \mu} (\partial_{X^{\mu}} \otimes \partial_{\pi_a} - \partial_{\pi_a} \otimes \partial_{X^{\mu}}) \notag \\
  & + (q F_{ab} + T^{\hat 0}_{\phantom{\hat 0} ab} \pi_{\hat 0}) \partial_{\pi_a} \otimes \partial_{\pi_b}, \label{eq:poisson}
\end{align}
and the Dyson equation in curved spacetime Eq.~\eqref{eq:dyson5} is modified as
\begin{equation}
  ({\cal L} - {\hb \Sigma}) \star {\hb G}(X, \pi)
  = 1. \label{eq:dyson6}
\end{equation}
See Appendix~\ref{app:moyal} for calculation details.
Although we do not derive the exact form of the star product,
Eqs.~\eqref{eq:poisson} and \eqref{eq:dyson6} complete the first-order perturbation theory of Green's functions with respect to static uniform electromagnetic fields and torsion.

\begin{widetext}
\section{Perturbation Theory with Respect to Electromagnetic Fields and Torsion} \label{sec:perturb}
This is the final section for reviews.
In this Section, we derive the first-order perturbation theory with respect to static uniform electromagnetic fields~\cite{PTP.116.61} and torsion.~\cite{Shitade01122014}
We can neglect the $X$ dependence in the Green's functions and the first term in Eq.~\eqref{eq:poisson}.
We also expand the Keldysh Green's function and the self-energy with respect to $F_{ab}$ and $-T^{\hat 0}_{\phantom{\hat 0} ab}$ as
\begin{subequations} \begin{align}
  \star
  = & 1 + i \hbar [q F_{ab} + (-T^{\hat 0}_{\phantom{\hat 0} ab}) (-\pi_{\hat 0})] \partial_{\pi_a} \otimes \partial_{\pi_b}/2, \label{eq:gem1a} \\
  {\hb G}
  = & {\hat G}_0 + \hbar F_{ab} {\hat G}_{F_{ab}}/2 + \hbar (-T^{\hat 0}_{\phantom{\hat 0} ab}) {\hat G}_{-T^{\hat 0}_{\phantom{\hat 0} ab}}/2, \label{eq:gem1b} \\
  {\hb \Sigma}
  = & {\hat \Sigma}_0 + \hbar F_{ab} {\hat \Sigma}_{F_{ab}}/2 + \hbar (-T^{\hat 0}_{\phantom{\hat 0} ab}) {\hat \Sigma}_{-T^{\hat 0}_{\phantom{\hat 0} ab}}/2. \label{eq:gem1c}
\end{align} \label{eq:gem1}\end{subequations}
By substituting these equations into Eq.~\eqref{eq:dyson6}, we obtain ${\hat G}_0 = ({\cal L} - {\hat \Sigma}_0)^{-1}$ and
\begin{subequations} \begin{align}
  {\hat G}_{F_{ab}}
  = & {\hat G}_0 {\hat \Sigma}_{F_{ab}} {\hat G}_0
  -q [{\hat G}_0 \partial_{\pi_a} {\hat G}_0^{-1} {\hat G}_0 \partial_{\pi_b} {\hat G}_0^{-1} {\hat G}_0 - (c \leftrightarrow d)]/2 i, \label{eq:gem2a} \\
  {\hat G}_{-T^{\hat 0}_{\phantom{\hat 0} ab}}
  = & {\hat G}_0 {\hat \Sigma}_{-T^{\hat 0}_{\phantom{\hat 0} ab}} {\hat G}_0
  - (-\pi_{\hat 0}) [{\hat G}_0 \partial_{\pi_a} {\hat G}_0^{-1} {\hat G}_0 \partial_{\pi_b} {\hat G}_0^{-1} {\hat G}_0 - (c \leftrightarrow d)]/2 i. \label{eq:gem2b}
\end{align} \label{eq:gem2}\end{subequations}
Obviously, the only difference in Eq.~\eqref{eq:gem2} is the charge; $q$ in Eq.~\eqref{eq:gem2a} and $-\pi_{\hat 0}$ in Eq.~\eqref{eq:gem2b}.
Below, we explicitly write down the results for electromagnetic fields~\cite{PTP.116.61} only,
because those for torsion~\cite{Shitade01122014} can be obtained by replacing $q \to -\pi_{\hat 0}$.

In order to calculate expectation values, we have to extract the lesser Green's function.
In the absence of electromagnetic fields or torsion, namely, in equilibrium,
we know $G_0^< = \pm (G_0^{\rm R} - G_0^{\rm A}) f(-\pi_{\hat 0})$ and $G_0^{-1 <} = \pm (G_0^{{\rm R} -1} - G_0^{{\rm A} -1}) f(-\pi_{\hat 0})$,
where $f(\xi) = (e^{\beta \xi} \mp 1)^{-1}$ is the distribution function.
By introducing
\begin{subequations} \begin{align}
  G_{F_{ab}}^<
  = & \pm [G_{F_{ab}}^{< (0)} f(-\pi_{\hat 0}) + G_{F_{ab}}^{< (1)} f^{\prime}(-\pi_{\hat 0})], \label{eq:gem3a} \\
  \Sigma_{F_{ab}}^<
  = & \pm [\Sigma_{F_{ab}}^{< (0)} f(-\pi_{\hat 0}) + \Sigma_{F_{ab}}^{< (1)} f^{\prime}(-\pi_{\hat 0})], \label{eq:gem3b}
\end{align} \label{eq:gem3}\end{subequations}
we obtain
\begin{subequations} \begin{align}
  G_{F_{ab}}^{\rm R}
  = & G_0^{\rm R} \Sigma_{F_{ab}}^{\rm R} G_0^{\rm R}
  -q [G_0^{\rm R} \partial_{\pi_a} G_0^{{\rm R} -1} G_0^{\rm R} \partial_{\pi_b} G_0^{{\rm R} -1} G_0^{\rm R} - (a \leftrightarrow b)]/2 i, \label{eq:gem4a} \\
  G_{F_{ab}}^{< (0)}
  = & G_{F_{ab}}^{\rm R} - G_{F_{ab}}^{\rm A}, \label{eq:gem4b} \\
  \Sigma_{F_{ab}}^{< (0)}
  = & \Sigma_{F_{ab}}^{\rm R} - \Sigma_{F_{ab}}^{\rm A}, \label{eq:gem4c} \\
  G_{F_{{\hat \jmath} {\hat 0}}}^{< (1)}
  = & G_0^{\rm R} \Sigma_{F_{{\hat \jmath} {\hat 0}}}^{< (1)} G_0^{\rm A}
  -q [G_0^{\rm R} \partial_{\pi_{\hat \jmath}} G_0^{{\rm R} -1} (G_0^{\rm R} - G_0^{\rm A}) - (G_0^{\rm R} - G_0^{\rm A}) \partial_{\pi_{\hat \jmath}} G_0^{{\rm A} -1} G_0^{\rm A}]/2 i. \label{eq:gem4d}
\end{align} \label{eq:gem4}\end{subequations}
The electric component of the self-energy $\Sigma_{F_{{\hat \jmath} {\hat 0}}}^{< (0)}$ is determined self-consistently together with $G_{F_{{\hat \jmath} {\hat 0}}}^{< (1)}$,
and the magnetic component is found to satisfy $G_{F_{{\hat \imath} {\hat \jmath}}}^{< (1)} = \Sigma_{F_{{\hat \imath} {\hat \jmath}}}^{< (1)} = 0$.
This is because systems remain in equilibrium even in the presence of a static uniform magnetic field.

First, we derive the Green's-function formula for the heat magnetization.~\cite{Shitade01122014}
Although we define the heat magnetization with the grand potential in Eq.~\eqref{eq:def1b},
we can also define the auxiliary heat magnetization with the total energy $K \equiv E - \mu N$.
In the Wigner representation, the total energy is represented by
\begin{equation}
  K
  = \pm \frac{i \hbar}{2} \int \frac{d^D \pi}{(2 \pi \hbar)^D} \tr [(-\pi_{\hat 0}) \star {\hb G}]^< + \cc \label{eq:en3}
\end{equation}
The auxiliary heat magnetization is defined and calculated as
\begin{subequations} \begin{align}
  {\tilde M}_{{\rm Q} {\hat k}}
  \equiv & -\frac{1}{2} \epsilon_{{\hat \imath} {\hat \jmath} {\hat k}} \frac{\partial K}{\partial (-T^{\hat 0}_{\phantom{\hat 0} {\hat \imath} {\hat \jmath}})} \notag \\
  = & -\frac{i \hbar^2}{2} \epsilon_{{\hat \imath} {\hat \jmath} {\hat k}} \int \frac{d^D \pi}{(2 \pi \hbar)^D} f(-\pi_{\hat 0}) (-\pi_{\hat 0})
  \tr G_{-T^{\hat 0}_{\phantom{\hat 0} {\hat \imath} {\hat \jmath}}}^{\rm R} + \cc \notag \\
  = & \frac{\hbar^2}{2} \epsilon_{{\hat \imath} {\hat \jmath} {\hat k}} \int \frac{d^D \pi}{(2 \pi \hbar)^D} f(-\pi_{\hat 0}) (-\pi_{\hat 0})^2
  \tr G_0^{\rm R} \partial_{\pi_{\hat \imath}} G_0^{{\rm R} -1} G_0^{\rm R} \partial_{\pi_{\hat \jmath}} G_0^{{\rm R} -1} G_0^{\rm R} + \cc \label{eq:auxhm1a} \\
  & - \frac{i \hbar^2}{2} \epsilon_{{\hat \imath} {\hat \jmath} {\hat k}} \int \frac{d^D \pi}{(2 \pi \hbar)^D} f(-\pi_{\hat 0}) (-\pi_{\hat 0})
  \tr G_0^{\rm R} \Sigma_{-T^{\hat 0}_{\phantom{\hat 0} {\hat \imath} {\hat \jmath}}}^{\rm R} G_0^{\rm R} + \cc, \label{eq:auxhm1b}
\end{align} \label{eq:auxhm1}\end{subequations}
which is obtained by using Eq.~\eqref{eq:gem4} and the cyclic property of trace.
The proper and auxiliary heat magnetizations are related as~\cite{PhysRevLett.107.236601}
\begin{equation}
  \beta^{-1} \frac{\partial (\beta^2 M_{{\rm Q} {\hat k}})}{\partial \beta}
  = {\tilde M}_{{\rm Q} {\hat k}}. \label{eq:transhm}
\end{equation}
This relation is derived as follows.
Following Ref.~\onlinecite{PhysRevLett.107.236601}, we introduce ${\vec M}_{\rm s} \equiv \beta {\vec M}_{\rm Q}$ and ${\vec B}_{\rm s} \equiv \beta^{-1} {\vec B}_{\rm g}$,
in which $B_{\rm g}^{\hat k} \equiv (1/2) \epsilon^{{\hat \imath} {\hat \jmath} {\hat k}} (-T^{\hat 0}_{\phantom{\hat 0} {\hat \imath} {\hat \jmath}})$ is a torsional magnetic field.
The thermodynamic relation in terms of the grand potential is expressed by $d \Omega = -S d T - {\vec M}_{\rm s} \cdot d {\vec B}_{\rm s} - N d \mu$,
which leads to $S = -(\partial \Omega/\partial T)$ and ${\vec M}_{\rm s} = -(\partial \Omega/\partial {\vec B}_{\rm s})$.
The latter is the definition of the heat magnetization in Eq.~\eqref{eq:def1b}.
By combining these two equations, we obtain Maxwell's relation $-(\partial^2 \Omega/\partial T \partial {\vec B}_{\rm s}) = \partial S/\partial {\vec B}_{\rm s} = \partial {\vec M}_{\rm s}/\partial T$.
Now we turn to the total energy $K = \Omega + T S$,
whose thermodynamic relation is expressed by $d K = T d S - {\vec M}_{\rm s} \cdot d {\vec B}_{\rm s} - N d \mu$.
From this relation, we obtain $-(\partial K/\partial {\vec B}_{\rm s}) = -T (\partial S/\partial {\vec B}_{\rm s}) + {\vec M}_{\rm s}
= -T (\partial {\vec M}_{\rm s}/\partial T) + {\vec M}_{\rm s} = \partial (\beta^2 {\vec M}_{\rm Q})/\partial \beta$.
This quantity is also expressed by $-(\partial K/\partial {\vec B}_{\rm s}) = \beta {\vec {\tilde M}}_{\rm Q}$, which results in Eq.~\eqref{eq:transhm}.

Second, we write down the Kubo formula for the electric and thermal (Hall) conductivities.~\cite{Shitade01122014}
The charge and heat currents are represented by
\begin{subequations} \begin{align}
  J^{\hat \imath}
  = & \pm \frac{i \hbar q}{2} \int \frac{d^D \pi}{(2 \pi \hbar)^D} \tr [(-\partial_{\pi_{\hat \imath}} {\hb G}^{-1}) \star {\hb G}]^< + \cc, \label{eq:j1a} \\
  J_{\rm Q}^{\hat \imath}
  = & \pm \frac{i \hbar}{2} \int \frac{d^D \pi}{(2 \pi \hbar)^D}
  \tr [(-\partial_{\pi_{\hat \imath}} {\hb G}^{-1}) \star {\hb G} \star (-\pi_{\hat 0})]^< + \cc, \label{eq:j1b}
\end{align} \label{eq:j1}\end{subequations}
in which $-\partial_{\pi_{\hat \imath}} {\hb G}^{-1}$ is the renormalized velocity to guarantee the Ward identity.
The electric (Hall) conductivity is defined and calculated as
\begin{subequations} \begin{align}
  \sigma^{{\hat \imath} {\hat \jmath}}
  \equiv & \frac{\partial J^{\hat \imath}}{\partial F_{{\hat \jmath} {\hat 0}}} \notag \\
  = & \frac{\hbar^2 q^2}{6} \epsilon^{{\hat \imath} {\hat \jmath} {\hat k}} \epsilon_{abc {\hat k}} \int \frac{d^D \pi}{(2 \pi \hbar)^D} f(-\pi_{\hat 0})
  \tr G_0^{\rm R} \partial_{\pi_a} G_0^{{\rm R} -1} G_0^{\rm R} \partial_{\pi_b} G_0^{{\rm R} -1}
  G_0^{\rm R} \partial_{\pi_c} G_0^{{\rm R} -1} + \cc \label{eq:sigma1a} \\
  & + \frac{\hbar^2 q^2}{4} \int \frac{d^D \pi}{(2 \pi \hbar)^D} f^{\prime}(-\pi_{\hat 0})
  \tr (G_0^{\rm R} - G_0^{\rm A}) \partial_{\pi_{\hat \imath}} (G_0^{{\rm R} -1} + G_0^{{\rm A} -1})
  G_0^{\rm R} \partial_{\pi_{\hat \jmath}} G_0^{{\rm R} -1} + \cc \label{eq:sigma1b} \\
  & + \frac{i \hbar^2 q}{2} \int \frac{d^D \pi}{(2 \pi \hbar)^D} f^{\prime}(-\pi_{\hat 0})
  \tr \Sigma_{F_{{\hat \jmath} {\hat 0}}}^{< (1)} (G_0^{\rm R} - G_0^{\rm A})
  \partial_{\pi_{\hat \imath}} G_0^{{\rm R} -1} G_0^{\rm R} + \cc \label{eq:sigma1c}
\end{align} \label{eq:sigma1}\end{subequations}
These terms are called ``int~II'', ``int~I'', and ``ext'', respectively.
Obviously, the Fermi-sea term denoted by ``int~II'' appears only in the Hall conductivity.
The Kubo formula for the thermal (Hall) conductivity is essentially the same as that for the electric (Hall) conductivity
and is given by
\begin{subequations} \begin{align}
  T {\tilde \kappa}^{{\hat \imath} {\hat \jmath}}
  \equiv & \frac{\partial J_{\rm Q}^{\hat \imath}}{\partial (-T^{\hat 0}_{\phantom{\hat 0} {\hat \jmath} {\hat 0}})} \notag \\
  = & \frac{\hbar^2}{6} \epsilon^{{\hat \imath} {\hat \jmath} {\hat k}} \epsilon_{abc {\hat k}}
  \int \frac{d^D \pi}{(2 \pi \hbar)^D} f(-\pi_{\hat 0}) (-\pi_{\hat 0})^2
  \tr G_0^{\rm R} \partial_{\pi_a} G_0^{{\rm R} -1} G_0^{\rm R} \partial_{\pi_b} G_0^{{\rm R} -1}
  G_0^{\rm R} \partial_{\pi_c} G_0^{{\rm R} -1} + \cc \label{eq:kappa1a} \\
  & + \frac{\hbar^2}{4} \int \frac{d^D \pi}{(2 \pi \hbar)^D} f^{\prime}(-\pi_{\hat 0}) (-\pi_{\hat 0})^2
  \tr (G_0^{\rm R} - G_0^{\rm A}) \partial_{\pi_{\hat \imath}} (G_0^{{\rm R} -1} + G_0^{{\rm A} -1})
  G_0^{\rm R} \partial_{\pi_{\hat \jmath}} G_0^{{\rm R} -1} + \cc \label{eq:kappa1b} \\
  & + \frac{i \hbar^2}{2} \int \frac{d^D \pi}{(2 \pi \hbar)^D} f^{\prime}(-\pi_{\hat 0}) (-\pi_{\hat 0})
  \tr \Sigma_{-T^{\hat 0}_{\phantom{\hat 0} {\hat \jmath} {\hat 0}}}^{< (1)} (G_0^{\rm R} - G_0^{\rm A})
  \partial_{\pi_{\hat \imath}} G_0^{{\rm R} -1} G_0^{\rm R} + \cc \label{eq:kappa1c}
\end{align} \label{eq:kappa1}\end{subequations}
Note that these expressions were already obtained by the Keldysh formalism in electromagnetic and gravitational fields,
and the electric and thermal transport properties in a disordered interacting Fermi liquid were deeply studied.~\cite{PhysRevB.80.115111,PhysRevB.80.214516}
As introduced above, the heat magnetization is necessary for calculating the proper thermal Hall conductivity,
whose expression was first obtained by gauging the time translation symmetry and developing the Keldysh formalism in curved spacetime.~\cite{Shitade01122014}
Equations~\eqref{eq:auxhm1}, \eqref{eq:sigma1}, and \eqref{eq:kappa1} can be applied to disordered or interacting systems,
because these are obtained by a perturbation theory with respect to field strengths but not with respect to the impurity or interaction strength.
\end{widetext}

\begin{figure*}
  \centering
  \includegraphics[clip,width=0.99\textwidth]{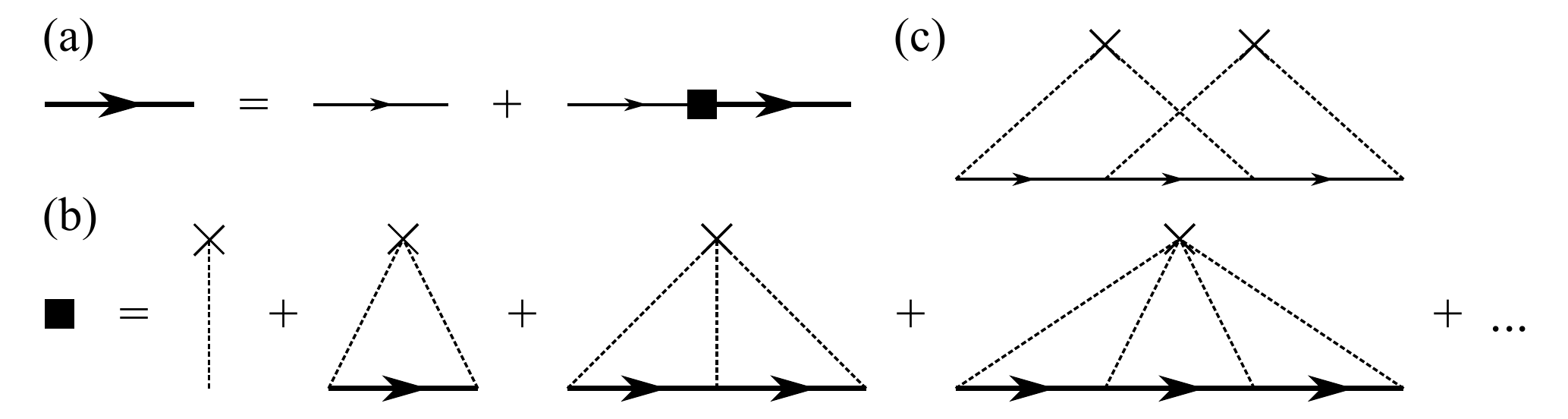}
  \caption{%
  Feynman diagrams of
  (a) the Dyson equation,
  (b) the self-energy within the self-consistent $T$-matrix approximation,
  and (c) the lowest-order self-energy which is not taken into account in the self-consistent $T$-matrix approximation.
  Thick and thin arrows indicate the perturbed and unperturbed Green's functions, respectively.
  Crosses and dotted lines indicate the impurity concentration $n_{\rm i}$ and the impurity strength $v_{\rm i}$, respectively.%
  } \label{fig:68050fig1}
\end{figure*}

Below we restrict ourselves to the self-consistent $T$-matrix approximation to deal with nonmagnetic impurities.
For $\delta$-function nonmagnetic impurities, the $T$ matrix is defined by ${\hb \Sigma} \equiv n_{\rm i} v_{\rm i} {\hb t}$ and is expressed by
\begin{equation}
  {\hb t}(-\pi_{\hat 0})
  = \left[1 - v_{\rm i} \int \frac{d^d \pi}{(2 \pi \hbar)^d} {\hb G}\right]^{-1}. \label{eq:tma1}
\end{equation}
Note that the $T$ matrix is defined by ${\hb \Sigma} \equiv n_{\rm i} {\hb T}$ for generic disordered systems.
In terms of Feynman diagrams, the self-consistent $T$-matrix approximation is expressed in Fig.~\ref{fig:68050fig1}.
The unperturbed and perturbed $T$-matrices are given by~\cite{PTP.116.61,PhysRevB.77.165103}
\begin{subequations} \begin{align}
  t_0^{\rm R}(-\pi_{\hat 0})
 = & \left[1 - v_{\rm i} \int \frac{d^d \pi}{(2 \pi \hbar)^d} G_0^{\rm R}\right]^{-1}, \label{eq:tma2a} \\
  t_{F_{cd}}^{\rm R}(-\pi_{\hat 0})
  = & v_{\rm i} t_0^{\rm R}(-\pi_{\hat 0}) \int \frac{d^d \pi}{(2 \pi \hbar)^d} G_{F_{cd}}^{\rm R} t_0^{\rm R}(-\pi_{\hat 0}), \label{eq:tma2b} \\
  t_{F_{j0}}^{< (1)}(-\pi_{\hat 0})
  = & v_{\rm i} t_0^{\rm R}(-\pi_{\hat 0}) \int \frac{d^d \pi}{(2 \pi \hbar)^d} G_{F_{j0}}^{< (1)} t_0^{\rm A}(-\pi_{\hat 0}). \label{eq:tma2c}
\end{align} \label{eq:tma2}\end{subequations}

\section{Electric and Thermal (Hall) Conductivities for a Disordered Weyl Ferromagnet} \label{sec:weyl}
Now that we have prepared all the Green's-function formulas to calculate the electric and thermal (Hall) conductivities, let us apply them to a disordered Weyl ferromagnet.
We study the linearized Weyl Hamiltonian without the time-reversal symmetry,
\begin{equation}
  {\cal H}({\vec p})
  = v \rho_x {\vec p} \cdot {\vec \sigma} + b \sigma_z, \label{eq:weyl}
\end{equation}
in which $v$, $b$, $\rho_x = \pm 1$, and ${\vec \sigma}$ indicate the Fermi velocity, the Zeeman interaction, chirality, and the Pauli matrices for spin, respectively.
The distance between Weyl nodes in the wave number space is denoted by $2 k_0 \equiv 2 b/\hbar v$, and the intrinsic contribution
is given by $-\sigma_0 \equiv -q^2 b/2 \pi^2 \hbar^2 v = -(q^2/2 \pi \hbar) (2 k_0/2 \pi)$.~\cite{PhysRevLett.107.127205,PhysRevB.86.115133,PhysRevB.88.245107,PhysRevLett.113.187202}
We deal with $\delta$-function nonmagnetic impurities within the self-consistent $T$-matrix approximation.

\begin{figure*}
  \centering
  \includegraphics[clip,width=0.99\textwidth]{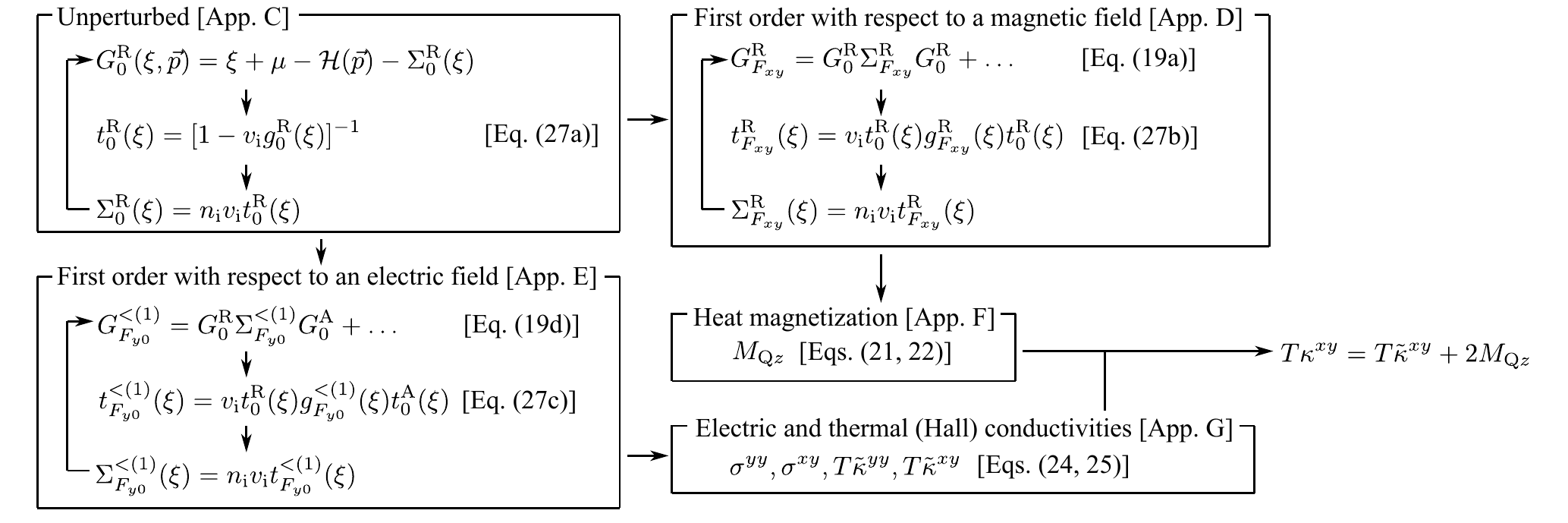}
  \caption{%
  Flowchart of the self-consistent $T$-matrix approximation to deal with $\delta$-function nonmagnetic impurities.
  $g_0^{\rm R}(\xi), g_{F_{xy}}^{\rm R}(\xi), g_{F_{y0}}^{< (1)}(\xi)$ with small letters represent the momentum integrals of
  $G_0^{\rm R}(\xi, {\vec p}), G_{F_{xy}}^{\rm R}(\xi, {\vec p}), G_{F_{y0}}^{< (1)}(\xi, {\vec p})$ with capital letters, respectively.
  See Appendices~\ref{app:g0r}-\ref{app:sigma} for the details.%
  } \label{fig:68050fig2}
\end{figure*}

Calculation is summarized in Fig.~\ref{fig:68050fig2}.
For simplicity, we omit the hat symbols for the locally flat coordinates and write $\pi_a = (-\xi, p_i)$.
We expand the Green's function and the self-energy with respect to the Pauli matrices, i.e., $A = A_0 + {\vec A} \cdot {\vec \sigma}$.
All the momentum integrals can be carried out analytically for each value of the energy $\xi$,
and the self-consistent calculation of the self-energy and the energy integrals are done numerically.
We introduce two cutoffs for the momentum integrals; $(v {\vec p}_{\perp})^2 < \Lambda_{\perp}^2$ and $|v p_z| < \Lambda_z$.
Here we emphasize that the intrinsic contribution strongly depends on ultraviolet regularizations.~\cite{PhysRevB.88.245107,PhysRevB.90.214418}
In particular, if we take the limit of $\Lambda_z \to \infty$ first,
the intrinsic contribution for a massive Dirac ferromagnet with mass $m > b$ is given by $\sigma_0$,~\cite{PhysRevB.90.214418}
although the system is not a Weyl semimetal but a trivial insulator.
In our case, the intrinsic contribution consists of the topological term $-\sigma_0$ and this additional term due to the incorrect regularization and hence vanishes.
We should keep $\Lambda_{\perp} \gg \Lambda_z \gg b$ to obtain the proper intrinsic contribution.
The chemical potential is determined self-consistently to fix the particle number without impurities at $T = 0$.
See Appendices~\ref{app:g0r}-\ref{app:sigma} for the details.

\begin{figure*}
  \centering
  \includegraphics[clip,width=0.99\textwidth]{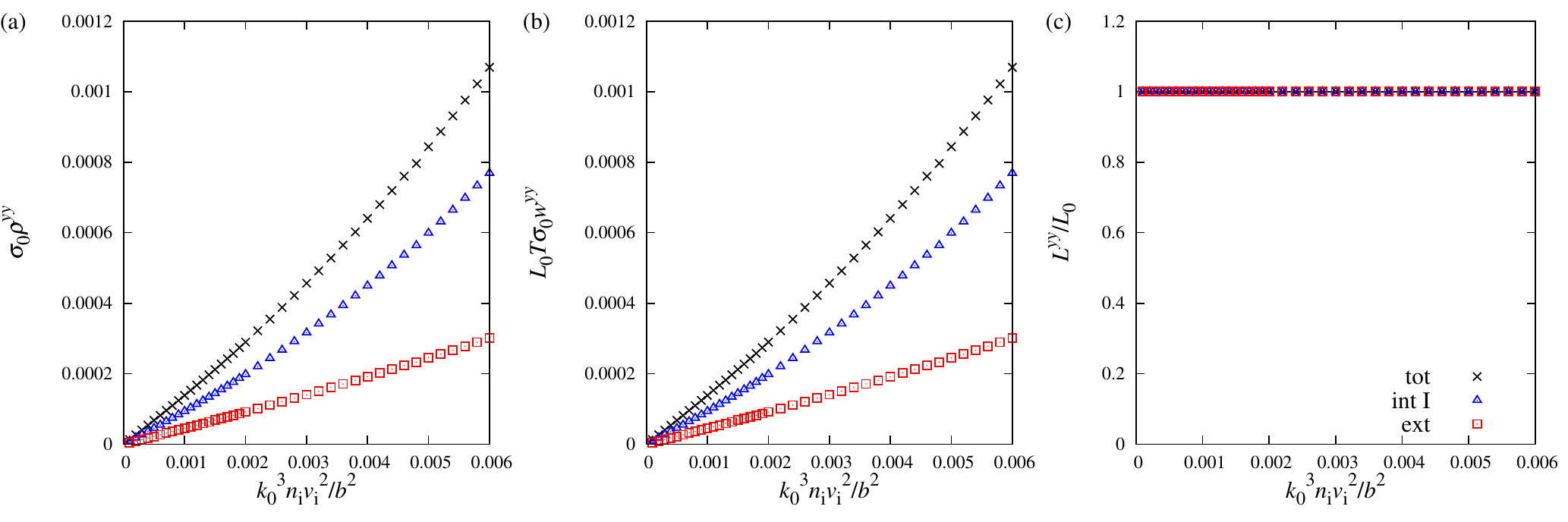}
  \caption{%
  (Color online)
  (a) Electric resistivity $\rho^{yy}$,
  (b) thermal resistivity $T w^{yy}$,
  and (c) Lorenz ratio $L^{yy}$ as functions of the impurity strength $n_{\rm i} v_{\rm i}^2$.
  ``Int~I'', ``ext'', and sum of them are indicated by blue triangles, red squares, and black crosses, respectively.
  We use $\mu_0/b = 2$, $k_0^3 v_{\rm i}/b = 0.01$, and $T/b = 0.01$.%
  } \label{fig:68050fig3}
\end{figure*}

First, we show the $n_{\rm i} v_{\rm i}^2$ dependences of the electric and thermal resistivities and the Lorenz ratio in Fig.~\ref{fig:68050fig3}.
The momentum cutoffs are given by $\Lambda_{\perp}/b = 300$ and $\Lambda_z/b = 30$,
and the energy interval $|\xi/b| < 305$ is divided into $2^{17}$ subintervals.
We use $\mu_0/b = 2$, $k_0^3 v_{\rm i}/b = 0.01$, and $T/b = 0.01$ for Fig.~\ref{fig:68050fig3}.
At the lowest temperature, $L^{yy}/L_0$ remains unity, and the Wiedemann-Franz law holds for each component.

\begin{figure*}
  \centering
  \includegraphics[clip,width=0.99\textwidth]{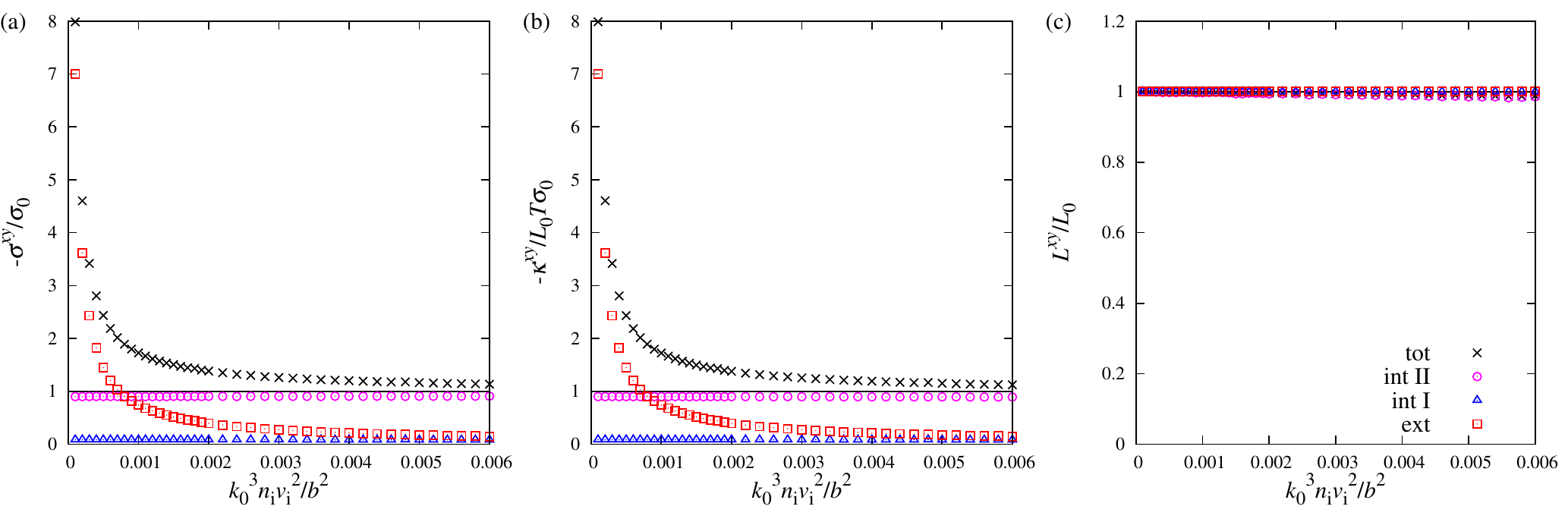}
  \caption{%
  (Color online)
  (a) Hall conductivity $-\sigma^{xy}$,
  (b) thermal Hall conductivity $-\kappa^{xy}/T$,
  and (c) Hall Lorenz ratio $L^{xy}$ as functions of the impurity strength $n_{\rm i} v_{\rm i}^2$.
  ``Int~II'', ``int~I'', ``ext'', and sum of them are indicated by magenta circles, blue triangles, red squares, and black crosses, respectively.
  The parameters are the same as in Fig.~\ref{fig:68050fig3}.%
  } \label{fig:68050fig4}
\end{figure*}

Next, we show the $n_{\rm i} v_{\rm i}^2$ dependences of the electric and thermal Hall conductivities and the Hall Lorenz ratio in Fig.~\ref{fig:68050fig4}.
Note that $T \kappa^{xy}_{\rm int~II}$ is defined by sum of $T {\tilde \kappa}^{xy}_{\rm int~II}$ and $2 M_{{\rm Q} z}$.
``Ext'' can be identified as the skew-scattering contribution~\cite{Smit1955877,Smit195839} since it is enhanced as $n_{\rm i} v_{\rm i}^2$ decreases.
In terms of the scaling, we find $\sigma^{xy}_{\rm ext} \propto \sigma^{xx}$.
``Int II'' remains constant as a function of $n_{\rm i} v_{\rm i}^2$, which is specific for the linearized Weyl Hamiltonian.
In fact, the intrinsic contribution is known to remain constant by changing $\mu$ or $T$,~\cite{PhysRevB.88.245107}
because the density of Berry curvature is finite only at $\Lambda_z - b < |\xi| < \Lambda_z + b$.
As a result, we do not find the nontrivial scaling relation $\sigma^{xy} \propto (\sigma^{xx})^{1.6}$
which was found in a two-dimensional disordered Rashba ferromagnet in the dirty regime.~\cite{PhysRevLett.97.126602,PhysRevB.77.165103,PhysRevB.79.195129}
We also note that $-\sigma^{xy}_{\rm int~II}/\sigma_0$ and $-\kappa^{xy}_{\rm int~II}/L_0 T \sigma_0$ are less than unity owing to the finite momentum cutoffs.
``Int I'' is small compared with ``int II'' but constant as a function of $n_{\rm i} v_{\rm i}^2$.
In the Berry-phase formula, i.e., in the clean limit, ``int I'' vanishes for the linearized Weyl Hamiltonian.
Therefore, this term is identified as the side-jump contribution.~\cite{PhysRevB.2.4559,PhysRevB.5.1862}
At the lowest temperature, $L^{xy}/L_0$ remains unity, which is consistent with the Wiedemann-Franz law.
Thus, the Keldysh formalism in curved spacetime provides a practical perturbation theory for calculating the electric and thermal (Hall) conductivities on an equal footing.

\begin{figure*}
  \centering
  \includegraphics[clip,width=0.99\textwidth]{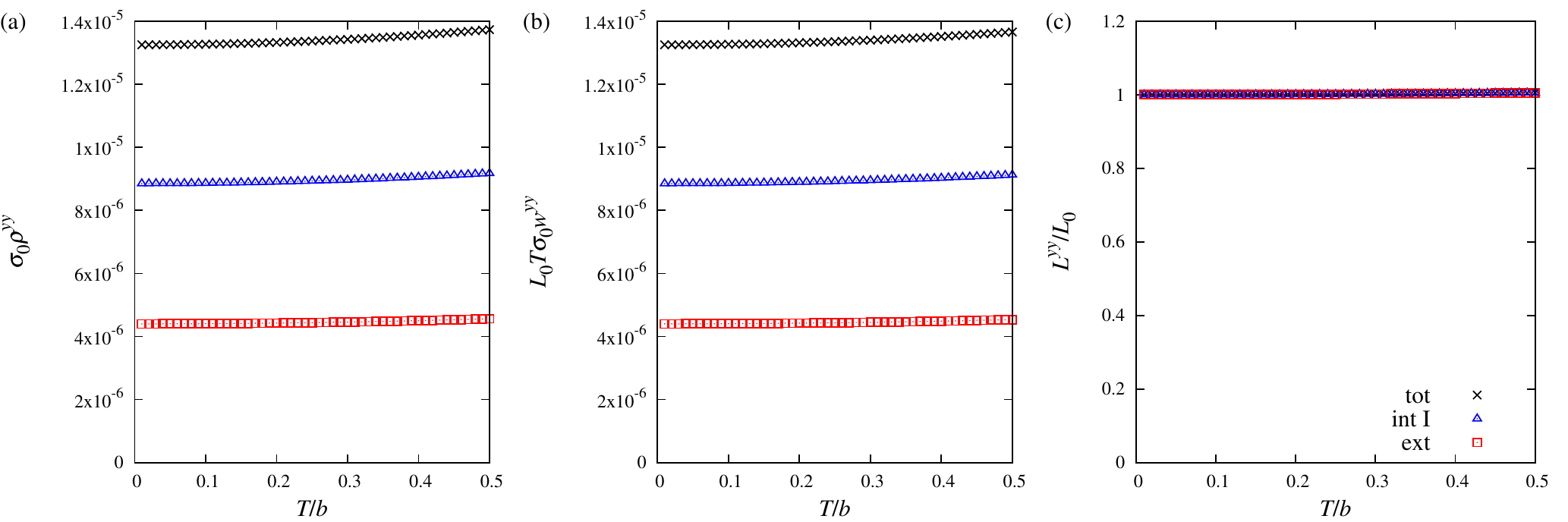}
  \caption{%
  (Color online)
  (a) Electric resistivity $\rho^{yy}$,
  (b) thermal resistivity $T w^{yy}$,
  and (c) Lorenz ratio $L^{yy}$ as functions of temperature $T$.
  The symbols are the same as in Fig.~\ref{fig:68050fig3}.
  We use $\mu_0/b = 2$, $k_0^3 v_{\rm i}/b = 0.01$, and $k_0^3 n_{\rm i} v_{\rm i}^2/b^2 = 0.0001$.%
  } \label{fig:68050fig5}
\end{figure*}

\begin{figure*}
  \centering
  \includegraphics[clip,width=0.99\textwidth]{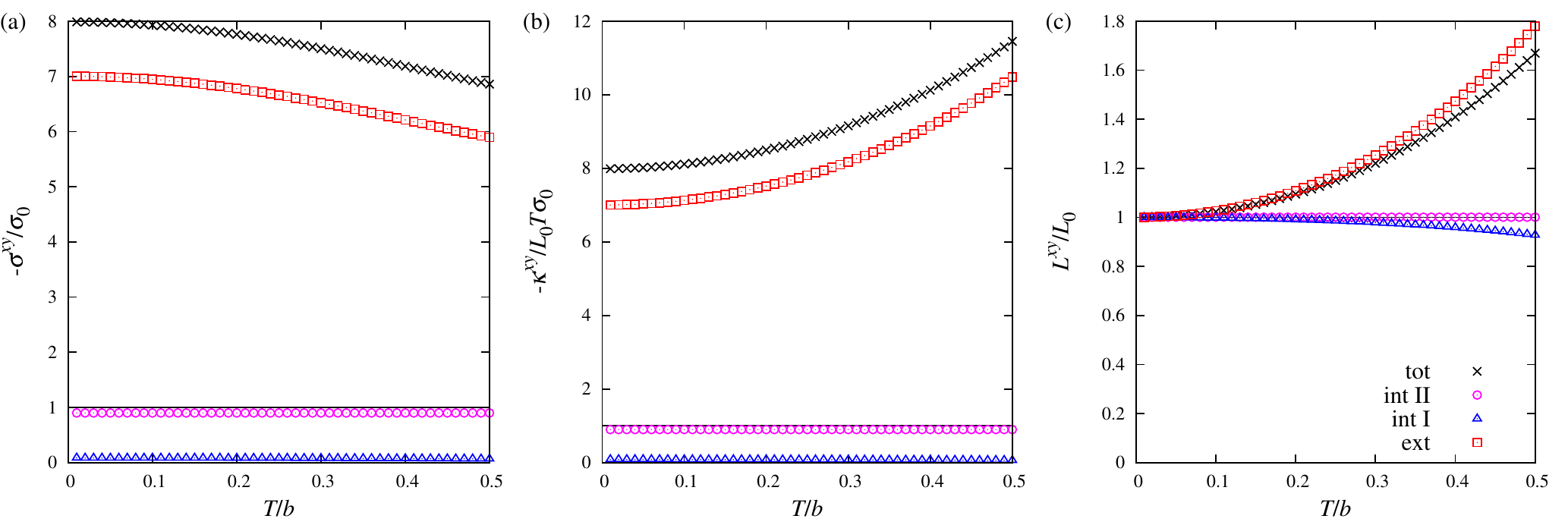}
  \caption{%
  (Color online)
  (a) Hall conductivity $-\sigma^{xy}$,
  (b) thermal Hall conductivity $-\kappa^{xy}/T$,
  and (c) Hall Lorenz ratio $L^{xy}$ as functions of temperature $T$.
  The symbols are the same as in Fig.~\ref{fig:68050fig4}, and the parameters are the same as in Fig.~\ref{fig:68050fig5}.%
  } \label{fig:68050fig6}
\end{figure*}

We also show the $T$ dependences of the electric and thermal resistivities and the Lorenz ratio in Fig.~\ref{fig:68050fig5}
and those of the electric and thermal Hall conductivities and the Hall Lorenz ratio in Fig.~\ref{fig:68050fig6}.
We use $\mu_0/b = 2$, $k_0^3 v_{\rm i}/b = 0.01$, and $k_0^3 n_{\rm i} v_{\rm i}^2/b^2 = 0.0001$.
The Lorenz ratio $L^{yy}/L_0$ remains unity as widely believed in the absence of inelastic scattering, while the Hall Lorenz ratio $L^{xy}/L_0$ does not.
Below, we discuss this temperature dependence more seriously.

\section{Discussion} \label{sec:discussion}
Here we discuss the temperature dependences of the Fermi-surface terms denoted by ``int~I'' and ``ext''.
To do this, we use the Sommerfeld expansion
\begin{align}
  \int d \epsilon [-f^{\prime}(\epsilon)] g(\epsilon)
  \simeq & g(\mu) + (1/6) (\pi T)^2 g^{(2)}(\mu) \notag \\
  & + (7/360) (\pi T)^4 g^{(4)}(\mu), \label{eq:sommer1}
\end{align}
for an infinitely differentiable function $g(\epsilon)$.
The Fermi-surface terms $\alpha = {\rm int~I}, {\rm ext}$ for the electric and thermal (Hall) conductivities can be approximated by
\begin{subequations} \begin{align}
  \sigma^{ij}_{\alpha}
  = & \int d \epsilon [-f^{\prime}(\epsilon - \mu)] s^{ij}_{\alpha}(\epsilon) \notag \\
  \simeq & s^{ij}_{\alpha}(\mu) + (1/6) (\pi T)^2 s^{ij(2)}_{\alpha}(\mu), \label{eq:fermi-surf1a} \\
  \kappa^{ij}_{\alpha}/L_0 T
  = & \frac{3}{(\pi T)^2} \int d \epsilon [-f^{\prime}(\epsilon - \mu)] (\epsilon - \mu)^2 s^{ij}_{\alpha}(\epsilon) \notag \\
  \simeq & s^{ij}_{\alpha}(\mu) + (7/10) (\pi T)^2 s^{ij (2)}_{\alpha}(\mu), \label{eq:fermi-surf1b} \\
  L^{ij}_{\alpha}/L_0
  = & \kappa^{ij}_{\alpha}/L_0 T \sigma^{ij}_{\alpha} \notag \\
  \simeq & 1 + (8/15) (\pi T)^2 s^{ij (2)}_{\alpha}(\mu)/s^{ij}_{\alpha}(\mu). \label{eq:fermi-surf1c}
\end{align} \label{eq:fermi-surf1}\end{subequations}
At finite temperature, the (Hall) Lorenz ratio $L^{ij}_{\alpha}/L_0$ can be more or less than unity depending on $s^{ij (2)}_{\alpha}(\mu)/s^{ij}_{\alpha}(\mu)$.
However, when the chemical potential is fixed,
the convexities of $\sigma^{ij}_{\alpha}$ and $\kappa^{ij}_{\alpha}/L_0 T$ depend only on the sign of $s^{ij(2)}_{\alpha}(\mu)$ and hence coincide.
As seen in Fig.~\ref{fig:68050fig6}, $-\sigma^{xy}_{\rm ext}$ is convex upward, while $-\kappa^{xy}_{\rm ext}/L_0 T$ is convex downward.
The temperature dependence of the skew-scattering contribution cannot be explained as it is.

\begin{figure*}
  \centering
  \includegraphics[clip,width=0.99\textwidth]{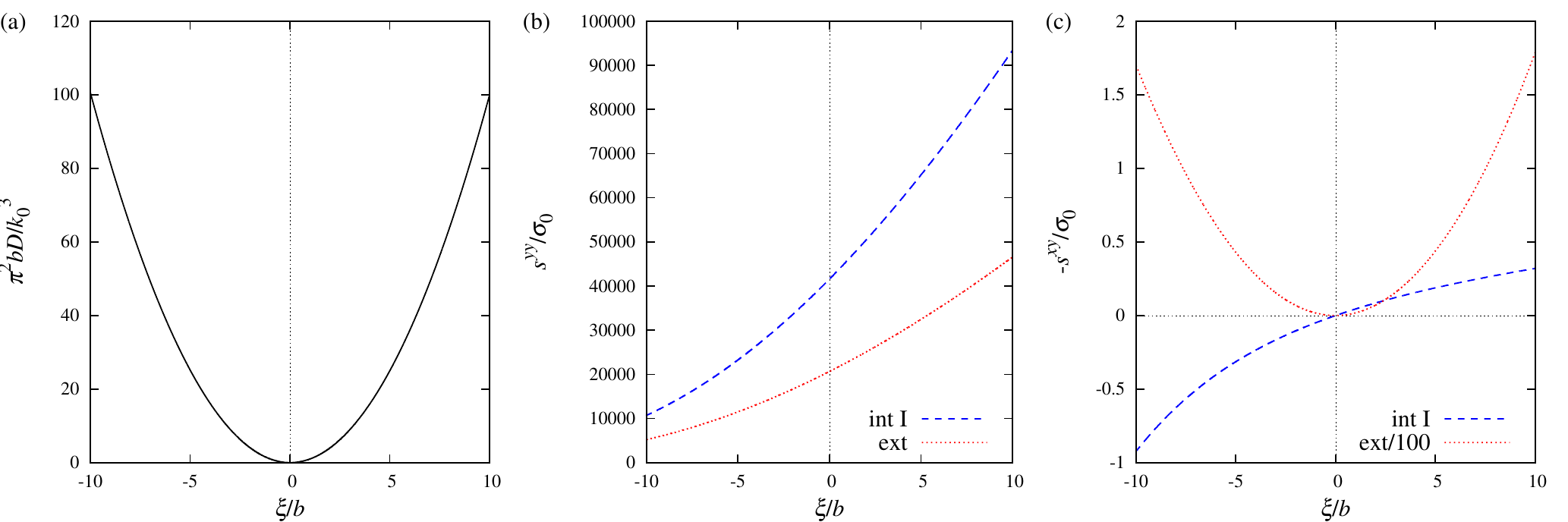}
  \caption{%
  (Color online)
  Energy dependences of (a) the density of states $D$,
  (b) the electric conductivity $s^{yy}_{\rm int~I/ext}$,
  and (c) the Hall conductivity $-s^{xy}_{\rm int~I/ext}$.
  In (b) and (c), ``int~I'' and ``ext'' are indicated by blue dashed and red dotted lines, respectively.
  The parameters are the same as in Fig.~\ref{fig:68050fig5}.%
  } \label{fig:68050fig7}
\end{figure*}

In our calculation, the chemical potential depends on temperature so as to fix the particle number.
Again by using Eq.~\eqref{eq:sommer1}, we obtain
\begin{equation}
  \mu
  \simeq \mu_0 - (1/6) (\pi T)^2 D^{\prime}(\mu_0)/D(\mu_0), \label{eq:sommer2}
\end{equation}
with $D(\mu_0)$ being the density of states.
We employ the Taylor expansion in Eq.~\eqref{eq:fermi-surf1} to obtain
\begin{subequations} \begin{align}
  \sigma^{ij}_{\alpha}
  \simeq & s^{ij}_{\alpha}(\mu_0) + (1/6) (\pi T)^2 \notag \\
  & \times \left[s^{ij (2)}_{\alpha}(\mu_0) - s^{ij \prime}_{\alpha}(\mu_0) \frac{D^{\prime}(\mu_0)}{D(\mu_0)}\right], \label{eq:fermi-surf2a} \\
  \kappa^{ij}_{\alpha}/L_0 T
  \simeq & s^{ij}_{\alpha}(\mu_0) + (1/6) (\pi T)^2 \notag \\
  & \times \left[\frac{21}{5} s^{ij (2)}_{\alpha}(\mu_0) - s^{ij \prime}_{\alpha}(\mu_0) \frac{D^{\prime}(\mu_0)}{D(\mu_0)}\right], \label{eq:fermi-surf2b} \\
  L^{ij}_{\alpha}/L_0
  \simeq & 1 + (8/15) (\pi T)^2 s^{ij (2)}_{\alpha}(\mu_0)/s^{ij}_{\alpha}(\mu_0). \label{eq:fermi-surf2c}
\end{align} \label{eq:fermi-surf2}\end{subequations}
Thus, the (Hall) Lorenz ratio $L^{ij}_{\alpha}/L_0$ can be more or less than unity depending on $s^{ij (2)}_{\alpha}(\mu_0)/s^{ij}_{\alpha}(\mu_0)$,
and the convexities of $\sigma^{ij}_{\alpha}$ and $\kappa^{ij}_{\alpha}/L_0 T$ do not always coincide.
In particular, for $-s^{xy (2)}_{\rm ext}(\mu_0) < -s^{xy \prime}_{\rm ext}(\mu_0) D^{\prime}(\mu_0)/D(\mu_0) < 21 [-s^{xy (2)}_{\rm ext}(\mu_0)]/5$,
$-\sigma^{xy}_{\rm ext}$ is convex upward, while $-\kappa^{xy}_{\rm ext}/L_0 T$ is convex downward.
We show the energy dependences of $D(\xi)$ and $s^{ij}_{\alpha}(\xi)$ in Fig.~\ref{fig:68050fig7}.
We find $D(\xi) \simeq c_1 \xi^2$ and $-s^{xy}_{\rm ext}(\xi) \simeq c_2 \xi^2$ with $c_1, c_2 > 0$, which satisfy the above inequalities.
The Lorenz ratio $L^{yy}_{\alpha}/L_0$ remains unity
because $s^{yy}_{\alpha}(\xi)$ is almost linear, and hence $s^{yy (2)}_{\alpha}(\mu_0)/s^{yy}_{\alpha}(\mu_0)$ is negligible.

Let us mention some impurity effects which we do not take into accout.
The self-consistent $T$-matrix approximation is valid in the dilute impurity limit and does not include the weak localization due to diffusons and Cooperons nor the Anderson localization.
A doped Weyl semimetal on which we focus is compressible and exhibits the Anderson transition.
When the chemical potential lies on Weyl nodes, where the density of states vanishes, two quantum phase transitions take place;
one is from a Weyl semimetal to a compressible metal, and the other is then to an Anderson insulator.~\cite{PhysRevB.33.3263,PhysRevLett.115.246603,PhysRevLett.116.066401}
As long as the electric (Hall) conductivity $\sigma^{ij}_{\alpha}$ is nonzero, the Lorenz ratio $L^{ij}_{\alpha}/L_0$ goes to unity for generic disorder.~\cite{0022-3719-10-12-021}
This can be also seen in our results Eqs.~\eqref{eq:sigma1} and \eqref{eq:kappa1}.
For elastic scattering, $\Sigma_{-T^0_{\phantom{0} y0}}^{< (1)}(\xi) = \xi \Sigma_{F_{y0}}^{< (1)}(\xi)/q$ holds true,
and these equations are expressed by Eq.~\eqref{eq:fermi-surf1a} and Eq.~\eqref{eq:fermi-surf1b}, respectively.
One counterexample is the extrinsic contribution to the Hall conductivity,
for which Eqs.~\eqref{eq:fermi-surf2a} and \eqref{eq:fermi-surf2b} start from $O(T^2)$ because of $-s^{xy}_{\rm ext}(\xi) \simeq c_2 \xi^2$.
As a result, the Lorenz ratio is given by $L^{xy}_{\rm ext}/L_0 = 21/5$.
Note that the chemical potential does not depend on temperature, and hence the density-of-states corrections in Eqs.~\eqref{eq:fermi-surf2a} and \eqref{eq:fermi-surf2b} are not necessary.

In realistic systems, we do not expect the above temperature dependences because of inelastic scattering.
Experimentally, the Hall Lorenz ratio in Ni and Co is almost constant while that in Fe rapidly decreases.~\cite{PhysRevLett.100.016601,PhysRevB.81.054414,PhysRevB.79.100404}
Since the intrinsic contribution is dominant in Ni and Co while the skew-scattering contribution is dominant in Fe,~\cite{PhysRevLett.99.086602}
it is concluded that the intrinsic mechanism is robust against inelastic scattering, but not the extrinsic mechanisms.
This conclusion was also supported experimentally by the (Hall) resistivity measurements~\cite{PhysRevLett.103.087206,PhysRevB.85.220403,0953-8984-24-48-482001} and theoretically.~\cite{JPSJ.81.083704}
It is an important future problem to calculate the electric and thermal (Hall) conductivities by taking into account both disorder and interactions.

To summarize, we investigated the transport properties for a doped Weyl ferromagnet with nonmagnetic impurities which exhibits the AHE.
We used the Keldysh formalism in curved spacetime which recently we developed~\cite{Shitade01122014}
and calculated the electric and thermal (Hall) conductivities on an equal footing within the self-consistent $T$-matrix approximation.
We successfully reproduced the Wiedemann-Franz law at low temperature
and found that the Lorenz ratio of the skew-scattering contribution $L^{xy}_{\rm ext}/L_0$ deviates from unity as temperature increases, which comes from the higher-order Sommerfeld expansion.
Note that this temperature dependnece is not expected in the presence of inelastic scattering caused by phonons and magnons.
Our results can be understood by the formula derived by Smr\v{c}ka and St\v{r}eda, which works at any temperature as far as elastic scattering is concerned.~\cite{0022-3719-10-12-021}
However, by using our practical perturbation theory, we can go beyond Smr\v{c}ka and St\v{r}eda and deal with both disorder and interactions.

\begin{acknowledgments}
  We discussed our numerical calculation with G.~Tatara and K.~Yamamoto and learned the cutoff dependence of the intrinsic contribution from J.~Fujimoto.
  We thank N.~Nagaosa for reading our manuscript and giving us comments.
\end{acknowledgments}
\appendix
\begin{widetext}
\section{Invariance of Tosion under Local Time Translation} \label{app:inv}
Here we explicitly show the invariance of $T^{\hat 0}_{\phantom{\hat 0} ab}$ under local time translation $x^{\prime 0} = x^0 + \epsilon^0(x)$ and $x^{\prime i} = x^i$.
By using
\begin{subequations} \begin{align}
  \frac{\partial x^{\prime \mu}}{\partial x^{\nu}}
  = & \begin{bmatrix}
        1 + \partial_0 \epsilon^0 & \partial_i \epsilon^0 \\
        0 & \delta^i_{\phantom{i}_j}
      \end{bmatrix}, \label{eq:transf1a} \\
  \frac{\partial x^{\nu}}{\partial x^{\prime \mu}}
  = & \begin{bmatrix}
        (1 + \partial_0 \epsilon^0)^{-1} & 0 \\
        -(1 + \partial_0 \epsilon^0)^{-1} \partial_i \epsilon^0 & \delta^i_{\phantom{i} j}
      \end{bmatrix}, \label{eq:transf1b}
\end{align} \label{eq:transf1}\end{subequations}
vielbein, its inverse, and torsion transform as
\begin{subequations} \begin{align}
  e^{\prime {\hat 0}}_{\phantom{\prime {\hat 0}} 0}(x^{\prime})
  = & (\partial x^{\nu}/\partial x^{\prime 0}) e^{\hat 0}_{\phantom{\hat 0} \nu}(x)
  = (1 + \partial_0 \epsilon^0)^{-1} e^{\hat 0}_{\phantom{\hat 0} 0}, \label{eq:transf2a} \\
  e^{\prime {\hat 0}}_{\phantom{\prime {\hat 0}} i}(x^{\prime})
  = & (\partial x^{\nu}/\partial x^{\prime i}) e^{\hat 0}_{\phantom{\hat 0} \nu}(x)
  = e^{\hat 0}_{\phantom{\hat 0} i} - (1 + \partial_0 \epsilon^0)^{-1} \partial_i \epsilon^0 e^{\hat 0}_{\phantom{\hat 0} 0}, \label{eq:transf2b}
\end{align} \label{eq:transf2}\end{subequations}
\begin{subequations} \begin{align}
  e_{\phantom{\prime} a}^{\prime \phantom{a} 0}(x^{\prime})
  = & (\partial x^{\prime 0}/\partial x^{\nu}) e_a^{\phantom{a} \nu}(x)
  = (1 + \partial_0 \epsilon^0) e_a^{\phantom{a} 0} + \partial_i \epsilon^0 e_a^{\phantom{a} i}, \label{eq:transf3a} \\
  e_{\phantom{\prime} a}^{\prime \phantom{a} i}(x^{\prime})
  = & (\partial x^{\prime i}/\partial x^{\nu}) e_a^{\phantom{a} \nu}(x)
  = e_a^{\phantom{a} i}, \label{eq:transf3b}
\end{align} \label{eq:transf3}\end{subequations}
and
\begin{subequations} \begin{align}
  T^{\prime {\hat 0}}_{\phantom{\prime {\hat 0}} j0}(x^{\prime})
  = & \partial_j^{\prime} e^{\prime {\hat 0}}_{\phantom{\prime {\hat 0}} 0}(x^{\prime}) - \partial_0^{\prime} e^{\prime {\hat 0}}_{\phantom{\prime {\hat 0}} j}(x^{\prime}) \notag \\
  = & [\partial_j - (1 + \partial_0 \epsilon^0)^{-1} \partial_j \epsilon^0 \partial_0] [(1 + \partial_0 \epsilon^0)^{-1} e^{\hat 0}_{\phantom{\hat 0} 0}]
  - (1 + \partial_0 \epsilon^0)^{-1} \partial_0 [e^{\hat 0}_{\phantom{\hat 0} j} - (1 + \partial_0 \epsilon^0)^{-1} \partial_j \epsilon^0 e^{\hat 0}_{\phantom{\hat 0} 0}] \notag \\
  = & (1 + \partial_0 \epsilon^0)^{-1} T^{\hat 0}_{\phantom{\hat 0} j0}, \label{eq:transf4a} \\
  T^{\prime {\hat 0}}_{\phantom{\prime {\hat 0}} ij}(x^{\prime})
  = & \partial_i^{\prime} e^{\prime {\hat 0}}_{\phantom{\prime {\hat 0}} j}(x^{\prime}) - \partial_j^{\prime} e^{\prime {\hat 0}}_{\phantom{\prime {\hat 0}} i}(x^{\prime}) \notag \\
  = & [\partial_i - (1 + \partial_0 \epsilon^0)^{-1} \partial_i \epsilon^0 \partial_0]
  [e^{\hat 0}_{\phantom{\hat 0} j} - (1 + \partial_0 \epsilon^0)^{-1} \partial_j \epsilon^0 e^{\hat 0}_{\phantom{\hat 0} 0}] - (i \leftrightarrow j) \notag \\
  = & T^{\hat 0}_{\phantom{\hat 0} ij} + (1 + \partial_0 \epsilon^0)^{-1} (\partial_i \epsilon^0 T^{\hat 0}_{\phantom{\hat 0} j0} - \partial_j \epsilon^0 T^{\hat 0}_{\phantom{\hat 0} i0}), \label{eq:transf4b}
\end{align} \label{eq:transf4}\end{subequations}
respectively.
Thus, $T^{\hat 0}_{\phantom{\hat 0} \mu \nu}$ is not invariant under local time translation.
On the other hand, $T^{\hat 0}_{\phantom{\hat 0} ab}$ is invariant as
\begin{align}
  T^{\prime {\hat 0}}_{\phantom{\prime {\hat 0}} ab}(x^{\prime})
  = & e_{\phantom{\prime} a}^{\prime \phantom{a} k}(x^{\prime}) e_{\phantom{\prime} b}^{\prime \phantom{b} 0}(x^{\prime}) T^{\prime {\hat 0}}_{\phantom{\prime {\hat 0}} k0}(x^{\prime})
  + e_{\phantom{\prime} a}^{\prime \phantom{a} 0}(x^{\prime}) e_{\phantom{\prime} b}^{\prime \phantom{b} l}(x^{\prime}) T^{\prime {\hat 0}}_{\phantom{\prime {\hat 0}} 0l}(x^{\prime})
  + e_{\phantom{\prime} a}^{\prime \phantom{a} k}(x^{\prime}) e_{\phantom{\prime} b}^{\prime \phantom{b} l}(x^{\prime}) T^{\prime {\hat 0}}_{\phantom{\prime {\hat 0}} kl}(x^{\prime}) \notag \\
  = & e_a^{\phantom{a} k} [(1 + \partial_0 \epsilon^0) e_b^{\phantom{b} 0} + \partial_l \epsilon^0 e_b^{\phantom{b} l}] (1 + \partial_0 \epsilon^0)^{-1} T^{\hat 0}_{\phantom{\hat 0} k0}
  + [(1 + \partial_0 \epsilon^0) e_a^{\phantom{a} 0} + \partial_k \epsilon^0 e_a^{\phantom{a} k}] e_b^{\phantom{b} l} (1 + \partial_0 \epsilon^0)^{-1} T^{\hat 0}_{\phantom{\hat 0} 0l} \notag \\
  & + e_a^{\phantom{a} k} e_b^{\phantom{b} l} [T^{\hat 0}_{\phantom{\hat 0} kl} + (1 + \partial_0 \epsilon^0)^{-1} (\partial_k \epsilon^0 T^{\hat 0}_{\phantom{\hat 0} l0} - \partial_l \epsilon^0 T^{\hat 0}_{\phantom{\hat 0} k0})] \notag \\
  = & e_a^{\phantom{a} k} e_b^{\phantom{b} 0} T^{\hat 0}_{\phantom{\hat 0} k0} + e_a^{\phantom{a} 0} e_b^{\phantom{b} l} T^{\hat 0}_{\phantom{\hat 0} 0l}
  + e_a^{\phantom{a} k} e_b^{\phantom{b} l} T^{\hat 0}_{\phantom{\hat 0} kl}
  = T^{\hat 0}_{\phantom{\hat 0} ab}. \label{eq:transf5}
\end{align}
We also note that Eq.~\eqref{eq:transf2} indicates the exact gauge transformation of a gravitational (vector) potential $e^{\hat 0}_{\phantom{\hat 0} \mu} = (1 + \phi_{\rm g}, -A_{{\rm g} i})$,
which is different from the U($1$) gauge transformation.
However, when $|\partial_{\mu} \epsilon^0| \ll 1$ and $|\phi_{\rm g}| \ll 1$, Eq.~\eqref{eq:transf2} is approximately expressed by the U($1$) gauge transformation
$\phi_{\rm g}^{\prime}(x^{\prime}) \simeq \phi_{\rm g}(x) - \partial_0 \epsilon^0(x)$ and $A_{{\rm g} i}^{\prime}(x^{\prime}) \simeq A_{{\rm g} i}(x) + \partial_i \epsilon^0(x)$.

\section{Derivation of the Moyal and Star Products} \label{app:moyal}
In this Appendix, we derive the Moyal product in flat spacetime and the star product in curved spacetime.
The Moyal product $\ast$ is defined by the Wigner representation of convolution and is calculated as
\begin{align}
  {\hat A} \ast {\hat B}(X, p)
  = & \int d^D x \int d^D y e^{-i p_a x^a/\hbar}  {\hat A}(X + x/2, y) {\hat B}(y, X - x/2) \notag \\
  = & \int d^D x \int d^D y \int \frac{d^D q}{(2 \pi \hbar)^D} \int \frac{d^D r}{(2 \pi \hbar)^D}
  e^{-i p_a x^a/\hbar} e^{i q_a (X - y + x/2)^a/\hbar} e^{i r_a (-X + y + x/2)^a/\hbar} \notag \\
  & \times {\hat A}(X/2 + y/2  +x/4, q) {\hat B}(X/2 + y/2 - x/4, r) \notag \\
  = & \int d^D x_1 \int d^D x_2 \int \frac{d^D p_1}{(2 \pi \hbar)^D} \int \frac{d^D p_2}{(2 \pi \hbar)^D}
  e^{-i (p_{1 a}  x_2^a - p_{2 a} x_1^a)/\hbar} {\hat A}(X + x_1/2, p + p_1) {\hat B}(X + x_2/2, p + p_2) \notag \\
  = & \int d^D x_1 \int d^D x_2 \int \frac{d^D p_1}{(2 \pi \hbar)^D} \int \frac{d^D p_2}{(2 \pi \hbar)^D}
  [e^{-i p_{1 a} (x_2^a + i \hbar \partial_{p_a})/\hbar} e^{x_1^a \partial_{X^a}/2} {\hat A}(X, p)] \notag \\
  & \times [e^{i p_{2 a} (x_1^a - i \hbar \partial_{p_a})/\hbar} e^{x_2^a \partial_{X^a}/2} {\hat B}(X, p)] \notag \\
  = & {\hat A}(X, p) e^{i \hbar {\cal F}_0/2} {\hat B}(X, p). \label{eq:moyal}
\end{align}
In the second line, we employ the inverse Fourier transformation.
In the third line, we change variables $x_1 = -X + y + x/2, x_2 = -X + y - x/2$ and introduce $p_1 = q - p, p_2 = r - p$.
As defined in the main text, ${\cal F}_0 = \partial_{X^a} \otimes \partial_{p_a} - \partial_{p_a} \otimes \partial_{X^a}$ is the Poisson bracket in flat spacetime.

The star product $\star$ is obtained by the Peierls substitution $\pi_a = e_a^{\phantom{a} \mu}(X) [p_{\mu} - q A_{\mu}(X)]$.
Strictly, we cannot employ the Peierls substitution for the Moyal product since the Moyal product is given by an exponential function of the Poisson bracket.
Instead, we employ it for the Poisson bracket, from which we construct the star product.
This procedure is known as deformation quantization in mathematics~\cite{kontsevich2003}
and was carried out to obtain the Keldysh formalism in dynamical or nonuniform electromagnetic fields.~\cite{PTP.117.415}
According to the chain rule, partial derivatives are modified as
$\partial_{X^a} \to e_a^{\phantom{a} \mu} \partial_{X^{\mu}}
- e_a^{\phantom{a} \mu} e_b^{\phantom{b} \nu} (q \partial_{X^{\mu}} A_{\nu} + \partial_{X^{\mu}} e^{\hat 0}_{\phantom{\hat 0} \nu} \pi_{\hat 0}) \partial_{\pi_b}$
and $\partial_{p_a} \to \partial_{\pi_a}$,
and the Poisson bracket is modified as
\begin{equation}
  {\cal F}_0
  = \partial_{X^a} \otimes \partial_{p_a} -  \partial_{p_a} \otimes \partial_{X^a}
  \to e_a^{\phantom{a} \mu} (\partial_{X^{\mu}} \otimes \partial_{\pi_a} - \partial_{\pi_a} \otimes \partial_{X^{\mu}})
  + e_a^{\phantom{a} \mu} e_b^{\phantom{b} \nu} (q F_{\mu \nu} + T^{\hat 0}_{\phantom{\hat 0} \mu \nu} \pi_{\hat 0}) \partial_{\pi_a} \otimes \partial_{\pi_b}
  = {\cal F}. \label{eq:chain2}
\end{equation}

\section{Unperturbed Self-Energy} \label{app:g0r}
We calculate the unperturbed retarded self-energy $\Sigma_0^{\rm R}(\xi)$.
We introduce $z^{\rm R}(\xi) \equiv \xi + \mu - \Sigma_{00}^{\rm R}(\xi)$ and $\eta^{\rm R}(\xi) \equiv b + \Sigma_{0z}^{\rm R}(\xi)$ and obtain
\begin{subequations} \begin{align}
  G_0^{{\rm R} -1}(\xi, {\vec p})
  = & \xi + \mu - {\cal H}({\vec p}) - \Sigma_0^{\rm R}(\xi)
  = z^{\rm R}(\xi)
  - [v \rho_x {\vec p}_{\perp} + {\vec \Sigma}_{0 \perp}^{\rm R}(\xi)] \cdot {\vec \sigma}_{\perp} - [v \rho_x p_z + \eta^{\rm R}(\xi)] \sigma_z, \label{eq:g0r1a} \\
  G_0^{\rm R}(\xi, {\vec p})
  = & \Gamma_0^{\rm R}(\xi, {\vec p}) \{z^{\rm R}(\xi)
  + [v \rho_x {\vec p}_{\perp} + {\vec \Sigma}_{0 \perp}^{\rm R}(\xi)] \cdot {\vec \sigma}_{\perp} + [v \rho_x p_z + \eta^{\rm R}(\xi)] \sigma_z\}, \label{eq:g0r1b} \\
  \Gamma_0^{\rm R}(\xi, {\vec p})
  \equiv & \det G_0^{\rm R}(\xi, {\vec p})
  = \{z^{{\rm R} 2}(\xi)
  - [v \rho_x {\vec p}_{\perp} + {\vec \Sigma}_{0 \perp}^{\rm R}(\xi)]^2 - [v \rho_x p_z + \eta^{\rm R}(\xi)]^2\}^{-1}. \label{eq:g0r1c}
\end{align} \label{eq:g0r1}\end{subequations}
In order to calculate the self-consistent $T$ matrix, we define $g_0^{\rm R}(\xi)$ by the momentum integral of $G_0^{\rm R}(\xi, {\vec p})$.
Explicitly, it is given by
\begin{subequations} \begin{align}
  g_0^{\rm R}(\xi)
  \equiv & \int \frac{d^3 p}{(2 \pi \hbar)^3} G_0^{\rm R}(\xi, {\vec p}), \label{eq:g0r2a} \\
  g_{00}^{\rm R}(\xi)
  = & z^{\rm R}(\xi) \int \frac{d^3 p}{(2 \pi \hbar)^3} \Gamma_0^{\rm R}(\xi, {\vec p}), \label{eq:g0r2b} \\
  {\vec g}_{0 \perp}^{\rm R}(\xi)
  = & \int \frac{d^3 p}{(2 \pi \hbar)^3} \Gamma_0^{\rm R}(\xi, {\vec p})
  [v \rho_x {\vec p}_{\perp} + {\vec \Sigma}_{0 \perp}^{\rm R}(\xi)], \label{eq:g0r2c} \\
  g_{0z}^{\rm R}(\xi)
  = & \int \frac{d^3 p}{(2 \pi \hbar)^3} \Gamma_0^{\rm R}(\xi, {\vec p}) [v \rho_x p_z + \eta^{\rm R}(\xi)], \label{eq:g0r2d}
\end{align} \label{eq:g0r2}\end{subequations}
from which we obtain
\begin{subequations} \begin{align}
  t_0^{\rm R}(\xi)
  = & [1 - v_{\rm i} g_0^{\rm R}(\xi)]^{-1}
  = \tau_0^{\rm R}(\xi) \{[1 - v_{\rm i} g_{00}^{\rm R}(\xi)] + v_{\rm i} {\vec g}_0^{\rm R}(\xi) \cdot {\vec \sigma}\}, \label{eq:g0r3a} \\
  \tau_0^{\rm R}(\xi)
  \equiv & \det t_0^{\rm R}(\xi)
  = \{[1 - v_{\rm i} g_{00}^{\rm R}(\xi)]^2 - [v_{\rm i} {\vec g}_0^{\rm R}(\xi)]^2\}^{-1}. \label{eq:g0r3b}
\end{align} \label{eq:g0r3}\end{subequations}
If we put ${\vec \Sigma}_{0 \perp}^{\rm R}(\xi) = 0$, $\Gamma_0^{\rm R}(\xi, {\vec p})$ is even with respect to ${\vec p}_{\perp}$,
which leads to ${\vec g}_{0 \perp}^{\rm R}(\xi) = 0$ in Eq.~\eqref{eq:g0r2c} and ${\vec t}_{0 \perp}^{\rm R}(\xi) = 0$ in Eq.~\eqref{eq:g0r3a}.
Therefore we obtain ${\vec \Sigma}_{0 \perp}^{\rm R}(\xi) = 0$ and
\begin{subequations} \begin{align}
  G_0^{{\rm R} -1}(\xi, {\vec p})
  = & z^{\rm R}(\xi) - v \rho_x {\vec p}_{\perp} \cdot {\vec \sigma}_{\perp} - [v \rho_x p_z + \eta^{\rm R}(\xi)] \sigma_z, \label{eq:g0r4a} \\
  G_0^{\rm R}(\xi, {\vec p})
  = & \Gamma_0^{\rm R}(\xi, {\vec p})
  \{z^{\rm R}(\xi) + v \rho_x {\vec p}_{\perp} \cdot {\vec \sigma}_{\perp} + [v \rho_x p_z + \eta^{\rm R}(\xi)] \sigma_z\}, \label{eq:g0r4b} \\
  \Gamma_0^{\rm R}(\xi, {\vec p})
  = & \{z^{{\rm R} 2}(\xi) - (v {\vec p}_{\perp})^2 - [v \rho_x p_z + \eta^{\rm R}(\xi)]^2\}^{-1}. \label{eq:g0r4c}
\end{align} \label{eq:g0r4}\end{subequations}
Here we define the integrals by
\begin{equation}
  I_{mn}^{\rm R}(\xi)
  \equiv \int_0^{\Lambda_{\perp}^2} d x \int_{-\Lambda_z}^{\Lambda_z} d y
  \frac{[y + \eta^{\rm R}(\xi)]^m}{\{z^{{\rm R} 2}(\xi) - x - [y + \eta^{\rm R}(\xi)]^2\}^n}, \label{eq:irmn}
\end{equation}
with $x \equiv (v {\vec p}_{\perp})^2$ and $y = v \rho_x p_z$.
Their explicit forms are given in Appendix~\ref{app:int}.
By introducing $g_0^{\rm R}(\xi) \equiv {\tilde g}_0^{\rm R}(\xi)/8 \pi^2 \hbar^3 v^3$,
the self-consistent equations which we solve are given by
\begin{subequations} \begin{align}
  z^{\rm R}(\xi)
  = & \xi + \mu - \Sigma_{00}^{\rm R}(\xi), &
  \eta^{\rm R}(\xi)
  = & b + \Sigma_{0z}^{\rm R}(\xi), \label{eq:g0r5a} \\
  {\tilde g}_{00}^{\rm R}(\xi)
  = & z^{\rm R}(\xi) I_{01}^{\rm R}(\xi), &
  {\tilde g}_{0z}^{\rm R}(\xi)
  = & I_{11}^{\rm R}(\xi), \label{eq:g0r5b} \\
  \tau_0^{\rm R}(\xi)
  = & \{[1 - (v_{\rm i}/8 \pi^2 \hbar^3 v^3) {\tilde g}_{00}^{\rm R}(\xi)]^2 - [(v_{\rm i}/8 \pi^2 \hbar^3 v^3) {\tilde g}_{0z}^{\rm R}(\xi)]^2\}^{-1}, \label{eq:g0r5c} \\
  t_{00}^{\rm R}(\xi)
  = & \tau_0^{\rm R}(\xi) [1 - (v_{\rm i}/8 \pi^2 \hbar^3 v^3) {\tilde g}_{00}^{\rm R}(\xi)], &
  t_{0z}^{\rm R}(\xi)
  = & \tau_0^{\rm R}(\xi) (v_{\rm i}/8 \pi^2 \hbar^3 v^3) {\tilde g}_{0z}^{\rm R}(\xi), \label{eq:g0r5d} \\
  \Sigma_{00}^{\rm R}(\xi)
  = & n_{\rm i} v_{\rm i} t_{00}^{\rm R}(\xi), &
  \Sigma_{0z}^{\rm R}(\xi)
  = & n_{\rm i} v_{\rm i} t_{0z}^{\rm R}(\xi). \label{eq:g0r5e}
\end{align} \label{eq:g0r5}\end{subequations}
The particle number to be fixed is calculated by
\begin{align}
  N
  = & \frac{i}{2 \pi} \sum_{\rho_x = \pm 1} \int d \xi f(\xi) \int \frac{d^3 p}{(2 \pi \hbar)^3} \tr G_0^{\rm R}(\xi, {\vec p}) + \cc
  = \frac{1}{\pi} \sum_{\rho_x = \pm 1} \int d \xi f(\xi) [i g_{00}^{\rm R}(\xi) + \cc] \notag \\
  = & \frac{1}{4 \pi^3 \hbar^3 v^3} \int d \xi f(\xi) [i {\tilde g}_{00}^{\rm R}(\xi) + \cc]. \label{eq:g0r6}
\end{align}

\section{First-Order Self-Energy with Respect to a Magnetic Field} \label{app:gbr}
Here we calculate the first-order retarded self-energy with respect to a magnetic field $\Sigma_{F_{xy}}^{\rm R}(\xi)$ from Eq.~\eqref{eq:gem4a}.
We obtain
\begin{subequations} \begin{align}
  G_{F_{xy} 0}^{\rm R}(\xi, {\vec p})
  = & \Sigma_{F_{xy} 0}^{\rm R}(\xi) \Gamma_0^{{\rm R} 2}(\xi, {\vec p}) \{z^{{\rm R} 2}(\xi) + (v {\vec p}_{\perp})^2 + [v \rho_x p_z + \eta^{\rm R}(\xi)]^2\}
  + 2 z^{\rm R}(\xi) {\vec \Sigma}_{F_{xy} \perp}^{\rm R}(\xi) \cdot \Gamma_0^{{\rm R} 2}(\xi, {\vec p}) v \rho_x {\vec p}_{\perp} \notag \\
  & + 2 z^{\rm R}(\xi) \Sigma_{F_{xy} z}^{\rm R}(\xi) \Gamma_0^{{\rm R} 2}(\xi, {\vec p}) [v \rho_x p_z + \eta^{\rm R}(\xi)]
  - q v^2 \Gamma_0^{{\rm R} 2}(\xi, {\vec p}) [v \rho_x p_z + \eta^{\rm R}(\xi)], \label{eq:gbr1a} \\
  G_{F_{xy} x}^{\rm R}(\xi, {\vec p})
  = & 2 z^{\rm R}(\xi) \Sigma_{F_{xy} 0}^{\rm R}(\xi) \Gamma_0^{{\rm R} 2}(\xi, {\vec p}) v \rho_x p_x \notag \\
  & + \Sigma_{F_{xy} x}^{\rm R}(\xi) \Gamma_0^{{\rm R} 2}(\xi, {\vec p}) \{z^{{\rm R} 2}(\xi) + v^2 (p_x^2 - p_y^2) - [v \rho_x p_z + \eta^{\rm R}(\xi)]^2\}
  + 2 \Sigma_{F_{xy} y}^{\rm R}(\xi) \Gamma_0^{{\rm R} 2}(\xi, {\vec p}) v^2 p_x p_y \notag \\
  & + 2 \Sigma_{F_{xy} z}^{\rm R}(\xi) \Gamma_0^{{\rm R} 2}(\xi, {\vec p}) v \rho_x p_x [v \rho_x p_z + \eta^{\rm R}(\xi)], \label{eq:gbr1b} \\
  G_{F_{xy} y}^{\rm R}(\xi, {\vec p})
  = & 2 z^{\rm R}(\xi) \Sigma_{F_{xy} 0}^{\rm R}(\xi) \Gamma_0^{{\rm R} 2}(\xi, {\vec p}) v \rho_x p_y \notag \\
  & + 2 \Sigma_{F_{xy} x}^{\rm R}(\xi) \Gamma_0^{{\rm R} 2}(\xi, {\vec p}) v^2 p_x p_y
  + \Sigma_{F_{xy} y}^{\rm R}(\xi) \Gamma_0^{{\rm R} 2}(\xi, {\vec p}) \{z^{{\rm R} 2}(\xi) - v^2 (p_x^2 - p_y^2) - [v \rho_x p_z + \eta^{\rm R}(\xi)]^2\} \notag \\
  & + 2 \Sigma_{F_{xy} z}^{\rm R}(\xi) \Gamma_0^{{\rm R} 2}(\xi, {\vec p}) v \rho_x p_y [v \rho_x p_z + \eta^{\rm R}(\xi)], \label{eq:gbr1c} \\
  G_{F_{xy} z}^{\rm R}(\xi, {\vec p})
  = & 2 z^{\rm R}(\xi) \Sigma_{F_{xy} 0}^{\rm R}(\xi) \Gamma_0^{{\rm R} 2}(\xi, {\vec p}) [v \rho_x p_z + \eta^{\rm R}(\xi)]
  + 2 {\vec \Sigma}_{F_{xy} \perp}^{\rm R}(\xi) \cdot \Gamma_0^{{\rm R} 2}(\xi, {\vec p}) v \rho_x {\vec p}_{\perp} [v \rho_x p_z + \eta^{\rm R}(\xi)] \notag \\
  & + \Sigma_{F_{xy} z}^{\rm R}(\xi) \Gamma_0^{{\rm R} 2}(\xi, {\vec p}) \{z^{{\rm R} 2}(\xi) - (v {\vec p}_{\perp})^2 + [v \rho_x p_z + \eta^{\rm R}(\xi)]^2\}
  - q v^2 z^{\rm R}(\xi) \Gamma_0^{{\rm R} 2}(\xi, {\vec p}), \label{eq:gbr1d}
\end{align} \label{eq:gbr1}\end{subequations}
and their momentum integrals denoted by $g_{F_{xy}}^{\rm R}(\xi)$ are given by
\begin{subequations} \begin{align}
  g_{F_{xy} 0}^{\rm R}(\xi)
  = & \Sigma_{F_{xy} 0}^{\rm R}(\xi) \int \frac{d^3 p}{(2 \pi \hbar)^3} \Gamma_0^{{\rm R} 2}(\xi, {\vec p}) \{z^{{\rm R} 2}(\xi) + (v {\vec p}_{\perp})^2 + [v \rho_x p_z + \eta^{\rm R}(\xi)]^2\} \notag \\
  & + 2 z^{\rm R}(\xi) \Sigma_{F_{xy} z}^{\rm R}(\xi) \int \frac{d^3 p}{(2 \pi \hbar)^3} \Gamma_0^{{\rm R} 2}(\xi, {\vec p}) [v \rho_x p_z + \eta^{\rm R}(\xi)] \notag \\
  & - q v^2 \int \frac{d^3 p}{(2 \pi \hbar)^3} \Gamma_0^{{\rm R} 2}(\xi, {\vec p}) [v \rho_x p_z + \eta^{\rm R}(\xi)], \label{eq:gbr2a} \\
  {\vec g}_{F_{xy} \perp}^{\rm R}(\xi)
  = & {\vec \Sigma}_{F_{xy} \perp}^{\rm R}(\xi) \int \frac{d^3 p}{(2 \pi \hbar)^3} \Gamma_0^{{\rm R} 2}(\xi, {\vec p}) \{z^{{\rm R} 2}(\xi) - [v \rho_x p_z + \eta^{\rm R}(\xi)]^2\}, \label{eq:gbr2b} \\
  g_{F_{xy} z}^{\rm R}(\xi)
  = & 2 z^{\rm R}(\xi) \Sigma_{F_{xy} 0}^{\rm R}(\xi) \int \frac{d^3 p}{(2 \pi \hbar)^3} \Gamma_0^{{\rm R} 2}(\xi, {\vec p}) [v \rho_x p_z + \eta^{\rm R}(\xi)] \notag \\
  & + \Sigma_{F_{xy} z}^{\rm R}(\xi) \int \frac{d^3 p}{(2 \pi \hbar)^3} \Gamma_0^{{\rm R} 2}(\xi, {\vec p}) \{z^{{\rm R} 2}(\xi) - (v {\vec p}_{\perp})^2 + [v \rho_x p_z + \eta^{\rm R}(\xi)]^2\} \notag \\
  & - q v^2 z^{\rm R}(\xi) \int \frac{d^3 p}{(2 \pi \hbar)^3} \Gamma_0^{{\rm R} 2}(\xi, {\vec p}). \label{eq:gbr2c}
\end{align} \label{eq:gbr2}\end{subequations}
We also obtain the self-consistent $T$ matrix Eq.~\eqref{eq:tma2b} as
\begin{subequations} \begin{align}
  t_{F_{xy}}^{\rm R}(\xi)
  = & v_{\rm i} t_0^{\rm R}(\xi) g_{F_{xy}}^{\rm R}(\xi) t_0^{\rm R}(\xi), \label{eq:gbr3a} \\
  t_{F_{xy} 0}^{\rm R}(\xi)
  = & v_{\rm i} [t_{00}^{{\rm R} 2}(\xi) + t_{0z}^{{\rm R} 2}(\xi)] g_{F_{xy} 0}^{\rm R}(\xi)
  + 2 v_{\rm i} t_{00}^{\rm R}(\xi) t_{0z}^{\rm R}(\xi) g_{F_{xy} z}^{\rm R}(\xi), \label{eq:gbr3b} \\
  {\vec t}_{F_{xy} \perp}^{\rm R}(\xi)
  = & v_{\rm i} [t_{00}^{{\rm R} 2}(\xi) - t_{0z}^{{\rm R} 2}(\xi)] {\vec g}_{F_{xy} \perp}^{\rm R}(\xi), \label{eq:gbr3c} \\
  t_{F_{xy} z}^{\rm R}(\xi)
  = & 2 v_{\rm i} t_{00}^{\rm R}(\xi) t_{0z}^{\rm R}(\xi) g_{F_{xy} 0}^{\rm R}(\xi)
  + v_{\rm i} [t_{00}^{{\rm R} 2}(\xi) + t_{0z}^{{\rm R} 2}(\xi)] g_{F_{xy} z}^{\rm R}(\xi), \label{eq:gbr3d}
\end{align} \label{eq:gbr3}\end{subequations}
Equations~\eqref{eq:gbr2b} and \eqref{eq:gbr3c} lead to ${\vec g}_{F_{xy} \perp}^{\rm R}(\xi) = {\vec t}_{F_{xy} \perp}^{\rm R}(\xi) = {\vec \Sigma}_{F_{xy} \perp}^{\rm R}(\xi) = 0$.
The remaining components are obtained
by introducing $\Sigma_{F_{xy}}^{\rm R}(\xi) \equiv q v^2 {\tilde \Sigma}_{q F_{xy}}^{\rm R}(\xi)$ and $g_{F_{xy}}^{\rm R}(\xi) \equiv q v^2 {\tilde g}_{q F_{xy}}^{\rm R}(\xi)/8 \pi^2 \hbar^3 v^3$
and solving
\begin{subequations} \begin{align}
  \begin{bmatrix}
    {\tilde g}_{q F_{xy} 0}^{\rm R}(\xi) \\
    {\tilde g}_{q F_{xy} z}^{\rm R}(\xi)
  \end{bmatrix}
  = & \begin{bmatrix}
    2 z^{{\rm R} 2}(\xi) I_{02}^{\rm R}(\xi) - I_{01}^{\rm R}(\xi) & 2 z^{\rm R}(\xi) I_{12}^{\rm R}(\xi) \\
    2 z^{\rm R}(\xi) I_{12}^{\rm R}(\xi) & 2 I_{22}^{\rm R}(\xi) + I_{01}^{\rm R}(\xi)
  \end{bmatrix}
  \begin{bmatrix}
    {\tilde \Sigma}_{q F_{xy} 0}^{\rm R}(\xi) \\
    {\tilde \Sigma}_{q F_{xy} z}^{\rm R}(\xi)
  \end{bmatrix}
  - \begin{bmatrix}
    I_{12}^{\rm R}(\xi) \\
    z^{\rm R}(\xi) I_{02}^{\rm R}(\xi)
  \end{bmatrix}, \label{eq:gbr4a} \\
  \begin{bmatrix}
    {\tilde \Sigma}_{q F_{xy} 0}^{\rm R}(\xi) \\
    {\tilde \Sigma}_{q F_{xy} z}^{\rm R}(\xi)
  \end{bmatrix}
  = & \frac{n_{\rm i} v_{\rm i}^2}{8 \pi^2 \hbar^3 v^3} \begin{bmatrix}
    t_{00}^{{\rm R} 2}(\xi) + t_{0z}^{{\rm R} 2}(\xi) & 2 t_{00}^{\rm R}(\xi) t_{0z}^{\rm R}(\xi) \\
    2 t_{00}^{\rm R}(\xi) t_{0z}^{\rm R}(\xi) & t_{00}^{{\rm R} 2}(\xi) + t_{0z}^{{\rm R} 2}(\xi)
  \end{bmatrix}
  \begin{bmatrix}
    {\tilde g}_{q F_{xy} 0}^{\rm R}(\xi) \\
    {\tilde g}_{q F_{xy} z}^{\rm R}(\xi)
  \end{bmatrix}. \label{eq:gbr4b}
\end{align} \label{eq:gbr4}\end{subequations}
The first-order self-energy with respect to a torsional magnetic field is obtained just by $\Sigma_{-T^0_{\phantom{0} xy}}^{\rm R}(\xi) = \xi \Sigma_{F_{xy}}^{\rm R}(\xi)/q$.

\section{First-Order Self-Energy with Respect to an Electric Field} \label{app:ge<1}
Next, we calculate the first-order lesser self-energy with respect to an electric field $\Sigma_{F_{y0}}^{< (1)}(\xi)$ from Eq.~\eqref{eq:gem4d}.
We obtain
\begin{subequations} \begin{align}
  G_{F_{y0} 0}^{< (1)}(\xi, {\vec p})
  = & \Sigma_{F_{y0} 0}^{< (1)}(\xi) |\Gamma_0^{\rm R}(\xi, {\vec p})|^2
  [|z^{\rm R}(\xi)|^2 + (v {\vec p}_{\perp})^2 + |v \rho_x p_z + \eta^{\rm R}(\xi)|^2] \notag \\
  & + \Sigma_{F_{y0} x}^{< (1)}(\xi) |\Gamma_0^{\rm R}(\xi, {\vec p})|^2
  \{[z^{\rm R}(\xi) + z^{\rm A}(\xi)] v \rho_x p_x + i [\eta^{\rm R}(\xi) - \eta^{\rm A}(\xi)] v \rho_x p_y\} \notag \\
  & + \Sigma_{F_{y0} y}^{< (1)}(\xi) |\Gamma_0^{\rm R}(\xi, {\vec p})|^2
  \{-i [\eta^{\rm R}(\xi) - \eta^{\rm A}(\xi)] v \rho_x p_x + [z^{\rm R}(\xi) + z^{\rm A}(\xi)] v \rho_x p_y\} \notag \\
  & + \Sigma_{F_{y0} z}^{< (1)}(\xi) |\Gamma_0^{\rm R}(\xi, {\vec p})|^2
  \{z^{\rm R}(\xi) [v \rho_x p_z + \eta^{\rm A}(\xi)] + z^{\rm A}(\xi) [v \rho_x p_z + \eta^{\rm R}(\xi)]\} \notag \\
  & + q v \rho_x |\Gamma_0^{\rm R}(\xi, {\vec p})|^2
  \{[\eta^{\rm R}(\xi) - \eta^{\rm A}(\xi)] v \rho_x p_x + i [z^{\rm R}(\xi) + z^{\rm A}(\xi)] v \rho_x p_y\} \notag \\
  & - i q v \rho_x [z^{\rm R}(\xi) \Gamma_0^{{\rm R} 2}(\xi, {\vec p}) v \rho_x p_y + \cc], \label{eq:ge<11a} \\
  G_{F_{y0} x}^{< (1)}(\xi, {\vec p})
  = & \Sigma_{F_{y0} 0}^{< (1)}(\xi) |\Gamma_0^{\rm R}(\xi, {\vec p})|^2
  \{[z^{\rm R}(\xi) + z^{\rm A}(\xi)] v \rho_x p_x - i [\eta^{\rm R}(\xi) - \eta^{\rm A}(\xi)] v \rho_x p_y\} \notag \\
  & + \Sigma_{F_{y0} x}^{< (1)}(\xi) |\Gamma_0^{\rm R}(\xi, {\vec p})|^2
  [|z^{\rm R}(\xi)|^2 + v^2 (p_x^2 - p_y^2) - |v \rho_x p_z + \eta^{\rm R}(\xi)|^2] \notag \\
  & + \Sigma_{F_{y0} y}^{< (1)}(\xi) |\Gamma_0^{\rm R}(\xi, {\vec p})|^2
  \{2 v^2 p_x p_y + i z^{\rm R}(\xi) [v \rho_x p_z + \eta^{\rm A}(\xi)] - i z^{\rm A}(\xi) [v \rho_x p_z + \eta^{\rm R}(\xi)]\} \notag \\
  & + \Sigma_{F_{y0} z}^{< (1)}(\xi) |\Gamma_0^{\rm R}(\xi, {\vec p})|^2
  \{v \rho_x p_x [2 v \rho_x p_z + \eta^{\rm R}(\xi) + \eta^{\rm A}(\xi)] - i [z^{\rm R}(\xi) - z^{\rm A}(\xi)] v \rho_x p_y\} \notag \\
  & + q v \rho_x |\Gamma_0^{\rm R}(\xi, {\vec p})|^2
  \{2 i v^2 p_x p_y - z^{\rm R}(\xi) [v \rho_x p_z + \eta^{\rm A}(\xi)] + z^{\rm A}(\xi) [v \rho_x p_z + \eta^{\rm R}(\xi)]\} \notag \\
  & - i q v \rho_x [\Gamma_0^{{\rm R} 2}(\xi, {\vec p}) v^2 p_x p_y + \cc], \label{eq:ge<11b} \\
  G_{F_{y0} y}^{< (1)}(\xi, {\vec p})
  = & \Sigma_{F_{y0} 0}^{< (1)}(\xi) |\Gamma_0^{\rm R}(\xi, {\vec p})|^2
  \{i [\eta^{\rm R}(\xi) - \eta^{\rm A}(\xi)] v \rho_x p_x + [z^{\rm R}(\xi) + z^{\rm A}(\xi)] v \rho_x p_y\} \notag \\
  & + \Sigma_{F_{y0} x}^{< (1)}(\xi) |\Gamma_0^{\rm R}(\xi, {\vec p})|^2
  \{2 v^2 p_x p_y - i z^{\rm R}(\xi) [v \rho_x p_z + \eta^{\rm A}(\xi)] + i z^{\rm A}(\xi) [v \rho_x p_z + \eta^{\rm R}(\xi)]\} \notag \\
  & + \Sigma_{F_{y0} y}^{< (1)}(\xi) |\Gamma_0^{\rm R}(\xi, {\vec p})|^2
  [|z^{\rm R}(\xi)|^2 - v^2 (p_x^2 - p_y^2) - |v \rho_x p_z + \eta^{\rm R}(\xi)|^2] \notag \\
  & + \Sigma_{F_{y0} z}^{< (1)}(\xi) |\Gamma_0^{\rm R}(\xi, {\vec p})|^2
  \{i [z^{\rm R}(\xi) - z^{\rm A}(\xi)] v \rho_x p_x + v \rho_x p_y [2 v \rho_x p_z + \eta^{\rm R}(\xi) + \eta^{\rm A}(\xi)]\} \notag \\
  & + i q v \rho_x |\Gamma_0^{\rm R}(\xi, {\vec p})|^2
  [|z^{\rm R}(\xi)|^2 - v^2 (p_x^2 - p_y^2) - |v \rho_x p_z + \eta^{\rm R}(\xi)|^2] \notag \\
  & - i q v \rho_x/2 (\Gamma_0^{{\rm R} 2}(\xi, {\vec p})
  \{z^{{\rm R} 2}(\xi) - v^2 (p_x^2 - p_y^2) - [v \rho_x p_z + \eta^{\rm R}(\xi)]^2\} + \cc), \label{eq:ge<11c} \\
  G_{F_{y0} z}^{< (1)}(\xi, {\vec p})
  = & \Sigma_{F_{y0} 0}^{< (1)}(\xi) |\Gamma_0^{\rm R}(\xi, {\vec p})|^2
  \{z^{\rm R}(\xi) [v \rho_x p_z + \eta^{\rm A}(\xi)] + z^{\rm A}(\xi) [v \rho_x p_z + \eta^{\rm R}(\xi)]\} \notag \\
  & + \Sigma_{F_{y0} x}^{< (1)}(\xi) |\Gamma_0^{\rm R}(\xi, {\vec p})|^2
  \{[2 v \rho_x p_z + \eta^{\rm R}(\xi) + \eta^{\rm A}(\xi)] v \rho_x p_x + i [z^{\rm R}(\xi) - z^{\rm A}(\xi)] v \rho_x p_y\} \notag \\
  & + \Sigma_{F_{y0} y}^{< (1)}(\xi) |\Gamma_0^{\rm R}(\xi, {\vec p})|^2
  \{- i [z^{\rm R}(\xi) - z^{\rm A}(\xi)] v \rho_x p_x + [2 v \rho_x p_z + \eta^{\rm R}(\xi) + \eta^{\rm A}(\xi)] v \rho_x p_y\} \notag \\
  & + \Sigma_{F_{y0} z}^{< (1)}(\xi) |\Gamma_0^{\rm R}(\xi, {\vec p})|^2
  [|z^{\rm R}(\xi)|^2 - (v {\vec p}_{\perp})^2 + |v \rho_x p_z + \eta^{\rm R}(\xi)|^2] \notag \\
  & + q v \rho_x |\Gamma_0^{\rm R}(\xi, {\vec p})|^2
  \{- i [z^{\rm R}(\xi) - z^{\rm A}(\xi)] v \rho_x p_x + [2 v \rho_x p_z + \eta^{\rm R}(\xi) + \eta^{\rm A}(\xi)] v \rho_x p_y\} \notag \\
  & - i q v \rho_x \{\Gamma_0^{{\rm R} 2}(\xi, {\vec p}) v \rho_x p_y [v \rho_x p_z + \eta^{\rm R}(\xi)] + \cc\}, \label{eq:ge<11d}
\end{align} \label{eq:ge<11}\end{subequations}
and their momentum integrals denoted by $g_{F_{y0}}^{< (1)}(\xi)$ are given by
\begin{subequations} \begin{align}
  g_{F_{y0} 0}^{< (1)}(\xi)
  = & \Sigma_{F_{y0} 0}^{< (1)}(\xi) \int \frac{d^3 p}{(2 \pi \hbar)^3} |\Gamma_0^{\rm R}(\xi, {\vec p})|^2
  [|z^{\rm R}(\xi)|^2 + (v {\vec p}_{\perp})^2 + |v \rho_x p_z + \eta^{\rm R}(\xi)|^2] \notag \\
  & + \Sigma_{F_{y0} z}^{< (1)}(\xi) \int \frac{d^3 p}{(2 \pi \hbar)^3} |\Gamma_0^{\rm R}(\xi, {\vec p})|^2
  \{z^{\rm R}(\xi) [v \rho_x p_z + \eta^{\rm A}(\xi)] + z^{\rm A}(\xi) [v \rho_x p_z + \eta^{\rm R}(\xi)]\}, \label{eq:ge<12a} \\
  g_{F_{y0} x}^{< (1)}(\xi)
  = & \Sigma_{F_{y0} x}^{< (1)}(\xi) \int \frac{d^3 p}{(2 \pi \hbar)^3} |\Gamma_0^{\rm R}(\xi, {\vec p})|^2
  [|z^{\rm R}(\xi)|^2 - |v \rho_x p_z + \eta^{\rm R}(\xi)|^2] \notag \\
  & + \Sigma_{F_{y0} y}^{< (1)}(\xi) \int \frac{d^3 p}{(2 \pi \hbar)^3} |\Gamma_0^{\rm R}(\xi, {\vec p})|^2
  \{i z^{\rm R}(\xi) [v \rho_x p_z + \eta^{\rm A}(\xi)] - i z^{\rm A}(\xi) [v \rho_x p_z + \eta^{\rm R}(\xi)]\} \notag \\
  & + i q v \rho_x \int \frac{d^3 p}{(2 \pi \hbar)^3} |\Gamma_0^{\rm R}(\xi, {\vec p})|^2
  \{i z^{\rm R}(\xi) [v \rho_x p_z + \eta^{\rm A}(\xi)] - i z^{\rm A}(\xi) [v \rho_x p_z + \eta^{\rm R}(\xi)]\}, \label{eq:ge<12b} \\
  g_{F_{y0} y}^{< (1)}(\xi)
  = & -\Sigma_{F_{y0} x}^{< (1)}(\xi) \int \frac{d^3 p}{(2 \pi \hbar)^3} |\Gamma_0^{\rm R}(\xi, {\vec p})|^2
  \{i z^{\rm R}(\xi) [v \rho_x p_z + \eta^{\rm A}(\xi)] - i z^{\rm A}(\xi) [v \rho_x p_z + \eta^{\rm R}(\xi)]\} \notag \\
  & + \Sigma_{F_{y0} y}^{< (1)}(\xi) \int \frac{d^3 p}{(2 \pi \hbar)^3} |\Gamma_0^{\rm R}(\xi, {\vec p})|^2
  [|z^{\rm R}(\xi)|^2 - |v \rho_x p_z + \eta^{\rm R}(\xi)|^2] \notag \\
  & + i q v \rho_x \int \frac{d^3 p}{(2 \pi \hbar)^3} |\Gamma_0^{\rm R}(\xi, {\vec p})|^2
  [|z^{\rm R}(\xi)|^2 - |v \rho_x p_z + \eta^{\rm R}(\xi)|^2] \notag \\
  & - \frac{1}{2} i q v \rho_x  \int \frac{d^3 p}{(2 \pi \hbar)^3}
  \left(\Gamma_0^{{\rm R} 2}(\xi, {\vec p}) \{z^{{\rm R} 2}(\xi) - [v \rho_x p_z + \eta^{\rm R}(\xi)]^2\} + \cc\right), \label{eq:ge<12c} \\
  g_{F_{y0} z}^{< (1)}(\xi)
  = & \Sigma_{F_{y0} 0}^{< (1)}(\xi) \int \frac{d^3 p}{(2 \pi \hbar)^3} |\Gamma_0^{\rm R}(\xi, {\vec p})|^2
  \{z^{\rm R}(\xi) [v \rho_x p_z + \eta^{\rm A}(\xi)] + z^{\rm A}(\xi) [v \rho_x p_z + \eta^{\rm R}(\xi)]\} \notag \\
  & + \Sigma_{F_{y0} z}^{< (1)}(\xi) \int \frac{d^3 p}{(2 \pi \hbar)^3} |\Gamma_0^{\rm R}(\xi, {\vec p})|^2
  [|z^{\rm R}(\xi)|^2 - (v {\vec p}_{\perp})^2 + |v \rho_x p_z + \eta^{\rm R}(\xi)|^2]. \label{eq:ge<12d}
\end{align} \label{eq:ge<12}\end{subequations}
The self-consistent $T$ matrix Eq.~\eqref{eq:tma2c} is expressed by
\begin{subequations} \begin{align}
  t_{F_{y0}}^{< (1)}(\xi)
  = & v_{\rm i} t_0^{\rm R}(\xi) g_{F_{y0}}^{< (1)}(\xi) t_0^{\rm A}(\xi), \label{eq:ge<13a} \\
  t_{F_{y0} 0}^{< (1)}(\xi)
  = & v_{\rm i} [|t_{00}^{\rm R}(\xi)|^2 + |t_{0z}^{\rm R}(\xi)|^2] g_{F_{y0} 0}^{< (1)}(\xi)
  + v_{\rm i} [t_{00}^{\rm R}(\xi) t_{0z}^{\rm A}(\xi) + t_{00}^{\rm A}(\xi) t_{0z}^{\rm R}(\xi)] g_{F_{y0} z}^{< (1)}(\xi), \label{eq:ge<13b} \\
  t_{F_{y0} x}^{< (1)}(\xi)
  = & v_{\rm i} [|t_{00}^{\rm R}(\xi)|^2 - |t_{0z}^{\rm R}(\xi)|^2] g_{F_{y0} x}^{< (1)}(\xi)
  + v_{\rm i} [i t_{00}^{\rm R}(\xi) t_{0z}^{\rm A}(\xi) - i t_{00}^{\rm A}(\xi) t_{0z}^{\rm R}(\xi)] g_{F_{y0} y}^{< (1)}(\xi), \label{eq:ge<13c} \\
  t_{F_{y0} y}^{< (1)}(\xi)
  = & - v_{\rm i} [i t_{00}^{\rm R}(\xi) t_{0z}^{\rm A}(\xi) - i t_{00}^{\rm A}(\xi) t_{0z}^{\rm R}(\xi)] g_{F_{y0} x}^{< (1)}(\xi)
  + v_{\rm i} [|t_{00}^{\rm R}(\xi)|^2 - |t_{0z}^{\rm R}(\xi)|^2] g_{F_{y0} y}^{< (1)}(\xi), \label{eq:ge<13d} \\
  t_{F_{y0} z}^{< (1)}(\xi)
  = & v_{\rm i} [t_{00}^{\rm R}(\xi) t_{0z}^{\rm A}(\xi) + t_{00}^{\rm A}(\xi) t_{0z}^{\rm R}(\xi)] g_{F_{y0} 0}^{< (1)}(\xi)
  + v_{\rm i} [|t_{00}^{\rm R}(\xi)|^2 + |t_{0z}^{\rm R}(\xi)|^2] g_{F_{y0} z}^{< (1)}(\xi), \label{eq:ge<13e}
\end{align} \label{eq:ge<13}\end{subequations}
leading to $g_{F_{y0} 0 (z)}^{< (1)}(\xi) = t_{F_{y0} 0 (z)}^{< (1)}(\xi) = \Sigma_{F_{y0} 0 (z)}^{< (1)}(\xi) = 0$.
The remaining components are obtained
by introducing $\Sigma_{F_{y0}}^{< (1)}(\xi) \equiv i q v \rho_x {\tilde \Sigma}_{q F_{y0}}^{< (1)}(\xi)$
and $g_{F_{y0}}^{< (1)}(\xi) \equiv i q v \rho_x {\tilde g}_{q F_{y0}}^{< (1)}(\xi)/8 \pi^2 \hbar^3 v^3$
and solving
\begin{subequations} \begin{align}
  \begin{bmatrix}
    {\tilde g}_{q F_{y0} x}^{< (1)}(\xi) \\
    {\tilde g}_{q F_{y0} y}^{< (1)}(\xi)
  \end{bmatrix}
  = & \begin{bmatrix}
    J_1^{< (1)}(\xi) & J_2^{< (1)}(\xi) \\
    -J_2^{< (1)}(\xi) & J_1^{< (1)}(\xi)
  \end{bmatrix}
  \begin{bmatrix}
    {\tilde \Sigma}_{q F_{y0} x}^{< (1)}(\xi) \\
    {\tilde \Sigma}_{q F_{y0} y}^{< (1)}(\xi)
  \end{bmatrix}
  + \begin{bmatrix}
    J_2^{< (1)}(\xi) \\
    J_1^{< (1)}(\xi) - [z^{{\rm R} 2}(\xi) I_{02}^{\rm R}(\xi) - I_{22}^{\rm R}(\xi)]/2 - \cc
  \end{bmatrix}, \label{eq:ge<14a} \\
  \begin{bmatrix}
    {\tilde \Sigma}_{q F_{y0} x}^{< (1)}(\xi) \\
    {\tilde \Sigma}_{q F_{y0} y}^{< (1)}(\xi)
  \end{bmatrix}
  = & \frac{n_{\rm i} v_{\rm i}^2}{8 \pi^2 \hbar^3 v^3} \begin{bmatrix}
    |t_{00}^{\rm R}(\xi)|^2 - |t_{0z}^{\rm R}(\xi)|^2
    & i t_{00}^{\rm R}(\xi) t_{0z}^{\rm A}(\xi) - i t_{00}^{\rm A}(\xi) t_{0z}^{\rm R}(\xi) \\
    -i t_{00}^{\rm R}(\xi) t_{0z}^{\rm A}(\xi) + i t_{00}^{\rm A}(\xi) t_{0z}^{\rm R}(\xi)
    & |t_{00}^{\rm R}(\xi)|^2 - |t_{0z}^{\rm R}(\xi)|^2
  \end{bmatrix}
  \begin{bmatrix}
    {\tilde g}_{q F_{y0} x}^{< (1)}(\xi) \\
    {\tilde g}_{q F_{y0} y}^{< (1)}(\xi)
  \end{bmatrix}, \label{eq:ge<14b} \\
  J_1^{< (1)}(\xi)
  \equiv & \int_0^{\Lambda_{\perp}^2} d x \int_{-\Lambda_z}^{\Lambda_z} d y
  \frac{|z^{\rm R}(\xi)|^2 - |y + \eta^{\rm R}(\xi)|^2}{|z^{{\rm R} 2}(\xi) - x - [y + \eta^{\rm R}(\xi)]^2|^2}, \label{eq:ge<14c} \\
  J_2^{< (1)}(\xi)
  \equiv & \int_0^{\Lambda_{\perp}^2} d x \int_{-\Lambda_z}^{\Lambda_z} d y
  \frac{i z^{\rm R}(\xi) [y + \eta^{\rm A}(\xi)] - i z^{\rm A}(\xi) [y + \eta^{\rm R}(\xi)]}{|z^{{\rm R} 2}(\xi) - x - [y + \eta^{\rm R}(\xi)]^2|^2}, \label{eq:ge<14d}
\end{align} \label{eq:ge<14}\end{subequations}
The explicit forms of $J_1^{< (1)}(\xi)$ and $J_2^{< (1)}(\xi)$ are given in Appendix~\ref{app:int}.
Again, the first-order self-energy with respect to a torsional electric field is obtained by $\Sigma_{-T^0_{\phantom{0} y0}}^{< (1)}(\xi) = \xi \Sigma_{F_{y0}}^{< (1)}(\xi)/q$.

\section{Heat Magnetization} \label{app:hm}
We calculate the auxiliary heat magnetization Eq.~\eqref{eq:auxhm1} and the proper one with Eq.~\eqref{eq:transhm}.
Since the heat magnetization depends on temperature through the distribution function $f(\xi)$ in the absence of interactions, Eq.~\eqref{eq:transhm} can be easily solved.
In fact, it is achieved by replacing $f(\xi) \to f^{(-2)}(\xi) \equiv \{-2 \beta \xi \ln (e^{\beta \xi} + 1) - \pi^2/6 + (\beta \xi)^2 - 2 \polyln_2(-e^{\beta \xi})\}/2 (\beta \xi)^2$.
Here $\polyln_2(z)$ is a dilogarithm function defined by
\begin{equation}
  \polyln_2(z)
  \equiv -\int_0^z d u \ln (1 - u)/u, \label{eq:dilog}
\end{equation}
and hence we can check $\partial [\beta^2 f^{(-2)}(\xi)]/\partial \beta = \beta f(\xi)$.
Therefore, the proper heat magnetization is obtained by
\begin{align}
  M_{{\rm Q} z}
  = & -\frac{i \hbar}{\pi q} \sum_{\rho_x = \pm 1} \int d \xi f^{(-2)}(\xi) \xi^2 \int \frac{d^3 p}{(2 \pi \hbar)^3} G_{F_{xy} 0}^{\rm R}(\xi, {\vec p}) + \cc \notag \\
  = & -\frac{i}{4 \pi^3 \hbar^2 v} \int d \xi f^{(-2)}(\xi) \xi^2 {\tilde g}_{q F_{xy} 0}^{\rm R}(\xi) + \cc \label{eq:mz1}
\end{align}

\section{Electric and Thermal (Hall) Conductivities} \label{app:sigma}
We also calculate the Kubo formulas for the electric and thermal (Hall) conductivities Eqs.~\eqref{eq:sigma1} and \eqref{eq:kappa1}.
By using the results in Appendix~\ref{app:ge<1}, we obtain the Hall conductivity
\begin{subequations} \begin{align}
  \sigma^{xy}_{\rm int~II}
  = & -\frac{i \hbar q^2 v^2}{\pi} \sum_{\rho_x = \pm 1} \int d \xi f(\xi)
  \int \frac{d^3 p}{(2 \pi \hbar)^3} \Gamma_0^{{\rm R} 2}(\xi, {\vec p})
  \{z^{\prime {\rm R}}(\xi) [v \rho_x p_z + \eta^{\rm R}(\xi)] - z^{\rm R}(\xi) \eta^{\prime {\rm R}}(\xi)\} + \cc \notag \\
  = & -\frac{i q^2}{4 \pi^3 \hbar^2 v} \int d \xi f(\xi)
  \{[1 - \Sigma_{00}^{\prime {\rm R}}(\xi)] I_{12}^{\rm R}(\xi)
  - z^{\rm R}(\xi) \Sigma_{0z}^{\prime {\rm R}}(\xi) I_{02}^{\rm R}(\xi)\} + \cc, \label{eq:sigmaxy1a} \\
  \sigma^{xy}_{\rm int~I}
  = & -\frac{\hbar q^2 v^2}{\pi} \sum_{\rho_x = \pm 1} \int d \xi f^{\prime}(\xi) \int \frac{d^3 p}{(2 \pi \hbar)^3} \notag \\
  & \times \left(|\Gamma_0^{\rm R}(\xi, {\vec p})|^2 \{2 v^2 p_x p_y + i z^{\rm R}(\xi) [v \rho_x p_z + \eta^{\rm A}(\xi)] - i z^{\rm A}(\xi) [v \rho_x + \eta^{\rm R}(\xi)]\}
  - [\Gamma_0^{{\rm R} 2}(\xi, {\vec p}) v^2 p_x p_y + \cc]\right) \notag \\
  = & \frac{q^2}{4 \pi^3 \hbar^2 v} \int d \xi [-f^{\prime}(\xi)] J_2^{< (1)}(\xi), \label{eq:sigmaxy1b} \\
  \sigma^{xy}_{\rm ext}
  = & \frac{i \hbar q v}{\pi} \sum_{\rho_x = \pm 1} \rho_x \int d \xi f^{\prime}(\xi) \Sigma_{F_{y0} x}^{< (1)}(\xi) \int \frac{d^3 p}{(2 \pi \hbar)^3}
  \left[|\Gamma_0^{\rm R}(\xi, {\vec p})|^2 [|z^{\rm R}(\xi)|^2 + v^2 (p_x^2 - p_y^2) - |v \rho_x p_z + \eta^{\rm R}(\xi)|^2]\right. \notag \\
  & \left.- \left(\Gamma_0^{{\rm R} 2}(\xi, {\vec p}) \{z^{{\rm R} 2}(\xi) + v^2 (p_x^2 - p_y^2) - [v \rho_x p_z + \eta^{\rm R}(\xi)]^2\} + \cc\right)/2\right] \notag \\
  & + \frac{i \hbar q v}{\pi} \sum_{\rho_x = \pm 1} \rho_x \int d \xi f^{\prime}(\xi) \Sigma_{F_{y0} y}^{< (1)}(\xi) \int \frac{d^3 p}{(2 \pi \hbar)^3} \notag \\
  & \times \left(|\Gamma_0^{\rm R}(\xi, {\vec p})|^2 \{2 v^2 p_x p_y + i z^{\rm R}(\xi) [v \rho_x p_z + \eta^{\rm A}(\xi)] - i z^{\rm A}(\xi) [v \rho_x p_z + \eta^{\rm R}(\xi)]\}
  - [\Gamma_0^{{\rm R} 2}(\xi, {\vec p}) v^2 p_x p_y + \cc]\right) \notag \\
  = & \frac{q^2}{4 \pi^3 \hbar^2 v} \int d \xi [-f^{\prime}(\xi)]
  \left({\tilde \Sigma}_{q F_{y0} x}^{< (1)}(\xi) \{J_1^{< (1)}(\xi) - [z^{{\rm R} 2}(\xi) I_{02}^{\rm R}(\xi) - I_{22}^{\rm R}(\xi) + \cc]/2\}
  + {\tilde \Sigma}_{q F_{y0} y}^{< (1)}(\xi) J_2^{< (1)}(\xi)\right), \label{eq:sigmaxy1c}
\end{align} \label{eq:sigmaxy1}\end{subequations}
and the electric conductivity
\begin{subequations} \begin{align}
  \sigma^{yy}_{\rm int~I}
  = & -\frac{\hbar q^2 v^2}{\pi} \sum_{\rho_x = \pm 1} \int d \xi f^{\prime}(\xi) \int \frac{d^3 p}{(2 \pi \hbar)^3}
  \left[|\Gamma_0^{\rm R}(\xi, {\vec p})|^2 [|z^{\rm R}(\xi)|^2 - v^2 (p_x^2 - p_y^2) - |v \rho_x p_z + \eta^{\rm R}(\xi)|^2]\right. \notag \\
  & \left.- \left(\Gamma_0^{{\rm R} 2}(\xi, {\vec p}) \{z^{{\rm R} 2}(\xi) - v^2 (p_x^2 - p_y^2) - [v \rho_x p_z + \eta^{\rm R}(\xi)]^2\} + \cc\right)/2\right] \notag \\
  = & \frac{q^2}{4 \pi^3 \hbar^2 v} \int d \xi [-f^{\prime}(\xi)]
  \{J_1^{< (1)}(\xi) - [z^{{\rm R} 2}(\xi) I_{02}^{\rm R}(\xi) - I_{22}^{\rm R}(\xi) + \cc]/2\}, \label{eq:sigmayy1b} \\
  \sigma^{yy}_{\rm ext}
  = & -\frac{i \hbar q v}{\pi} \sum_{\rho_x = \pm 1} \rho_x \int d \xi f^{\prime}(\xi) \Sigma_{F_{y0} x}^{< (1)}(\xi) \int \frac{d^3 p}{(2 \pi \hbar)^3} \notag \\
  & \times \left(|\Gamma_0^{\rm R}(\xi, {\vec p})|^2 \{-2 v^2 p_x p_y + i z^{\rm R}(\xi) [v \rho_x p_z + \eta^{\rm A}(\xi)] - i z^{\rm A}(\xi) [v \rho_x p_z + \eta^{\rm R}(\xi)]\}
  + [\Gamma_0^{{\rm R} 2}(\xi, {\vec p}) v^2 p_x p_y + \cc]\right) \notag \\
  & + \frac{i \hbar q v}{\pi} \sum_{\rho_x = \pm 1} \rho_x \int d \xi f^{\prime}(\xi) \Sigma_{F_{y0} y}^{< (1)}(\xi) \int \frac{d^3 p}{(2 \pi \hbar)^3}
  \left[|\Gamma_0^{\rm R}(\xi, {\vec p})|^2 [|z^{\rm R}(\xi)|^2 - v^2 (p_x^2 - p_y^2) - |v \rho_x p_z + \eta^{\rm R}(\xi)|^2]\right. \notag \\
  & \left.- \left(\Gamma_0^{{\rm R} 2}(\xi, {\vec p}) \{z^{{\rm R} 2}(\xi) - v^2 (p_x^2 - p_y^2) - [v \rho_x p_z + \eta^{\rm R}(\xi)]^2\} + \cc\right)/2\right] \notag \\
  = & \frac{q^2}{4 \pi^3 \hbar^2 v} \int d \xi [-f^{\prime}(\xi)]
  \left(-{\tilde \Sigma}_{q F_{y0} x}^{< (1)}(\xi) J_2^{< (1)}(\xi)
  + {\tilde \Sigma}_{q F_{y0} y}^{< (1)}(\xi) \{J_1^{< (1)}(\xi) - [z^{{\rm R} 2}(\xi) I_{02}^{\rm R}(\xi) - I_{22}^{\rm R}(\xi) + \cc]/2\}\right). \label{eq:sigmayy1c}
\end{align} \label{eq:sigmayy1}\end{subequations}
The Kubo formula for the thermal (Hall) conductivity is obtained by replacing $q \to \xi$ in the above expressions.

\section{Momentum Integrals} \label{app:int}
We give the explicit forms of the momentum integrals defined above.
Eq.~\eqref{eq:irmn} is expressed by
\begin{subequations} \begin{align}
  I_{01}^{\rm R}(\xi)
  = & -[\sqrt{z^{{\rm R} 2}(\xi) - x} + y + \eta^{\rm R}(\xi)] \ln [\sqrt{z^{{\rm R} 2}(\xi) - x} + y + \eta^{\rm R}(\xi)] \notag \\
  & \left.\left.+ [\sqrt{z^{{\rm R} 2}(\xi) - x} - y - \eta^{\rm R}(\xi)] \ln [\sqrt{z^{{\rm R} 2}(\xi) - x} - y - \eta^{\rm R}(\xi)]
  \right|_{x = 0}^{x = \Lambda_{\perp}^2}\right|_{y = -\Lambda_z}^{y = \Lambda_z}, \label{eq:i01r} \\
  I_{11}^{\rm R}(\xi)
  = & \left.\left.
  -\frac{1}{2} \{[y + \eta^{\rm R}(\xi)]^2 - z^{{\rm R} 2}(\xi) + x\} \ln \{[y + \eta^{\rm R}(\xi)]^2 - z^{{\rm R} 2}(\xi) + x\}
  \right|_{x = 0}^{x = \Lambda_{\perp}^2}\right|_{y = -\Lambda_z}^{y = \Lambda_z}, \label{eq:i11r} \\
  I_{02}^{\rm R}(\xi)
  = & \left.\left.
  \frac{1}{2 \sqrt{z^{{\rm R} 2}(\xi) - x}}
  \ln \frac{y + \eta^{\rm R}(\xi) + \sqrt{z^{{\rm R} 2}(\xi) - x}}{y + \eta^{\rm R}(\xi) - \sqrt{z^{{\rm R} 2}(\xi) - x}}
  \right|_{x = 0}^{x = \Lambda_{\perp}^2}\right|_{y = -\Lambda_z}^{y = \Lambda_z}, \label{eq:i02r} \\
  I_{12}^{\rm R}(\xi)
  = & \left.\left.
  -\frac{1}{2} \ln \{[y + \eta^{\rm R}(\xi)]^2 - z^{{\rm R} 2}(\xi) + x\}
  \right|_{x = 0}^{x = \Lambda_{\perp}^2}\right|_{y = -\Lambda_z}^{y = \Lambda_z}, \label{eq:i12r} \\
  I_{22}^{\rm R}(\xi)
  = & \left.\left.
  \frac{1}{2} \sqrt{z^{{\rm R} 2}(\xi) - x}
  \ln \frac{y + \eta^{\rm R}(\xi) + \sqrt{z^{{\rm R} 2}(\xi) - x}}{y + \eta^{\rm R}(\xi) - \sqrt{z^{{\rm R} 2}(\xi) - x}}
  \right|_{x = 0}^{x = \Lambda_{\perp}^2}\right|_{y = -\Lambda_z}^{y = \Lambda_z}. \label{eq:i22r}
\end{align} \label{eq:ir}\end{subequations}
with $A(\xi, x, y)|_{x = 0}^{x = \Lambda_{\perp}^2}|_{y = -\Lambda_z}^{y = \Lambda_z}
\equiv A(\xi, \Lambda_{\perp}^2, \Lambda_z) - A(\xi, \Lambda_{\perp}^2, -\Lambda_z) - A(\xi, 0, \Lambda_z) + A(\xi, 0, -\Lambda_z)$.
Equations~\eqref{eq:ge<14c} and \eqref{eq:ge<14d} are expressed as
\begin{subequations} \begin{align}
  J_1^{< (1)}(\xi)
  = & \left.\left.a_1(\xi) A(\xi, x, y) + [b_1(\xi) - c_1(\xi) r(\xi)] B(\xi, x, y) + c_1(\xi) C(\xi, x, y)\right|_{x = 0}^{x = \Lambda_{\perp}^2}\right|_{y = -\Lambda_z}^{y = \Lambda_z}, \label{eq:varxi1<1} \\
  J_2^{< (1)}(\xi)
  = & \left.\left.a_2(\xi) A(\xi, x, y) + b_2(\xi) B(\xi, x, y)\right|_{x = 0}^{x = \Lambda_{\perp}^2}\right|_{y = -\Lambda_z}^{y = \Lambda_z}, \label{eq:varxi2<1} \\
  A(\xi, x, y)
  \equiv & -\frac{1}{2 i} \left[\polyln_2\left(\frac{r(\xi) - y}{r(\xi) - q_+^{\rm R}(\xi, x)}\right) + \polyln_2\left(\frac{r(\xi) - y}{r(\xi) - q_-^{\rm R}(\xi, x)}\right)\right] + \cc, \label{eq:A<1} \\
  B(\xi, x, y)
  \equiv & \{[y - q_+^{\rm R}(\xi, x)] \ln [y - q_+^{\rm R}(\xi, x)] + [y - q_-^{\rm R}(\xi, x)] \ln [y - q_-^{\rm R}(\xi, x)]\}/2 i + \cc, \label{eq:B<1} \\
  C(\xi, x, y)
  \equiv & \{[y^2 - q_+^{{\rm R} 2}(\xi, x)] \ln [y - q_+^{\rm R}(\xi, x)] + [y^2 - q_-^{{\rm R} 2}(\xi, x)] \ln [y - q_-^{\rm R}(\xi, x)]\}/4 i + \cc, \label{eq:C<1} \\
  q_{\pm}^{\rm R}(\xi, x)
  \equiv & -\eta^{\rm R}(\xi) \pm \sqrt{z^{{\rm R} 2}(\xi) - x}, \label{eq:q<1} \\
  r(\xi)
  \equiv & [z^{{\rm R} 2}(\xi) - \eta^{{\rm R} 2}(\xi) - z^{{\rm A} 2}(\xi) + \eta^{{\rm A} 2}(\xi)]/2 [\eta^{\rm R}(\xi) - \eta^{\rm A}(\xi)], \label{eq:r<1} \\
  a_1(\xi)
  \equiv & -\{[z^{\rm R}(\xi) + z^{\rm A}(\xi)]^2 + [\eta^{\rm R}(\xi) - \eta^{\rm A}(\xi)]^2\} \notag \\
  & \times [z^{\rm R}(\xi) + \eta^{\rm R}(\xi) - z^{\rm A}(\xi) - \eta^{\rm A}(\xi)] [z^{\rm R}(\xi) - \eta^{\rm R}(\xi) - z^{\rm A}(\xi) + \eta^{\rm A}(\xi)]
  /4 i [\eta^{\rm R}(\xi) - \eta^{\rm A}(\xi)]^3, \label{eq:a1<1} \\
  b_1(\xi)
  \equiv & -[z^{{\rm R} 2}(\xi) - z^{{\rm A} 2}(\xi)]/i [\eta^{\rm R}(\xi) - \eta^{\rm A}(\xi)]^2, \label{eq:b1<1} \\
  c_1(\xi)
  \equiv & -1/i [\eta^{\rm R}(\xi) - \eta^{\rm A}(\xi)], \label{eq:c1<1} \\
  a_2(\xi)
  \equiv & [z^{\rm R}(\xi) + z^{\rm A}(\xi)] \notag \\
  & \times [z^{\rm R}(\xi) + \eta^{\rm R}(\xi) - z^{\rm A}(\xi) - \eta^{\rm A}(\xi)] [z^{\rm R}(\xi) - \eta^{\rm R}(\xi) - z^{\rm A}(\xi) + \eta^{\rm A}(\xi)]
  /2 [\eta^{\rm R}(\xi) - \eta^{\rm A}(\xi)]^2, \label{eq:a2<1} \\
  b_2(\xi)
  \equiv & [z^{\rm R}(\xi) - z^{\rm A}(\xi)]/[\eta^{\rm R}(\xi) - \eta^{\rm A}(\xi)], \label{eq:b2<1}  
\end{align} \label{eq:varxi<1}\end{subequations}
in which $\polyln_2(z)$ is a dilogarithm function defined in Eq.~\eqref{eq:dilog}.
\end{widetext}
%

\begin{thebibliography}{66}%
\makeatletter
\providecommand \@ifxundefined [1]{%
 \@ifx{#1\undefined}
}%
\providecommand \@ifnum [1]{%
 \ifnum #1\expandafter \@firstoftwo
 \else \expandafter \@secondoftwo
 \fi
}%
\providecommand \@ifx [1]{%
 \ifx #1\expandafter \@firstoftwo
 \else \expandafter \@secondoftwo
 \fi
}%
\providecommand \natexlab [1]{#1}%
\providecommand \enquote  [1]{``#1''}%
\providecommand \bibnamefont  [1]{#1}%
\providecommand \bibfnamefont [1]{#1}%
\providecommand \citenamefont [1]{#1}%
\providecommand \href@noop [0]{\@secondoftwo}%
\providecommand \href [0]{\begingroup \@sanitize@url \@href}%
\providecommand \@href[1]{\@@startlink{#1}\@@href}%
\providecommand \@@href[1]{\endgroup#1\@@endlink}%
\providecommand \@sanitize@url [0]{\catcode `\\12\catcode `\$12\catcode
  `\&12\catcode `\#12\catcode `\^12\catcode `\_12\catcode `\%12\relax}%
\providecommand \@@startlink[1]{}%
\providecommand \@@endlink[0]{}%
\providecommand \url  [0]{\begingroup\@sanitize@url \@url }%
\providecommand \@url [1]{\endgroup\@href {#1}{\urlprefix }}%
\providecommand \urlprefix  [0]{URL }%
\providecommand \Eprint [0]{\href }%
\providecommand \doibase [0]{http://dx.doi.org/}%
\providecommand \selectlanguage [0]{\@gobble}%
\providecommand \bibinfo  [0]{\@secondoftwo}%
\providecommand \bibfield  [0]{\@secondoftwo}%
\providecommand \translation [1]{[#1]}%
\providecommand \BibitemOpen [0]{}%
\providecommand \bibitemStop [0]{}%
\providecommand \bibitemNoStop [0]{.\EOS\space}%
\providecommand \EOS [0]{\spacefactor3000\relax}%
\providecommand \BibitemShut  [1]{\csname bibitem#1\endcsname}%
\let\auto@bib@innerbib\@empty
\bibitem [{\citenamefont {Onose}\ \emph {et~al.}(2008)\citenamefont {Onose},
  \citenamefont {Shiomi},\ and\ \citenamefont
  {Tokura}}]{PhysRevLett.100.016601}%
  \BibitemOpen
  \bibfield  {author} {\bibinfo {author} {\bibfnamefont {Y.}~\bibnamefont
  {Onose}}, \bibinfo {author} {\bibfnamefont {Y.}~\bibnamefont {Shiomi}}, \
  and\ \bibinfo {author} {\bibfnamefont {Y.}~\bibnamefont {Tokura}},\ }\href
  {\doibase 10.1103/PhysRevLett.100.016601} {\bibfield  {journal} {\bibinfo
  {journal} {Phys. Rev. Lett.}\ }\textbf {\bibinfo {volume} {100}},\ \bibinfo
  {pages} {016601} (\bibinfo {year} {2008})}\BibitemShut {NoStop}%
\bibitem [{\citenamefont {Shiomi}\ \emph {et~al.}(2010)\citenamefont {Shiomi},
  \citenamefont {Onose},\ and\ \citenamefont {Tokura}}]{PhysRevB.81.054414}%
  \BibitemOpen
  \bibfield  {author} {\bibinfo {author} {\bibfnamefont {Y.}~\bibnamefont
  {Shiomi}}, \bibinfo {author} {\bibfnamefont {Y.}~\bibnamefont {Onose}}, \
  and\ \bibinfo {author} {\bibfnamefont {Y.}~\bibnamefont {Tokura}},\ }\href
  {\doibase 10.1103/PhysRevB.81.054414} {\bibfield  {journal} {\bibinfo
  {journal} {Phys. Rev. B}\ }\textbf {\bibinfo {volume} {81}},\ \bibinfo
  {pages} {054414} (\bibinfo {year} {2010})}\BibitemShut {NoStop}%
\bibitem [{\citenamefont {Shiomi}\ \emph {et~al.}(2009)\citenamefont {Shiomi},
  \citenamefont {Onose},\ and\ \citenamefont {Tokura}}]{PhysRevB.79.100404}%
  \BibitemOpen
  \bibfield  {author} {\bibinfo {author} {\bibfnamefont {Y.}~\bibnamefont
  {Shiomi}}, \bibinfo {author} {\bibfnamefont {Y.}~\bibnamefont {Onose}}, \
  and\ \bibinfo {author} {\bibfnamefont {Y.}~\bibnamefont {Tokura}},\ }\href
  {\doibase 10.1103/PhysRevB.79.100404} {\bibfield  {journal} {\bibinfo
  {journal} {Phys. Rev. B}\ }\textbf {\bibinfo {volume} {79}},\ \bibinfo
  {pages} {100404} (\bibinfo {year} {2009})}\BibitemShut {NoStop}%
\bibitem [{\citenamefont {Onose}\ \emph {et~al.}(2010)\citenamefont {Onose},
  \citenamefont {Ideue}, \citenamefont {Katsura}, \citenamefont {Shiomi},
  \citenamefont {Nagaosa},\ and\ \citenamefont {Tokura}}]{Onose16072010}%
  \BibitemOpen
  \bibfield  {author} {\bibinfo {author} {\bibfnamefont {Y.}~\bibnamefont
  {Onose}}, \bibinfo {author} {\bibfnamefont {T.}~\bibnamefont {Ideue}},
  \bibinfo {author} {\bibfnamefont {H.}~\bibnamefont {Katsura}}, \bibinfo
  {author} {\bibfnamefont {Y.}~\bibnamefont {Shiomi}}, \bibinfo {author}
  {\bibfnamefont {N.}~\bibnamefont {Nagaosa}}, \ and\ \bibinfo {author}
  {\bibfnamefont {Y.}~\bibnamefont {Tokura}},\ }\href {\doibase
  10.1126/science.1188260} {\bibfield  {journal} {\bibinfo  {journal}
  {Science}\ }\textbf {\bibinfo {volume} {329}},\ \bibinfo {pages} {297}
  (\bibinfo {year} {2010})}\BibitemShut {NoStop}%
\bibitem [{\citenamefont {Ideue}\ \emph {et~al.}(2012)\citenamefont {Ideue},
  \citenamefont {Onose}, \citenamefont {Katsura}, \citenamefont {Shiomi},
  \citenamefont {Ishiwata}, \citenamefont {Nagaosa},\ and\ \citenamefont
  {Tokura}}]{PhysRevB.85.134411}%
  \BibitemOpen
  \bibfield  {author} {\bibinfo {author} {\bibfnamefont {T.}~\bibnamefont
  {Ideue}}, \bibinfo {author} {\bibfnamefont {Y.}~\bibnamefont {Onose}},
  \bibinfo {author} {\bibfnamefont {H.}~\bibnamefont {Katsura}}, \bibinfo
  {author} {\bibfnamefont {Y.}~\bibnamefont {Shiomi}}, \bibinfo {author}
  {\bibfnamefont {S.}~\bibnamefont {Ishiwata}}, \bibinfo {author}
  {\bibfnamefont {N.}~\bibnamefont {Nagaosa}}, \ and\ \bibinfo {author}
  {\bibfnamefont {Y.}~\bibnamefont {Tokura}},\ }\href {\doibase
  10.1103/PhysRevB.85.134411} {\bibfield  {journal} {\bibinfo  {journal} {Phys.
  Rev. B}\ }\textbf {\bibinfo {volume} {85}},\ \bibinfo {pages} {134411}
  (\bibinfo {year} {2012})}\BibitemShut {NoStop}%
\bibitem [{\citenamefont {Hirschberger}\ \emph
  {et~al.}(2015{\natexlab{a}})\citenamefont {Hirschberger}, \citenamefont
  {Krizan}, \citenamefont {Cava},\ and\ \citenamefont
  {Ong}}]{Hirschberger03042015}%
  \BibitemOpen
  \bibfield  {author} {\bibinfo {author} {\bibfnamefont {M.}~\bibnamefont
  {Hirschberger}}, \bibinfo {author} {\bibfnamefont {J.~W.}\ \bibnamefont
  {Krizan}}, \bibinfo {author} {\bibfnamefont {R.~J.}\ \bibnamefont {Cava}}, \
  and\ \bibinfo {author} {\bibfnamefont {N.~P.}\ \bibnamefont {Ong}},\ }\href
  {\doibase 10.1126/science.1257340} {\bibfield  {journal} {\bibinfo  {journal}
  {Science}\ }\textbf {\bibinfo {volume} {348}},\ \bibinfo {pages} {106}
  (\bibinfo {year} {2015}{\natexlab{a}})}\BibitemShut {NoStop}%
\bibitem [{\citenamefont {Hirschberger}\ \emph
  {et~al.}(2015{\natexlab{b}})\citenamefont {Hirschberger}, \citenamefont
  {Chisnell}, \citenamefont {Lee},\ and\ \citenamefont
  {Ong}}]{PhysRevLett.115.106603}%
  \BibitemOpen
  \bibfield  {author} {\bibinfo {author} {\bibfnamefont {M.}~\bibnamefont
  {Hirschberger}}, \bibinfo {author} {\bibfnamefont {R.}~\bibnamefont
  {Chisnell}}, \bibinfo {author} {\bibfnamefont {Y.~S.}\ \bibnamefont {Lee}}, \
  and\ \bibinfo {author} {\bibfnamefont {N.~P.}\ \bibnamefont {Ong}},\ }\href
  {\doibase 10.1103/PhysRevLett.115.106603} {\bibfield  {journal} {\bibinfo
  {journal} {Phys. Rev. Lett.}\ }\textbf {\bibinfo {volume} {115}},\ \bibinfo
  {pages} {106603} (\bibinfo {year} {2015}{\natexlab{b}})}\BibitemShut
  {NoStop}%
\bibitem [{\citenamefont {Watanabe}\ \emph {et~al.}(2016)\citenamefont
  {Watanabe}, \citenamefont {Sugii}, \citenamefont {Shimozawa}, \citenamefont
  {Suzuki}, \citenamefont {Yajima}, \citenamefont {Ishikawa}, \citenamefont
  {Hiroi}, \citenamefont {Shibauchi}, \citenamefont {Matsuda},\ and\
  \citenamefont {Yamashita}}]{Watanabe02082016}%
  \BibitemOpen
  \bibfield  {author} {\bibinfo {author} {\bibfnamefont {D.}~\bibnamefont
  {Watanabe}}, \bibinfo {author} {\bibfnamefont {K.}~\bibnamefont {Sugii}},
  \bibinfo {author} {\bibfnamefont {M.}~\bibnamefont {Shimozawa}}, \bibinfo
  {author} {\bibfnamefont {Y.}~\bibnamefont {Suzuki}}, \bibinfo {author}
  {\bibfnamefont {T.}~\bibnamefont {Yajima}}, \bibinfo {author} {\bibfnamefont
  {H.}~\bibnamefont {Ishikawa}}, \bibinfo {author} {\bibfnamefont
  {Z.}~\bibnamefont {Hiroi}}, \bibinfo {author} {\bibfnamefont
  {T.}~\bibnamefont {Shibauchi}}, \bibinfo {author} {\bibfnamefont
  {Y.}~\bibnamefont {Matsuda}}, \ and\ \bibinfo {author} {\bibfnamefont
  {M.}~\bibnamefont {Yamashita}},\ }\href {\doibase 10.1073/pnas.1524076113}
  {\bibfield  {journal} {\bibinfo  {journal} {Proc. Natl. Acad. Sci. USA}\
  }\textbf {\bibinfo {volume} {113}},\ \bibinfo {pages} {8653} (\bibinfo {year}
  {2016})}\BibitemShut {NoStop}%
\bibitem [{\citenamefont {Luttinger}(1964)}]{PhysRev.135.A1505}%
  \BibitemOpen
  \bibfield  {author} {\bibinfo {author} {\bibfnamefont {J.~M.}\ \bibnamefont
  {Luttinger}},\ }\href {\doibase 10.1103/PhysRev.135.A1505} {\bibfield
  {journal} {\bibinfo  {journal} {Phys. Rev.}\ }\textbf {\bibinfo {volume}
  {135}},\ \bibinfo {pages} {A1505} (\bibinfo {year} {1964})}\BibitemShut
  {NoStop}%
\bibitem [{\citenamefont {Smr\v{c}ka}\ and\ \citenamefont
  {St\v{r}eda}(1977)}]{0022-3719-10-12-021}%
  \BibitemOpen
  \bibfield  {author} {\bibinfo {author} {\bibfnamefont {L.}~\bibnamefont
  {Smr\v{c}ka}}\ and\ \bibinfo {author} {\bibfnamefont {P.}~\bibnamefont
  {St\v{r}eda}},\ }\href {\doibase 10.1088/0022-3719/10/12/021} {\bibfield
  {journal} {\bibinfo  {journal} {J. Phys. C: Solid State Phys.}\ }\textbf
  {\bibinfo {volume} {10}},\ \bibinfo {pages} {2153} (\bibinfo {year}
  {1977})}\BibitemShut {NoStop}%
\bibitem [{\citenamefont {Cooper}\ \emph {et~al.}(1997)\citenamefont {Cooper},
  \citenamefont {Halperin},\ and\ \citenamefont {Ruzin}}]{PhysRevB.55.2344}%
  \BibitemOpen
  \bibfield  {author} {\bibinfo {author} {\bibfnamefont {N.~R.}\ \bibnamefont
  {Cooper}}, \bibinfo {author} {\bibfnamefont {B.~I.}\ \bibnamefont
  {Halperin}}, \ and\ \bibinfo {author} {\bibfnamefont {I.~M.}\ \bibnamefont
  {Ruzin}},\ }\href {\doibase 10.1103/PhysRevB.55.2344} {\bibfield  {journal}
  {\bibinfo  {journal} {Phys. Rev. B}\ }\textbf {\bibinfo {volume} {55}},\
  \bibinfo {pages} {2344} (\bibinfo {year} {1997})}\BibitemShut {NoStop}%
\bibitem [{\citenamefont {Matsumoto}\ and\ \citenamefont
  {Murakami}(2011{\natexlab{a}})}]{PhysRevLett.106.197202}%
  \BibitemOpen
  \bibfield  {author} {\bibinfo {author} {\bibfnamefont {R.}~\bibnamefont
  {Matsumoto}}\ and\ \bibinfo {author} {\bibfnamefont {S.}~\bibnamefont
  {Murakami}},\ }\href {\doibase 10.1103/PhysRevLett.106.197202} {\bibfield
  {journal} {\bibinfo  {journal} {Phys. Rev. Lett.}\ }\textbf {\bibinfo
  {volume} {106}},\ \bibinfo {pages} {197202} (\bibinfo {year}
  {2011}{\natexlab{a}})}\BibitemShut {NoStop}%
\bibitem [{\citenamefont {Matsumoto}\ and\ \citenamefont
  {Murakami}(2011{\natexlab{b}})}]{PhysRevB.84.184406}%
  \BibitemOpen
  \bibfield  {author} {\bibinfo {author} {\bibfnamefont {R.}~\bibnamefont
  {Matsumoto}}\ and\ \bibinfo {author} {\bibfnamefont {S.}~\bibnamefont
  {Murakami}},\ }\href {\doibase 10.1103/PhysRevB.84.184406} {\bibfield
  {journal} {\bibinfo  {journal} {Phys. Rev. B}\ }\textbf {\bibinfo {volume}
  {84}},\ \bibinfo {pages} {184406} (\bibinfo {year}
  {2011}{\natexlab{b}})}\BibitemShut {NoStop}%
\bibitem [{\citenamefont {Qin}\ \emph {et~al.}(2011)\citenamefont {Qin},
  \citenamefont {Niu},\ and\ \citenamefont {Shi}}]{PhysRevLett.107.236601}%
  \BibitemOpen
  \bibfield  {author} {\bibinfo {author} {\bibfnamefont {T.}~\bibnamefont
  {Qin}}, \bibinfo {author} {\bibfnamefont {Q.}~\bibnamefont {Niu}}, \ and\
  \bibinfo {author} {\bibfnamefont {J.}~\bibnamefont {Shi}},\ }\href {\doibase
  10.1103/PhysRevLett.107.236601} {\bibfield  {journal} {\bibinfo  {journal}
  {Phys. Rev. Lett.}\ }\textbf {\bibinfo {volume} {107}},\ \bibinfo {pages}
  {236601} (\bibinfo {year} {2011})}\BibitemShut {NoStop}%
\bibitem [{\citenamefont {Shitade}(2014)}]{Shitade01122014}%
  \BibitemOpen
  \bibfield  {author} {\bibinfo {author} {\bibfnamefont {A.}~\bibnamefont
  {Shitade}},\ }\href {\doibase 10.1093/ptep/ptu162} {\bibfield  {journal}
  {\bibinfo  {journal} {Prog. Theor. Exp. Phys.}\ }\textbf {\bibinfo {volume}
  {2014}},\ \bibinfo {pages} {123I01} (\bibinfo {year} {2014})}\BibitemShut
  {NoStop}%
\bibitem [{\citenamefont {Fukuyama}\ \emph {et~al.}(1969)\citenamefont
  {Fukuyama}, \citenamefont {Ebisawa},\ and\ \citenamefont
  {Wada}}]{Fukuyama01091969}%
  \BibitemOpen
  \bibfield  {author} {\bibinfo {author} {\bibfnamefont {H.}~\bibnamefont
  {Fukuyama}}, \bibinfo {author} {\bibfnamefont {H.}~\bibnamefont {Ebisawa}}, \
  and\ \bibinfo {author} {\bibfnamefont {Y.}~\bibnamefont {Wada}},\ }\href
  {\doibase 10.1143/PTP.42.494} {\bibfield  {journal} {\bibinfo  {journal}
  {Prog. Theor. Phys.}\ }\textbf {\bibinfo {volume} {42}},\ \bibinfo {pages}
  {494} (\bibinfo {year} {1969})}\BibitemShut {NoStop}%
\bibitem [{\citenamefont {Fukuyama}(1969)}]{Fukuyama01121969}%
  \BibitemOpen
  \bibfield  {author} {\bibinfo {author} {\bibfnamefont {H.}~\bibnamefont
  {Fukuyama}},\ }\href {\doibase 10.1143/PTP.42.1284} {\bibfield  {journal}
  {\bibinfo  {journal} {Prog. Theor. Phys.}\ }\textbf {\bibinfo {volume}
  {42}},\ \bibinfo {pages} {1284} (\bibinfo {year} {1969})}\BibitemShut
  {NoStop}%
\bibitem [{\citenamefont {Nagaosa}\ \emph {et~al.}(2010)\citenamefont
  {Nagaosa}, \citenamefont {Sinova}, \citenamefont {Onoda}, \citenamefont
  {MacDonald},\ and\ \citenamefont {Ong}}]{RevModPhys.82.1539}%
  \BibitemOpen
  \bibfield  {author} {\bibinfo {author} {\bibfnamefont {N.}~\bibnamefont
  {Nagaosa}}, \bibinfo {author} {\bibfnamefont {J.}~\bibnamefont {Sinova}},
  \bibinfo {author} {\bibfnamefont {S.}~\bibnamefont {Onoda}}, \bibinfo
  {author} {\bibfnamefont {A.~H.}\ \bibnamefont {MacDonald}}, \ and\ \bibinfo
  {author} {\bibfnamefont {N.~P.}\ \bibnamefont {Ong}},\ }\href {\doibase
  10.1103/RevModPhys.82.1539} {\bibfield  {journal} {\bibinfo  {journal} {Rev.
  Mod. Phys.}\ }\textbf {\bibinfo {volume} {82}},\ \bibinfo {pages} {1539}
  (\bibinfo {year} {2010})}\BibitemShut {NoStop}%
\bibitem [{\citenamefont {Karplus}\ and\ \citenamefont
  {Luttinger}(1954)}]{PhysRev.95.1154}%
  \BibitemOpen
  \bibfield  {author} {\bibinfo {author} {\bibfnamefont {R.}~\bibnamefont
  {Karplus}}\ and\ \bibinfo {author} {\bibfnamefont {J.~M.}\ \bibnamefont
  {Luttinger}},\ }\href {\doibase 10.1103/PhysRev.95.1154} {\bibfield
  {journal} {\bibinfo  {journal} {Phys. Rev.}\ }\textbf {\bibinfo {volume}
  {95}},\ \bibinfo {pages} {1154} (\bibinfo {year} {1954})}\BibitemShut
  {NoStop}%
\bibitem [{\citenamefont {Xiao}\ \emph {et~al.}(2010)\citenamefont {Xiao},
  \citenamefont {Chang},\ and\ \citenamefont {Niu}}]{RevModPhys.82.1959}%
  \BibitemOpen
  \bibfield  {author} {\bibinfo {author} {\bibfnamefont {D.}~\bibnamefont
  {Xiao}}, \bibinfo {author} {\bibfnamefont {M.-C.}\ \bibnamefont {Chang}}, \
  and\ \bibinfo {author} {\bibfnamefont {Q.}~\bibnamefont {Niu}},\ }\href
  {\doibase 10.1103/RevModPhys.82.1959} {\bibfield  {journal} {\bibinfo
  {journal} {Rev. Mod. Phys.}\ }\textbf {\bibinfo {volume} {82}},\ \bibinfo
  {pages} {1959} (\bibinfo {year} {2010})}\BibitemShut {NoStop}%
\bibitem [{\citenamefont {Smit}(1955)}]{Smit1955877}%
  \BibitemOpen
  \bibfield  {author} {\bibinfo {author} {\bibfnamefont {J.}~\bibnamefont
  {Smit}},\ }\href {\doibase 10.1016/S0031-8914(55)92596-9} {\bibfield
  {journal} {\bibinfo  {journal} {Physica}\ }\textbf {\bibinfo {volume} {21}},\
  \bibinfo {pages} {877 } (\bibinfo {year} {1955})}\BibitemShut {NoStop}%
\bibitem [{\citenamefont {Smit}(1958)}]{Smit195839}%
  \BibitemOpen
  \bibfield  {author} {\bibinfo {author} {\bibfnamefont {J.}~\bibnamefont
  {Smit}},\ }\href {\doibase 10.1016/S0031-8914(58)93541-9} {\bibfield
  {journal} {\bibinfo  {journal} {Physica}\ }\textbf {\bibinfo {volume} {24}},\
  \bibinfo {pages} {39 } (\bibinfo {year} {1958})}\BibitemShut {NoStop}%
\bibitem [{\citenamefont {Berger}(1970)}]{PhysRevB.2.4559}%
  \BibitemOpen
  \bibfield  {author} {\bibinfo {author} {\bibfnamefont {L.}~\bibnamefont
  {Berger}},\ }\href {\doibase 10.1103/PhysRevB.2.4559} {\bibfield  {journal}
  {\bibinfo  {journal} {Phys. Rev. B}\ }\textbf {\bibinfo {volume} {2}},\
  \bibinfo {pages} {4559} (\bibinfo {year} {1970})}\BibitemShut {NoStop}%
\bibitem [{\citenamefont {Berger}(1972)}]{PhysRevB.5.1862}%
  \BibitemOpen
  \bibfield  {author} {\bibinfo {author} {\bibfnamefont {L.}~\bibnamefont
  {Berger}},\ }\href {\doibase 10.1103/PhysRevB.5.1862} {\bibfield  {journal}
  {\bibinfo  {journal} {Phys. Rev. B}\ }\textbf {\bibinfo {volume} {5}},\
  \bibinfo {pages} {1862} (\bibinfo {year} {1972})}\BibitemShut {NoStop}%
\bibitem [{\citenamefont {Onoda}\ \emph
  {et~al.}(2006{\natexlab{a}})\citenamefont {Onoda}, \citenamefont {Sugimoto},\
  and\ \citenamefont {Nagaosa}}]{PTP.116.61}%
  \BibitemOpen
  \bibfield  {author} {\bibinfo {author} {\bibfnamefont {S.}~\bibnamefont
  {Onoda}}, \bibinfo {author} {\bibfnamefont {N.}~\bibnamefont {Sugimoto}}, \
  and\ \bibinfo {author} {\bibfnamefont {N.}~\bibnamefont {Nagaosa}},\ }\href
  {\doibase 10.1143/PTP.116.61} {\bibfield  {journal} {\bibinfo  {journal}
  {Prog. Theor. Phys.}\ }\textbf {\bibinfo {volume} {116}},\ \bibinfo {pages}
  {61} (\bibinfo {year} {2006}{\natexlab{a}})}\BibitemShut {NoStop}%
\bibitem [{\citenamefont {Sugimoto}\ \emph {et~al.}(2007)\citenamefont
  {Sugimoto}, \citenamefont {Onoda},\ and\ \citenamefont
  {Nagaosa}}]{PTP.117.415}%
  \BibitemOpen
  \bibfield  {author} {\bibinfo {author} {\bibfnamefont {N.}~\bibnamefont
  {Sugimoto}}, \bibinfo {author} {\bibfnamefont {S.}~\bibnamefont {Onoda}}, \
  and\ \bibinfo {author} {\bibfnamefont {N.}~\bibnamefont {Nagaosa}},\ }\href
  {\doibase 10.1143/PTP.117.415} {\bibfield  {journal} {\bibinfo  {journal}
  {Prog. Theor. Phys.}\ }\textbf {\bibinfo {volume} {117}},\ \bibinfo {pages}
  {415} (\bibinfo {year} {2007})}\BibitemShut {NoStop}%
\bibitem [{\citenamefont {Onoda}\ \emph
  {et~al.}(2006{\natexlab{b}})\citenamefont {Onoda}, \citenamefont {Sugimoto},\
  and\ \citenamefont {Nagaosa}}]{PhysRevLett.97.126602}%
  \BibitemOpen
  \bibfield  {author} {\bibinfo {author} {\bibfnamefont {S.}~\bibnamefont
  {Onoda}}, \bibinfo {author} {\bibfnamefont {N.}~\bibnamefont {Sugimoto}}, \
  and\ \bibinfo {author} {\bibfnamefont {N.}~\bibnamefont {Nagaosa}},\ }\href
  {\doibase 10.1103/PhysRevLett.97.126602} {\bibfield  {journal} {\bibinfo
  {journal} {Phys. Rev. Lett.}\ }\textbf {\bibinfo {volume} {97}},\ \bibinfo
  {pages} {126602} (\bibinfo {year} {2006}{\natexlab{b}})}\BibitemShut
  {NoStop}%
\bibitem [{\citenamefont {Onoda}\ \emph {et~al.}(2008)\citenamefont {Onoda},
  \citenamefont {Sugimoto},\ and\ \citenamefont
  {Nagaosa}}]{PhysRevB.77.165103}%
  \BibitemOpen
  \bibfield  {author} {\bibinfo {author} {\bibfnamefont {S.}~\bibnamefont
  {Onoda}}, \bibinfo {author} {\bibfnamefont {N.}~\bibnamefont {Sugimoto}}, \
  and\ \bibinfo {author} {\bibfnamefont {N.}~\bibnamefont {Nagaosa}},\ }\href
  {\doibase 10.1103/PhysRevB.77.165103} {\bibfield  {journal} {\bibinfo
  {journal} {Phys. Rev. B}\ }\textbf {\bibinfo {volume} {77}},\ \bibinfo
  {pages} {165103} (\bibinfo {year} {2008})}\BibitemShut {NoStop}%
\bibitem [{\citenamefont {Kovalev}\ \emph {et~al.}(2009)\citenamefont
  {Kovalev}, \citenamefont {Tserkovnyak}, \citenamefont {V\'{y}born\'{y}},\
  and\ \citenamefont {Sinova}}]{PhysRevB.79.195129}%
  \BibitemOpen
  \bibfield  {author} {\bibinfo {author} {\bibfnamefont {A.~A.}\ \bibnamefont
  {Kovalev}}, \bibinfo {author} {\bibfnamefont {Y.}~\bibnamefont
  {Tserkovnyak}}, \bibinfo {author} {\bibfnamefont {K.}~\bibnamefont
  {V\'{y}born\'{y}}}, \ and\ \bibinfo {author} {\bibfnamefont {J.}~\bibnamefont
  {Sinova}},\ }\href {\doibase 10.1103/PhysRevB.79.195129} {\bibfield
  {journal} {\bibinfo  {journal} {Phys. Rev. B}\ }\textbf {\bibinfo {volume}
  {79}},\ \bibinfo {pages} {195129} (\bibinfo {year} {2009})}\BibitemShut
  {NoStop}%
\bibitem [{\citenamefont {Hosur}\ and\ \citenamefont
  {Qi}(2013)}]{Hosur2013857}%
  \BibitemOpen
  \bibfield  {author} {\bibinfo {author} {\bibfnamefont {P.}~\bibnamefont
  {Hosur}}\ and\ \bibinfo {author} {\bibfnamefont {X.}~\bibnamefont {Qi}},\
  }\href {\doibase 10.1016/j.crhy.2013.10.010} {\bibfield  {journal} {\bibinfo
  {journal} {Comptes Rendus Physique}\ }\textbf {\bibinfo {volume} {14}},\
  \bibinfo {pages} {857 } (\bibinfo {year} {2013})}\BibitemShut {NoStop}%
\bibitem [{\citenamefont {Burkov}(2015)}]{0953-8984-27-11-113201}%
  \BibitemOpen
  \bibfield  {author} {\bibinfo {author} {\bibfnamefont {A.~A.}\ \bibnamefont
  {Burkov}},\ }\href {\doibase 10.1088/0953-8984/27/11/113201} {\bibfield
  {journal} {\bibinfo  {journal} {J. Phys.: Condens. Matter}\ }\textbf
  {\bibinfo {volume} {27}},\ \bibinfo {pages} {113201} (\bibinfo {year}
  {2015})}\BibitemShut {NoStop}%
\bibitem [{\citenamefont {Wan}\ \emph {et~al.}(2011)\citenamefont {Wan},
  \citenamefont {Turner}, \citenamefont {Vishwanath},\ and\ \citenamefont
  {Savrasov}}]{PhysRevB.83.205101}%
  \BibitemOpen
  \bibfield  {author} {\bibinfo {author} {\bibfnamefont {X.}~\bibnamefont
  {Wan}}, \bibinfo {author} {\bibfnamefont {A.~M.}\ \bibnamefont {Turner}},
  \bibinfo {author} {\bibfnamefont {A.}~\bibnamefont {Vishwanath}}, \ and\
  \bibinfo {author} {\bibfnamefont {S.~Y.}\ \bibnamefont {Savrasov}},\ }\href
  {\doibase 10.1103/PhysRevB.83.205101} {\bibfield  {journal} {\bibinfo
  {journal} {Phys. Rev. B}\ }\textbf {\bibinfo {volume} {83}},\ \bibinfo
  {pages} {205101} (\bibinfo {year} {2011})}\BibitemShut {NoStop}%
\bibitem [{\citenamefont {Burkov}\ and\ \citenamefont
  {Balents}(2011)}]{PhysRevLett.107.127205}%
  \BibitemOpen
  \bibfield  {author} {\bibinfo {author} {\bibfnamefont {A.~A.}\ \bibnamefont
  {Burkov}}\ and\ \bibinfo {author} {\bibfnamefont {L.}~\bibnamefont
  {Balents}},\ }\href {\doibase 10.1103/PhysRevLett.107.127205} {\bibfield
  {journal} {\bibinfo  {journal} {Phys. Rev. Lett.}\ }\textbf {\bibinfo
  {volume} {107}},\ \bibinfo {pages} {127205} (\bibinfo {year}
  {2011})}\BibitemShut {NoStop}%
\bibitem [{\citenamefont {Murakami}(2007)}]{1367-2630-9-9-356}%
  \BibitemOpen
  \bibfield  {author} {\bibinfo {author} {\bibfnamefont {S.}~\bibnamefont
  {Murakami}},\ }\href {\doibase 10.1088/1367-2630/9/9/356} {\bibfield
  {journal} {\bibinfo  {journal} {New J. Phys.}\ }\textbf {\bibinfo {volume}
  {9}},\ \bibinfo {pages} {356} (\bibinfo {year} {2007})}\BibitemShut {NoStop}%
\bibitem [{\citenamefont {Hal\'asz}\ and\ \citenamefont
  {Balents}(2012)}]{PhysRevB.85.035103}%
  \BibitemOpen
  \bibfield  {author} {\bibinfo {author} {\bibfnamefont {G.~B.}\ \bibnamefont
  {Hal\'asz}}\ and\ \bibinfo {author} {\bibfnamefont {L.}~\bibnamefont
  {Balents}},\ }\href {\doibase 10.1103/PhysRevB.85.035103} {\bibfield
  {journal} {\bibinfo  {journal} {Phys. Rev. B}\ }\textbf {\bibinfo {volume}
  {85}},\ \bibinfo {pages} {035103} (\bibinfo {year} {2012})}\BibitemShut
  {NoStop}%
\bibitem [{\citenamefont {Zyuzin}\ and\ \citenamefont
  {Burkov}(2012)}]{PhysRevB.86.115133}%
  \BibitemOpen
  \bibfield  {author} {\bibinfo {author} {\bibfnamefont {A.~A.}\ \bibnamefont
  {Zyuzin}}\ and\ \bibinfo {author} {\bibfnamefont {A.~A.}\ \bibnamefont
  {Burkov}},\ }\href {\doibase 10.1103/PhysRevB.86.115133} {\bibfield
  {journal} {\bibinfo  {journal} {Phys. Rev. B}\ }\textbf {\bibinfo {volume}
  {86}},\ \bibinfo {pages} {115133} (\bibinfo {year} {2012})}\BibitemShut
  {NoStop}%
\bibitem [{\citenamefont {Goswami}\ and\ \citenamefont
  {Tewari}(2013)}]{PhysRevB.88.245107}%
  \BibitemOpen
  \bibfield  {author} {\bibinfo {author} {\bibfnamefont {P.}~\bibnamefont
  {Goswami}}\ and\ \bibinfo {author} {\bibfnamefont {S.}~\bibnamefont
  {Tewari}},\ }\href {\doibase 10.1103/PhysRevB.88.245107} {\bibfield
  {journal} {\bibinfo  {journal} {Phys. Rev. B}\ }\textbf {\bibinfo {volume}
  {88}},\ \bibinfo {pages} {245107} (\bibinfo {year} {2013})}\BibitemShut
  {NoStop}%
\bibitem [{\citenamefont {Burkov}(2014)}]{PhysRevLett.113.187202}%
  \BibitemOpen
  \bibfield  {author} {\bibinfo {author} {\bibfnamefont {A.~A.}\ \bibnamefont
  {Burkov}},\ }\href {\doibase 10.1103/PhysRevLett.113.187202} {\bibfield
  {journal} {\bibinfo  {journal} {Phys. Rev. Lett.}\ }\textbf {\bibinfo
  {volume} {113}},\ \bibinfo {pages} {187202} (\bibinfo {year}
  {2014})}\BibitemShut {NoStop}%
\bibitem [{\citenamefont {Lundgren}\ \emph {et~al.}(2014)\citenamefont
  {Lundgren}, \citenamefont {Laurell},\ and\ \citenamefont
  {Fiete}}]{PhysRevB.90.165115}%
  \BibitemOpen
  \bibfield  {author} {\bibinfo {author} {\bibfnamefont {R.}~\bibnamefont
  {Lundgren}}, \bibinfo {author} {\bibfnamefont {P.}~\bibnamefont {Laurell}}, \
  and\ \bibinfo {author} {\bibfnamefont {G.~A.}\ \bibnamefont {Fiete}},\ }\href
  {\doibase 10.1103/PhysRevB.90.165115} {\bibfield  {journal} {\bibinfo
  {journal} {Phys. Rev. B}\ }\textbf {\bibinfo {volume} {90}},\ \bibinfo
  {pages} {165115} (\bibinfo {year} {2014})}\BibitemShut {NoStop}%
\bibitem [{\citenamefont {Sharma}\ \emph {et~al.}(2016)\citenamefont {Sharma},
  \citenamefont {Goswami},\ and\ \citenamefont {Tewari}}]{PhysRevB.93.035116}%
  \BibitemOpen
  \bibfield  {author} {\bibinfo {author} {\bibfnamefont {G.}~\bibnamefont
  {Sharma}}, \bibinfo {author} {\bibfnamefont {P.}~\bibnamefont {Goswami}}, \
  and\ \bibinfo {author} {\bibfnamefont {S.}~\bibnamefont {Tewari}},\ }\href
  {\doibase 10.1103/PhysRevB.93.035116} {\bibfield  {journal} {\bibinfo
  {journal} {Phys. Rev. B}\ }\textbf {\bibinfo {volume} {93}},\ \bibinfo
  {pages} {035116} (\bibinfo {year} {2016})}\BibitemShut {NoStop}%
\bibitem [{\citenamefont {Shi}\ \emph {et~al.}(2007)\citenamefont {Shi},
  \citenamefont {Vignale}, \citenamefont {Xiao},\ and\ \citenamefont
  {Niu}}]{PhysRevLett.99.197202}%
  \BibitemOpen
  \bibfield  {author} {\bibinfo {author} {\bibfnamefont {J.}~\bibnamefont
  {Shi}}, \bibinfo {author} {\bibfnamefont {G.}~\bibnamefont {Vignale}},
  \bibinfo {author} {\bibfnamefont {D.}~\bibnamefont {Xiao}}, \ and\ \bibinfo
  {author} {\bibfnamefont {Q.}~\bibnamefont {Niu}},\ }\href {\doibase
  10.1103/PhysRevLett.99.197202} {\bibfield  {journal} {\bibinfo  {journal}
  {Phys. Rev. Lett.}\ }\textbf {\bibinfo {volume} {99}},\ \bibinfo {pages}
  {197202} (\bibinfo {year} {2007})}\BibitemShut {NoStop}%
\bibitem [{\citenamefont {Chen}\ and\ \citenamefont
  {Lee}(2011)}]{PhysRevB.84.205137}%
  \BibitemOpen
  \bibfield  {author} {\bibinfo {author} {\bibfnamefont {K.-T.}\ \bibnamefont
  {Chen}}\ and\ \bibinfo {author} {\bibfnamefont {P.~A.}\ \bibnamefont {Lee}},\
  }\href {\doibase 10.1103/PhysRevB.84.205137} {\bibfield  {journal} {\bibinfo
  {journal} {Phys. Rev. B}\ }\textbf {\bibinfo {volume} {84}},\ \bibinfo
  {pages} {205137} (\bibinfo {year} {2011})}\BibitemShut {NoStop}%
\bibitem [{\citenamefont {Zhu}\ \emph {et~al.}(2012)\citenamefont {Zhu},
  \citenamefont {Yang}, \citenamefont {Fang}, \citenamefont {Liu},\ and\
  \citenamefont {Yao}}]{PhysRevB.86.214415}%
  \BibitemOpen
  \bibfield  {author} {\bibinfo {author} {\bibfnamefont {G.}~\bibnamefont
  {Zhu}}, \bibinfo {author} {\bibfnamefont {S.~A.}\ \bibnamefont {Yang}},
  \bibinfo {author} {\bibfnamefont {C.}~\bibnamefont {Fang}}, \bibinfo {author}
  {\bibfnamefont {W.~M.}\ \bibnamefont {Liu}}, \ and\ \bibinfo {author}
  {\bibfnamefont {Y.}~\bibnamefont {Yao}},\ }\href {\doibase
  10.1103/PhysRevB.86.214415} {\bibfield  {journal} {\bibinfo  {journal} {Phys.
  Rev. B}\ }\textbf {\bibinfo {volume} {86}},\ \bibinfo {pages} {214415}
  (\bibinfo {year} {2012})}\BibitemShut {NoStop}%
\bibitem [{\citenamefont {Nourafkan}\ \emph {et~al.}(2014)\citenamefont
  {Nourafkan}, \citenamefont {Kotliar},\ and\ \citenamefont
  {Tremblay}}]{PhysRevB.90.125132}%
  \BibitemOpen
  \bibfield  {author} {\bibinfo {author} {\bibfnamefont {R.}~\bibnamefont
  {Nourafkan}}, \bibinfo {author} {\bibfnamefont {G.}~\bibnamefont {Kotliar}},
  \ and\ \bibinfo {author} {\bibfnamefont {A.-M.~S.}\ \bibnamefont
  {Tremblay}},\ }\href {\doibase 10.1103/PhysRevB.90.125132} {\bibfield
  {journal} {\bibinfo  {journal} {Phys. Rev. B}\ }\textbf {\bibinfo {volume}
  {90}},\ \bibinfo {pages} {125132} (\bibinfo {year} {2014})}\BibitemShut
  {NoStop}%
\bibitem [{\citenamefont {Hughes}\ \emph {et~al.}(2011)\citenamefont {Hughes},
  \citenamefont {Leigh},\ and\ \citenamefont
  {Fradkin}}]{PhysRevLett.107.075502}%
  \BibitemOpen
  \bibfield  {author} {\bibinfo {author} {\bibfnamefont {T.~L.}\ \bibnamefont
  {Hughes}}, \bibinfo {author} {\bibfnamefont {R.~G.}\ \bibnamefont {Leigh}}, \
  and\ \bibinfo {author} {\bibfnamefont {E.}~\bibnamefont {Fradkin}},\ }\href
  {\doibase 10.1103/PhysRevLett.107.075502} {\bibfield  {journal} {\bibinfo
  {journal} {Phys. Rev. Lett.}\ }\textbf {\bibinfo {volume} {107}},\ \bibinfo
  {pages} {075502} (\bibinfo {year} {2011})}\BibitemShut {NoStop}%
\bibitem [{\citenamefont {Hidaka}\ \emph {et~al.}(2013)\citenamefont {Hidaka},
  \citenamefont {Hirono}, \citenamefont {Kimura},\ and\ \citenamefont
  {Minami}}]{Hidaka01012013}%
  \BibitemOpen
  \bibfield  {author} {\bibinfo {author} {\bibfnamefont {Y.}~\bibnamefont
  {Hidaka}}, \bibinfo {author} {\bibfnamefont {Y.}~\bibnamefont {Hirono}},
  \bibinfo {author} {\bibfnamefont {T.}~\bibnamefont {Kimura}}, \ and\ \bibinfo
  {author} {\bibfnamefont {Y.}~\bibnamefont {Minami}},\ }\href {\doibase
  10.1093/ptep/pts063} {\bibfield  {journal} {\bibinfo  {journal} {Prog. Theor.
  Exp. Phys.}\ }\textbf {\bibinfo {volume} {2013}},\ \bibinfo {pages} {013A02}
  (\bibinfo {year} {2013})}\BibitemShut {NoStop}%
\bibitem [{\citenamefont {Hughes}\ \emph {et~al.}(2013)\citenamefont {Hughes},
  \citenamefont {Leigh},\ and\ \citenamefont {Parrikar}}]{PhysRevD.88.025040}%
  \BibitemOpen
  \bibfield  {author} {\bibinfo {author} {\bibfnamefont {T.~L.}\ \bibnamefont
  {Hughes}}, \bibinfo {author} {\bibfnamefont {R.~G.}\ \bibnamefont {Leigh}}, \
  and\ \bibinfo {author} {\bibfnamefont {O.}~\bibnamefont {Parrikar}},\ }\href
  {\doibase 10.1103/PhysRevD.88.025040} {\bibfield  {journal} {\bibinfo
  {journal} {Phys. Rev. D}\ }\textbf {\bibinfo {volume} {88}},\ \bibinfo
  {pages} {025040} (\bibinfo {year} {2013})}\BibitemShut {NoStop}%
\bibitem [{\citenamefont {Parrikar}\ \emph {et~al.}(2014)\citenamefont
  {Parrikar}, \citenamefont {Hughes},\ and\ \citenamefont
  {Leigh}}]{PhysRevD.90.105004}%
  \BibitemOpen
  \bibfield  {author} {\bibinfo {author} {\bibfnamefont {O.}~\bibnamefont
  {Parrikar}}, \bibinfo {author} {\bibfnamefont {T.~L.}\ \bibnamefont
  {Hughes}}, \ and\ \bibinfo {author} {\bibfnamefont {R.~G.}\ \bibnamefont
  {Leigh}},\ }\href {\doibase 10.1103/PhysRevD.90.105004} {\bibfield  {journal}
  {\bibinfo  {journal} {Phys. Rev. D}\ }\textbf {\bibinfo {volume} {90}},\
  \bibinfo {pages} {105004} (\bibinfo {year} {2014})}\BibitemShut {NoStop}%
\bibitem [{\citenamefont {Shitade}\ and\ \citenamefont
  {Kimura}(2014)}]{PhysRevB.90.134510}%
  \BibitemOpen
  \bibfield  {author} {\bibinfo {author} {\bibfnamefont {A.}~\bibnamefont
  {Shitade}}\ and\ \bibinfo {author} {\bibfnamefont {T.}~\bibnamefont
  {Kimura}},\ }\href {\doibase 10.1103/PhysRevB.90.134510} {\bibfield
  {journal} {\bibinfo  {journal} {Phys. Rev. B}\ }\textbf {\bibinfo {volume}
  {90}},\ \bibinfo {pages} {134510} (\bibinfo {year} {2014})}\BibitemShut
  {NoStop}%
\bibitem [{\citenamefont {Sumiyoshi}\ and\ \citenamefont
  {Fujimoto}(2016)}]{PhysRevLett.116.166601}%
  \BibitemOpen
  \bibfield  {author} {\bibinfo {author} {\bibfnamefont {H.}~\bibnamefont
  {Sumiyoshi}}\ and\ \bibinfo {author} {\bibfnamefont {S.}~\bibnamefont
  {Fujimoto}},\ }\href {\doibase 10.1103/PhysRevLett.116.166601} {\bibfield
  {journal} {\bibinfo  {journal} {Phys. Rev. Lett.}\ }\textbf {\bibinfo
  {volume} {116}},\ \bibinfo {pages} {166601} (\bibinfo {year}
  {2016})}\BibitemShut {NoStop}%
\bibitem [{\citenamefont {Rammer}(2007)}]{9780521874991}%
  \BibitemOpen
  \bibfield  {author} {\bibinfo {author} {\bibfnamefont {J.}~\bibnamefont
  {Rammer}},\ }\href@noop {} {\emph {\bibinfo {title} {Quantum Field Theory of
  Non-equilibrium States}}}\ (\bibinfo  {publisher} {Cambridge University
  Press},\ \bibinfo {address} {New York},\ \bibinfo {year} {2007})\BibitemShut
  {NoStop}%
\bibitem [{\citenamefont {Kamenev}(2011)}]{9780521760829}%
  \BibitemOpen
  \bibfield  {author} {\bibinfo {author} {\bibfnamefont {A.}~\bibnamefont
  {Kamenev}},\ }\href@noop {} {\emph {\bibinfo {title} {Field Theory of
  Non-Equilibrium Systems}}}\ (\bibinfo  {publisher} {Cambridge University
  Press},\ \bibinfo {address} {New York},\ \bibinfo {year} {2011})\BibitemShut
  {NoStop}%
\bibitem [{\citenamefont {Birrell}\ and\ \citenamefont
  {Davies}(1984)}]{9780521278584}%
  \BibitemOpen
  \bibfield  {author} {\bibinfo {author} {\bibfnamefont {N.~D.}\ \bibnamefont
  {Birrell}}\ and\ \bibinfo {author} {\bibfnamefont {P.~C.~W.}\ \bibnamefont
  {Davies}},\ }\href@noop {} {\emph {\bibinfo {title} {Quantum Fields in Curved
  Space}}}\ (\bibinfo  {publisher} {Cambridge University Press},\ \bibinfo
  {address} {New York},\ \bibinfo {year} {1984})\BibitemShut {NoStop}%
\bibitem [{\citenamefont {Parker}\ and\ \citenamefont
  {Toms}(2009)}]{9780521877879}%
  \BibitemOpen
  \bibfield  {author} {\bibinfo {author} {\bibfnamefont {L.}~\bibnamefont
  {Parker}}\ and\ \bibinfo {author} {\bibfnamefont {D.}~\bibnamefont {Toms}},\
  }\href@noop {} {\emph {\bibinfo {title} {Quantum Field Theory in Curved
  Spacetime: Quantized Fields and Gravity}}}\ (\bibinfo  {publisher} {Cambridge
  University Press},\ \bibinfo {address} {New York},\ \bibinfo {year}
  {2009})\BibitemShut {NoStop}%
\bibitem [{\citenamefont {Michaeli}\ and\ \citenamefont
  {Finkel'stein}(2009{\natexlab{a}})}]{PhysRevB.80.115111}%
  \BibitemOpen
  \bibfield  {author} {\bibinfo {author} {\bibfnamefont {K.}~\bibnamefont
  {Michaeli}}\ and\ \bibinfo {author} {\bibfnamefont {A.~M.}\ \bibnamefont
  {Finkel'stein}},\ }\href {\doibase 10.1103/PhysRevB.80.115111} {\bibfield
  {journal} {\bibinfo  {journal} {Phys. Rev. B}\ }\textbf {\bibinfo {volume}
  {80}},\ \bibinfo {pages} {115111} (\bibinfo {year}
  {2009}{\natexlab{a}})}\BibitemShut {NoStop}%
\bibitem [{\citenamefont {Michaeli}\ and\ \citenamefont
  {Finkel'stein}(2009{\natexlab{b}})}]{PhysRevB.80.214516}%
  \BibitemOpen
  \bibfield  {author} {\bibinfo {author} {\bibfnamefont {K.}~\bibnamefont
  {Michaeli}}\ and\ \bibinfo {author} {\bibfnamefont {A.~M.}\ \bibnamefont
  {Finkel'stein}},\ }\href {\doibase 10.1103/PhysRevB.80.214516} {\bibfield
  {journal} {\bibinfo  {journal} {Phys. Rev. B}\ }\textbf {\bibinfo {volume}
  {80}},\ \bibinfo {pages} {214516} (\bibinfo {year}
  {2009}{\natexlab{b}})}\BibitemShut {NoStop}%
\bibitem [{\citenamefont {Fujimoto}\ and\ \citenamefont
  {Kohno}(2014)}]{PhysRevB.90.214418}%
  \BibitemOpen
  \bibfield  {author} {\bibinfo {author} {\bibfnamefont {J.}~\bibnamefont
  {Fujimoto}}\ and\ \bibinfo {author} {\bibfnamefont {H.}~\bibnamefont
  {Kohno}},\ }\href {\doibase 10.1103/PhysRevB.90.214418} {\bibfield  {journal}
  {\bibinfo  {journal} {Phys. Rev. B}\ }\textbf {\bibinfo {volume} {90}},\
  \bibinfo {pages} {214418} (\bibinfo {year} {2014})}\BibitemShut {NoStop}%
\bibitem [{\citenamefont {Fradkin}(1986)}]{PhysRevB.33.3263}%
  \BibitemOpen
  \bibfield  {author} {\bibinfo {author} {\bibfnamefont {E.}~\bibnamefont
  {Fradkin}},\ }\href {\doibase 10.1103/PhysRevB.33.3263} {\bibfield  {journal}
  {\bibinfo  {journal} {Phys. Rev. B}\ }\textbf {\bibinfo {volume} {33}},\
  \bibinfo {pages} {3263} (\bibinfo {year} {1986})}\BibitemShut {NoStop}%
\bibitem [{\citenamefont {Chen}\ \emph {et~al.}(2015)\citenamefont {Chen},
  \citenamefont {Song}, \citenamefont {Jiang}, \citenamefont {Sun},
  \citenamefont {Wang},\ and\ \citenamefont {Xie}}]{PhysRevLett.115.246603}%
  \BibitemOpen
  \bibfield  {author} {\bibinfo {author} {\bibfnamefont {C.-Z.}\ \bibnamefont
  {Chen}}, \bibinfo {author} {\bibfnamefont {J.}~\bibnamefont {Song}}, \bibinfo
  {author} {\bibfnamefont {H.}~\bibnamefont {Jiang}}, \bibinfo {author}
  {\bibfnamefont {Q.-f.}\ \bibnamefont {Sun}}, \bibinfo {author} {\bibfnamefont
  {Z.}~\bibnamefont {Wang}}, \ and\ \bibinfo {author} {\bibfnamefont {X.~C.}\
  \bibnamefont {Xie}},\ }\href {\doibase 10.1103/PhysRevLett.115.246603}
  {\bibfield  {journal} {\bibinfo  {journal} {Phys. Rev. Lett.}\ }\textbf
  {\bibinfo {volume} {115}},\ \bibinfo {pages} {246603} (\bibinfo {year}
  {2015})}\BibitemShut {NoStop}%
\bibitem [{\citenamefont {Liu}\ \emph {et~al.}(2016)\citenamefont {Liu},
  \citenamefont {Ohtsuki},\ and\ \citenamefont
  {Shindou}}]{PhysRevLett.116.066401}%
  \BibitemOpen
  \bibfield  {author} {\bibinfo {author} {\bibfnamefont {S.}~\bibnamefont
  {Liu}}, \bibinfo {author} {\bibfnamefont {T.}~\bibnamefont {Ohtsuki}}, \ and\
  \bibinfo {author} {\bibfnamefont {R.}~\bibnamefont {Shindou}},\ }\href
  {\doibase 10.1103/PhysRevLett.116.066401} {\bibfield  {journal} {\bibinfo
  {journal} {Phys. Rev. Lett.}\ }\textbf {\bibinfo {volume} {116}},\ \bibinfo
  {pages} {066401} (\bibinfo {year} {2016})}\BibitemShut {NoStop}%
\bibitem [{\citenamefont {Miyasato}\ \emph {et~al.}(2007)\citenamefont
  {Miyasato}, \citenamefont {Abe}, \citenamefont {Fujii}, \citenamefont
  {Asamitsu}, \citenamefont {Onoda}, \citenamefont {Onose}, \citenamefont
  {Nagaosa},\ and\ \citenamefont {Tokura}}]{PhysRevLett.99.086602}%
  \BibitemOpen
  \bibfield  {author} {\bibinfo {author} {\bibfnamefont {T.}~\bibnamefont
  {Miyasato}}, \bibinfo {author} {\bibfnamefont {N.}~\bibnamefont {Abe}},
  \bibinfo {author} {\bibfnamefont {T.}~\bibnamefont {Fujii}}, \bibinfo
  {author} {\bibfnamefont {A.}~\bibnamefont {Asamitsu}}, \bibinfo {author}
  {\bibfnamefont {S.}~\bibnamefont {Onoda}}, \bibinfo {author} {\bibfnamefont
  {Y.}~\bibnamefont {Onose}}, \bibinfo {author} {\bibfnamefont
  {N.}~\bibnamefont {Nagaosa}}, \ and\ \bibinfo {author} {\bibfnamefont
  {Y.}~\bibnamefont {Tokura}},\ }\href {\doibase 10.1103/PhysRevLett.99.086602}
  {\bibfield  {journal} {\bibinfo  {journal} {Phys. Rev. Lett.}\ }\textbf
  {\bibinfo {volume} {99}},\ \bibinfo {pages} {086602} (\bibinfo {year}
  {2007})}\BibitemShut {NoStop}%
\bibitem [{\citenamefont {Tian}\ \emph {et~al.}(2009)\citenamefont {Tian},
  \citenamefont {Ye},\ and\ \citenamefont {Jin}}]{PhysRevLett.103.087206}%
  \BibitemOpen
  \bibfield  {author} {\bibinfo {author} {\bibfnamefont {Y.}~\bibnamefont
  {Tian}}, \bibinfo {author} {\bibfnamefont {L.}~\bibnamefont {Ye}}, \ and\
  \bibinfo {author} {\bibfnamefont {X.}~\bibnamefont {Jin}},\ }\href {\doibase
  10.1103/PhysRevLett.103.087206} {\bibfield  {journal} {\bibinfo  {journal}
  {Phys. Rev. Lett.}\ }\textbf {\bibinfo {volume} {103}},\ \bibinfo {pages}
  {087206} (\bibinfo {year} {2009})}\BibitemShut {NoStop}%
\bibitem [{\citenamefont {Ye}\ \emph {et~al.}(2012)\citenamefont {Ye},
  \citenamefont {Tian}, \citenamefont {Jin},\ and\ \citenamefont
  {Xiao}}]{PhysRevB.85.220403}%
  \BibitemOpen
  \bibfield  {author} {\bibinfo {author} {\bibfnamefont {L.}~\bibnamefont
  {Ye}}, \bibinfo {author} {\bibfnamefont {Y.}~\bibnamefont {Tian}}, \bibinfo
  {author} {\bibfnamefont {X.}~\bibnamefont {Jin}}, \ and\ \bibinfo {author}
  {\bibfnamefont {D.}~\bibnamefont {Xiao}},\ }\href {\doibase
  10.1103/PhysRevB.85.220403} {\bibfield  {journal} {\bibinfo  {journal} {Phys.
  Rev. B}\ }\textbf {\bibinfo {volume} {85}},\ \bibinfo {pages} {220403}
  (\bibinfo {year} {2012})}\BibitemShut {NoStop}%
\bibitem [{\citenamefont {Hou}\ \emph {et~al.}(2012)\citenamefont {Hou},
  \citenamefont {Li}, \citenamefont {Wei}, \citenamefont {Tian}, \citenamefont
  {Wu},\ and\ \citenamefont {Jin}}]{0953-8984-24-48-482001}%
  \BibitemOpen
  \bibfield  {author} {\bibinfo {author} {\bibfnamefont {D.}~\bibnamefont
  {Hou}}, \bibinfo {author} {\bibfnamefont {Y.}~\bibnamefont {Li}}, \bibinfo
  {author} {\bibfnamefont {D.}~\bibnamefont {Wei}}, \bibinfo {author}
  {\bibfnamefont {D.}~\bibnamefont {Tian}}, \bibinfo {author} {\bibfnamefont
  {L.}~\bibnamefont {Wu}}, \ and\ \bibinfo {author} {\bibfnamefont
  {X.}~\bibnamefont {Jin}},\ }\href {\doibase 10.1088/0953-8984/24/48/482001}
  {\bibfield  {journal} {\bibinfo  {journal} {J. Phys.: Condens. Matter}\
  }\textbf {\bibinfo {volume} {24}},\ \bibinfo {pages} {482001} (\bibinfo
  {year} {2012})}\BibitemShut {NoStop}%
\bibitem [{\citenamefont {Shitade}\ and\ \citenamefont
  {Nagaosa}(2012)}]{JPSJ.81.083704}%
  \BibitemOpen
  \bibfield  {author} {\bibinfo {author} {\bibfnamefont {A.}~\bibnamefont
  {Shitade}}\ and\ \bibinfo {author} {\bibfnamefont {N.}~\bibnamefont
  {Nagaosa}},\ }\href {\doibase 10.1143/JPSJ.81.083704} {\bibfield  {journal}
  {\bibinfo  {journal} {J. Phys. Soc. Jpn.}\ }\textbf {\bibinfo {volume}
  {81}},\ \bibinfo {pages} {083704} (\bibinfo {year} {2012})}\BibitemShut
  {NoStop}%
\bibitem [{\citenamefont {Kontsevich}(2003)}]{kontsevich2003}%
  \BibitemOpen
  \bibfield  {author} {\bibinfo {author} {\bibfnamefont {M.}~\bibnamefont
  {Kontsevich}},\ }\href {\doibase 10.1023/B:MATH.0000027508.00421.bf}
  {\bibfield  {journal} {\bibinfo  {journal} {Lett. Math. Phys.}\ }\textbf
  {\bibinfo {volume} {66}},\ \bibinfo {pages} {157} (\bibinfo {year}
  {2003})}\BibitemShut {NoStop}%
\end{thebibliography}
\end{document}